\UseRawInputEncoding
\documentclass[11pt,a4paper]{article}
\usepackage{jcappub}

\usepackage{aasmacros}

\usepackage[english]{babel}               

\usepackage{lmodern}        
\usepackage[T1]{fontenc}    
\usepackage{mathpazo}                                            

\usepackage{tabu}

\usepackage{geometry}     
\usepackage[final, babel]{microtype} 

\usepackage[textwidth=2.4cm]{todonotes}%
\usepackage{doi}
\usepackage{hyperref}
\hypersetup{
  colorlinks=true,        
  linkcolor=blue,         
  citecolor=cyan,         
}

\usepackage{graphicx}
\usepackage{dcolumn}
\usepackage{bm}
\usepackage{color}

\usepackage{amsmath}
\usepackage{amssymb}
\usepackage{float}

\usepackage{soul,xcolor}
\setstcolor{red}
\definecolor{maroon}{rgb}{0.5, 0.0, 0.0}

\title{Charged black hole in $4D$ Einstein-Gauss-Bonnet gravity: Particle motion, plasma effect on weak gravitational lensing and centre-of-mass energy }

\author[a,b,c]{Farruh Atamurotov,}
\author[c,b,d,e,f]{Sanjar Shaymatov,}
\author[g]{Pankaj Sheoran,}
\author[g]{Sanjay Siwach}

\affiliation[a]{Inha University in Tashkent, Ziyolilar 9, Tashkent 100170, Uzbekistan}
\affiliation[b]{Akfa University, Kichik Halqa Yuli Street 17,  Tashkent 100095, Uzbekistan}
\affiliation[c]{Ulugh Beg Astronomical Institute, Astronomy St. 33, Tashkent 100052, Uzbekistan}

\affiliation[d]{Institute for Theoretical Physics and Cosmology, Zheijiang University of Technology, Hangzhou 310023, China}
\affiliation[e]{National University of Uzbekistan, Tashkent 100174, Uzbekistan}
\affiliation[f]{Tashkent Institute of Irrigation and Agricultural Mechanization Engineers,\\ Kori Niyoziy 39, Tashkent 100000, Uzbekistan }
\affiliation[g]{Department of Physics, Institute of Science, Banaras Hindu University, Varanasi-221005, India}

\emailAdd{atamurotov@yahoo.com,sanjar@astrin.uz,
hukmipankaj@gmail.com,sksiwach@hotmail.com}
\date{\today}
\abstract{We study the motion of charged and spinning particles and photons in the $4D$ charged Einstein-Gauss-Bonnet (EGB) black hole vicinity. We determine the radius of the innermost stable circular orbit (ISCO) for test particles. We show that the combined effect of the Gauss-Bonnet (GB) coupling parameter and black hole charge decreases the ISCO and the radius of the photon sphere. 
Further, we study the gravitational deflection angle and show that the impact of GB term and black hole charge on it is quite noticeable. We also consider the effect of plasma and find the analytical form of the deflection angle in the case of a uniform and non-uniform plasma. Interestingly we find that the deflection angle becomes larger when uniform plasma is considered in comparison to the case of non-uniform plasma. We also study the center of mass energy ($E_{C.M.}$) obtained by collision process for non-spinning particles and show that the impact of GB parameter and black hole charge leads to high energy collision. In addition, we also study the $E_{C.M.}$ for the case of spinning particles and show that if the two spinning particles collide near the horizon of $4D$ charged EGB BH, the $E_{C.M.}$ becomes infinitely high which is in disparity with the non-spinning particles counterpart where $E_{C.M.}$ never grows infinitely. To achieve this, an important role is played by the spinning particle known as the \textit{near-critical} particle (i.e. a particle with fine-tuned parameters). In order to achieve the unbounded $E_{C.M}$ from the collision of two spinning particles, the energy per unit mass must be less than unity for a \textit{near-critical} particle, which means such a particle starts from some intermediate position $r>r_{h}$ and not from infinity.}


\begin{document}

\maketitle


\section{Introduction}
\label{introduction}

In general relativity (GR) black holes {(BHs)} have been known as a generic result of Einstein gravity as their geometric properties are described by simple mathematical equations. {It is worth to note that the recent experiments and astrophysical observations related to the gravitational waves~\cite{Abbott16a,Abbott16b} and the first image of the supermassive BH shadows~\cite{Akiyama19L1,Akiyama19L6} can allow to understand the nature of the geometry and test the different gravity models in the strong field regime.} {Although}, Einstein theory has the limits of applicability where it can lose its predictive power. Thus, for its validity and applicability higher order theories are proposed for possible extensions of GR~\cite{Dadhich12c}. In this context, the Lovelock theory can be considered as a generalization of Einstein's theory~\cite{Lovelock1971} as it is a higher order theory which exist in $D$ dimensions. 
Gauss-Bonnet (GB)/Lovelock gravity with the quadratic order gives contribution to the equation of motion only in $D>4$. However, it was recently proposed {that} the novel Einstein-Gauss-Bonnet (EGB) theory could exist even in $D=4$ dimensions, thus allowing to avoid the Lovelock's theorem by rescaling the GB coupling constant~\cite{Glavan20prl}. However, the resulting theory seems to have more degrees of freedom than Gauss-Bonnet gravity in $4D$ and is argued to be a scalar-tensor theory \cite{Lu:2020iav,Bonifacio:2020vbk,Fernandes:2020nbq,Kobayashi:2020wqy,Hennigar:2020lsl}. The problem was revisited by Aoki, Gorji and Mukhohyama (AGM)  \cite{Aoki20egb} and they obtained a consistent theory of $4D$ Gauss-Bonnet gravity by dimensionally reducing $D=d+1$ theory. The theory breaks temporal diffeomorphism invariance but still preserves spatial diffeomorphism invariance in the limit $d\rightarrow 3$.

Unlike Einstein theory, in $4D$ EGB gravity the causal structure departs from its counterpart that the region around singularity becomes time-like, while it is space-like for Einstein gravity~\cite{Dadhich20egb}. Notice that there were three main arguments against the $4D$ EGB: (\textit{i}) regarding the redefinition of the GB term which may be impossible for given system ~\cite{Hennigar20egb,Arrechea20egb}, (\textit{ii}) the one coming from tree-level graviton scattering amplitudes \cite{Bonifacio:2020vbk} and,(\textit{iii}) regarding the action for new theory~\cite{Gurses20egb,Mahapatra20egb}. These objections may mark the limit of the $4D$ EGB theory's applicability. Howbeit, the new $4D$ EGB theory as an alternative to Einstein's theory attracted strong attention. In this framework, there is an extensive body of work devoted to understanding the nature of new $4D$ EGB theory~\cite{Liu20egb,Guo20egb,Wei20egb,Kumar20egb,Konoplya20egb,Churilova20egb,Malafarina20egb,Aragon20egb,Mansoori20egb,Ge20egb,Rayimbaev20egb,Chakraborty20egb,Odintsov20egb,Odintsov20plb,Lin20egb,Aoki20egb,Shaymatov20egb,Islam20egb,Singh20-egb,EslamPanah:2020hoj}. Very recently much of the analyses involved the impact of GB term on the superradiance~\cite{Zhang20aegb}, the motion of spinning particle~\cite{Zhang20egb}, the scalar and electromagnetic perturbations in testing the strong cosmic censorship conjecture~\cite{Mishra:2020gce}, charged particle and epicyclic motions~\cite{Shaymatov20egb} and Bondi-Hoyle accretion around $4D$ EGB BH~\cite{Donmez2021}. Later, following \cite{Glavan20prl} the charged \cite{Fernandes20plb} and rotating \cite{Kumar20egb} analogues were obtained for new $4D$ EGB theory. It is worth noting that some properties of GB BH in higher $D$ dimensions were also investigated in Refs.~\cite{Abdujabbarov15a,Aguilar19,Shaymatov20-pl,Dadhich21,Wu:2021zyl}.

In fact, recently, the first image of the supermassive BH at the center of {Messier} 87 galaxy has been captured by Event horizon telescope (EHT) project by using very long baseline interferometer (VLBI)~\cite{Akiyama19L1,Akiyama19L6} as a result of observations over the past two decades. It is well known that the BH shadow gives rise to the effect of gravitational lensing. BH shadow has also been widely analyzed in different gravity models ~\cite{Hioki09,Atamurotov13,Atamurotov13b,Abdujabbarov13aa,Atamurotov2015a,Atamurotov2016a,Papnoi2015,Abdujabbarov15a,Babar:2020a,Rahul:2020a, Cunha20a,Cunha17a,Atamurotov21b}. Gravitational lensing is considered as one of the most important phenomenon in GR, by which one can get information about a compact gravitational object and source of light as well. In this context, strong gravitational lensing was also theoretically studied by Virbhadra and Ellis~\cite{Virbha:2000a} and large amount of work on these lines has been done in various frameworks in vacuum~\cite[see,
e.g.][]{Bozza:2001a,Bozza:2002b,Zhao:2017a,Vazquez04,Eiroa:2002b,Eiroa:2004a,Chak:2017a,Perlick04,Babar2021c,Abu:2017a,Islam20egb,Virbha:2002a,Kumar2020a}. Here, we focus on the effect of gravitational lensing through the deflection angle and bring out the effect of uniform and non-uniform plasma on the deflection angle surrounding the charged $4D$ EGB BH. {There have been several investigations~\cite{Kogan10,Tsupko12,Bisnovatyi15,Babar2021a,Synge:1960b} exploring gravitational lensing in the weak-field regime as a consequence of the presence of plasma medium. It was later extended to various gravity models~\cite{Hakimov2016a,Rog:2015a,Turi:2019a,Far:2021a,Car:2018a,Chak:2018a,Abu:2017aa,Li2020}, thus addressing the plasma effect on the gravitational lensing.} 

Gravitational lensing in the presence of plasma is fascinating for two reasons: first, photons within cosmos mostly pass via this medium, and second, the presence of plasma results in various angular positions of an equivalent image when examined at different wavelengths. It is argued in \cite{nla.cat-vn2032203} that photons in a non-uniform plasma move along a curved path due to the dispersive nature of plasma with a permitivity tensor depending upon its density. In a non-uniform dispersive medium, motion is solely determined by photon frequency and has no relation to gravity. However, the effect of non-uniform plasma on the motion of photons in the curved spacetime (i.e., gravity) can be non-trivial.The important effects which are predicted in this context are: \textit{(i)} vacuum deflection due to gravitation, and \textit{(ii)} the deflection due to non-homogeneity of the medium. \cite{PhysRevLett_24_1377,1975pbrg.book.....L,B-Minakov}. The first effect is achromatic in nature means it is independent of the photon frequency (or energy), while the later one is dispersive in nature and depends on photon frequency, but vanishes for the case of uniform medium \cite{Bisnovatyi-Kogan:2010flt}.

In this work we investigate the effect of  plasma on the gravitational lensing in the background of the $4D$ charged EGB BH beside the effect of Gauss-Bonnet coupling parameter $\alpha$ and charge parameter $Q$ of BH. In the presence of plasma, the main characteristic of gravitational lensing is that the deflection angles shift, resulting in chromatic deflection. This results in angular differences in picture positions at different wave frequencies. However, such effects may be important only for very long radio waves propagating through cosmic plasma. Beside this limitation, it is still important to investigate these chromatic effects as they can reveal details about the source structure \cite{1986A&A...166...36K,1991AJ....102..864W}.

An extensive analysis has been done to study the particle motion around BHs in the strong gravitational field regime; here we give some representative references~\cite[see,
e.g.][]{
Frolov10,Aliev02,Abdujabbarov10,
Shaymatov14,Toshmatov15d,Shaymatov15,Pavlovic19,Shaymatov19b,Jamil15,Hussain17,Shaymatov20a,Toshmatov19d,Rayimbaev20c,Shaymatov20b,Narzilloev20a,Narzilloev20b,Stuchlik20,Shaymatov21-b,Shaymatov21c}. Note that these productive studies of particle motion as a useful tool were devoted to probe spacetime properties and astrophysical events such as winds and jets from active galactic nuclei (AGN)~\cite{Fender04mnrs,Auchettl17ApJ,IceCube17b}. {These} observational phenomenona have been modeled by a number of energetic mechanisms of which well accepted one is the BSW mechanism proposed by Banados, Silk and West~\cite{Banados09} and Penrose process~\cite{Penrose69}.  Following~\cite{Banados09}, there have been several investigations on these lines~\cite[see,
e.g.][]{Grib11,Jacobson10,Harada11b,Wei10,
Zaslavskii11b,Zaslavskii11c,Kimura11,Banados11,Frolov12,
Abdujabbarov13a,Liu11,Atamurotov13a,Stuchlik11a,
Stuchlik12a,Igata12,Shaymatov13,Tursunov13,Ghosh:2014mea,Shaymatov18a,Babar2021b,Josh:2016a}. Similarly, the Penrose process has also been widely studied in Refs.~\cite{Dadhich18,Abdujabbarov11,Okabayashi20,Ghosh:2013ona}. 

Contrary to non-spinning particles, the work on spinning particles in the curved spacetime is limited and mainly devoted to two topics i.e. the study of inner most stable circular orbits (ISCOs) \cite{Jefremov:2015gza,Armaza:2016vfz,Zhang:2017nhl,ZHANG2019393,PhysRevD.98.084023,Hojman:2018evi,PhysRevD.99.104059,2018mgm..conf.3715J,101142S0218271820501217,Toshmatov:2020wky,Toshmatov:2019bda,Nucamendi:2019qsn,Zhang:2020qew}, and collision of spinning particles \cite{Armaza:2015eha,Zhang:2016btg,Zaslavskii:2016dfh,Guo:2016vbt,Zhang:2020cpu,Sheoran:2020kmn,Yuan:2019dih,Liu:2019wvp,Okabayashi:2019wjs,Zhang:2018ocv,Zhang:2018gpn,Maeda:2018hfi,Liu:2018myg,Mukherjee:2018kju} besides these spinning particles also appear in the studies \cite{Zalaquett:2014eia,Zhang:2018omr,Chakraborty:2018tvy,Deriglazov:2017jub,Chakraborty:2016mhx,Hojman:2016mox,Deriglazov:2016mhk,Deriglazov:2015wde}. As a result, it would be intriguing to learn more about spinning particles in additional BH spacetimes. Here, we investigate  the high-energy collision process for non-spinning and spinning particles, and how parameters like $\alpha$, $Q$, and $S$ (where $S"$ is the particle's spin parameter) affect it. $S=0$ and $S\neq0$ for non-spinning and spinning particles, respectively. We begin by considering the motion of charged and spinning particles, and photons around the $4D$ charged EGB BH to fulfill the goals outlined above.

The paper is organized as follows: In Sec.~\ref{Sec:metic} we briefly describe $4D$ charged EGB BH metric. We study the particle geodesics around $4D$ charged EGB BH in Sec.~\ref{Sec:circular}, which is followed by discussion of analysis of gravitational lensing around BH in Sec.~\ref{Sec:lensing}. 
We analyze the BSW effect (i.e, can charged $4D$ EGB BH act a particle accelerator?) in Sec.~\ref{Sec:energy} for the case of both non-spinning and spinning particles and bring out the effect of parameters $\alpha$ (Gauss-Bonnet coupling constant), $Q$ (BH charge) and $\tilde{S}$ (spin of particle). We numerically shows the parameter space region between particle spin $\tilde{S}$ and $r$ for different combinations $Q$, $\alpha$, conserved  energy $E$ and conserved total angular momentum $J$ for which the square of spinning particle four velocity lies in subluminal region (i.e. $u_{\mu}u^{\mu}<0$). We end up our concluding remarks in the Sec.~\ref{Sec:conclusion}. In order to make the paper self-sufficient, we show the general equations for the spinning particles in the curved spacetime in appendix \ref{EOM_spin_particles}, and calculate the components of four-momentum and four velocity for the charged $4D$ EGB BH in appendix \ref{four_p_u}. Throughout we use a system of units in which $G=c=1$ {and choose the sign conventions $(-,+,+,+)$}. 

\section{\label{Sec:metic} $4D$ charged Einstein-Gauss-Bonnet BH metric }

According to the Lovelock's theorem, the GB term contributes to the gravitational dynamics in $D>4$ only. It is however shown by Glavan and Lin \cite{Glavan20prl} that the one way to consider this non-trivial contribution in $D=4$ is to rescale the GB coupling constant, i.e. $\alpha\rightarrow \alpha/(D-4)$ with the limit $D\rightarrow4$ in the GB term. By this way Glavan and Lin \cite{Glavan20prl} were able to avoid the Lovelock's theorem and obtain $4D$ EGB BH solution. 
  
It has been noticed recently that the limiting procedure of Glavan and Lin is not sufficient to account for the degrees of freedom of Gauss-Boonet gravity and additional scalar degrees of freedom is required in the framework \cite{Lu:2020iav,Bonifacio:2020vbk,Fernandes:2020nbq,Kobayashi:2020wqy,Hennigar:2020lsl}. Thus, a pure gravity theory with Gauss-Bonnet terms seems elusive. However, the problem was overcome by Aoki, Gorji and Mukhohyama (AGM) \cite{Aoki20egb}  and a consistent theory of Gauss-Bonnet gravity in $4D$ can be obtained by dimensional reduction of higher dimensional Gauss-Bonnet gravity given by the action,
\begin{eqnarray}
S&=&\frac{1}{2\kappa^2}\int d^Dx\sqrt{-g} \left[R+\alpha R^2_{GB}\right],
\end{eqnarray}
where,
$ R^2_{GB}=R_{\mu\nu\lambda\delta}R^{\mu\nu\lambda \delta} - 4R_{\mu\nu}R^{\mu\nu}+R^2$ is the Gauss-Bonnet combination of curvature squared terms, $g_{\mu\nu}$ is $D=d+1$ dimensional metric and $\kappa^2$ is related with the  Newton's constant in $D$ dimension. The $d\rightarrow 3$ limit is shown to be smooth by AGM provided some additional constraints are imposed. The gravitational solutions of $D=d+1$ theory which obey these constraints also become the solutions of $4D$ Gauss-Bonnet gravity in the limit $d \rightarrow 3$.  

The spherically symmetric $4D$ charged EGB BH solution of $4D$ Gauss-Bonnet gravity was obtained in \cite{Fernandes20plb} with the line element of the form, 
\begin{eqnarray}\label{solution}
ds^2=-F(r)dt^2+\frac{dr^2}{F(r)}+r^2d\Omega^2,
\end{eqnarray}
with
\begin{eqnarray}\label{Eq:lapse}
F(r)=1+\frac{r^2}{2\alpha}\left[1\pm\sqrt{1+4\alpha\left(\frac{2 M}{r^3}-\frac{Q^2}{r^4}\right)}\right],
\end{eqnarray}
with mass $M$ and charge $Q$.

It has been argued that the static black hole solutions obtained by Glavan and Lin procedure also happen to be the solutions of the AGM theory \cite{Jafarzade_2021} and the charged black hole metric (\ref{solution}) can be regarded as a consistent solution of $4D$ Gauss-Bonnet gravity coupled to Maxwell's electrodynamics.

\begin{figure}
\begin{tabular}{c c}
 \includegraphics[scale=0.55]{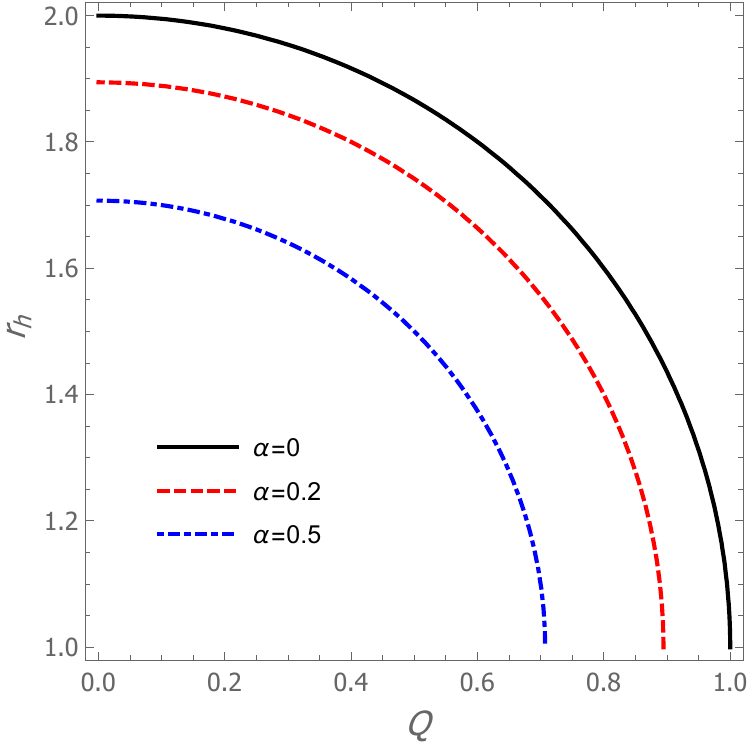}
  \includegraphics[scale=0.55]{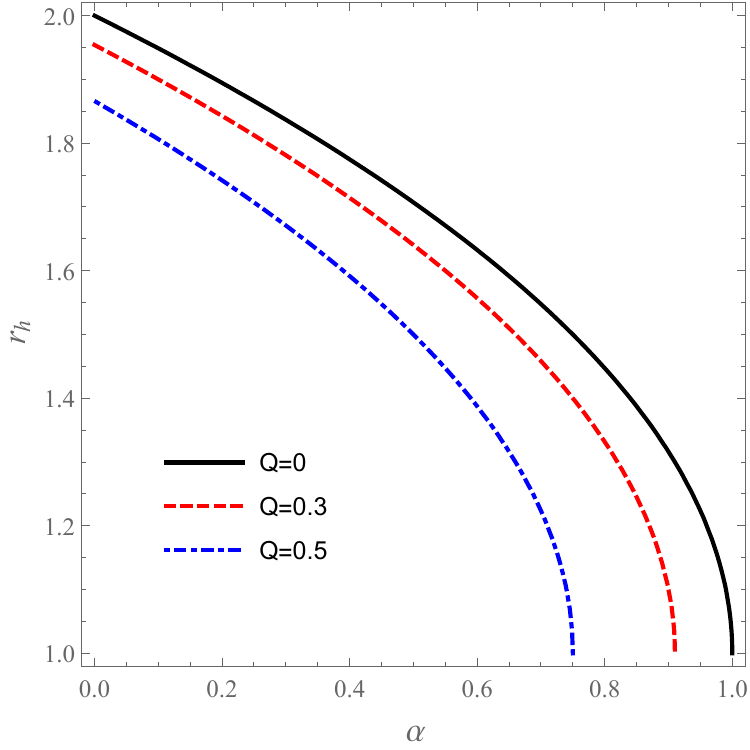}
\end{tabular}
\caption{The dependence of the horizon radius $r_{h}$ on the BH charge $Q$ for different values of GB parameter $\alpha$ (left panel) and on the GB coupling constant $\alpha$ for different values of $Q$ (right panel). Note that the horizon behaviour is similar to the one in Einstein gravity in case one of two parameters vanishes.  }
\label{fig:horizon}
\end{figure}
From Eq.~(\ref{Eq:lapse}) there exist two branches of BH solutions. However, we shall restrict ourselves to the minus sign of solution as it exhibits an attractive massive source~\cite{Dadhich07gb,Torii05}. We must ensure that this is the right branch, thus we need to expand Eq.~(\ref{Eq:lapse}) in small $\alpha$  
\begin{equation}
\lim_{\alpha\rightarrow 0} F(r) = 1-\frac{2M}{r}+\frac{Q^2}{r^2}+\frac{(Q^2-2Mr)^2}{r^6}\alpha+... 
\end{equation}
This clearly shows that it exhibits a Reissner-Nordstr\"{o}m solution in the case when $\alpha \rightarrow 0$. 

Let us then turn to the horizon of $4D$ charged EGB BH. It is given by positive root of $F(r)=0$, and that gives 
\begin{eqnarray}\label{Eq:hor}
r_{h}=M + \sqrt{M^2-Q^2-\alpha }\, .
\end{eqnarray}
As shown from above equation this clearly shows that an extremal BH is given by with horizon $r=M$ in case $\alpha+Q^2=M^{2}$ is satisfied. If the term $\alpha+Q^2$ dominates over $M^2$ the BH then turns into a naked singularity. This issue for $4D$ charged EGB BH was addressed by the authors of \cite{Yang20b}.  
The dependence of BH horizon on $\alpha$ and $Q$ is shown in Fig.~\ref{fig:horizon}. As shown in Fig.~\ref{fig:horizon}, the GB coupling constant and BH charge exhibit a similar effect that the outer horizon shifts towards to smaller $r$, i.e. to the central singularity. This is in agreement with the fact that the GB coupling constant exhibits a repulsive gravitational effect~\cite{Dadhich20egb,Shaymatov20egb}.

\section{\label{Sec:circular}
Particle geodesics around $4D$ charged EGB BH}

\subsection{\label{mass}
massive particle}

Here we first study charged particle motion by considering the Hamiltonian of the system~\cite{Misner73} around the static and spherically symmetric charged BH spacetime in $4D$ EGB 
gravity 
\begin{eqnarray}
 H  \equiv \frac{1}{2}g^{\mu\nu}(\pi_\mu - q A_\mu)(\pi_\nu - q A_\nu)\, ,
\label{Eq:H}
\end{eqnarray}
where $\pi_\mu$ is the canonical four momentum of a charged particle and $A_\mu$ is the electromagnetic four-vector potential for $4D$ charged EGB BH and given by
\begin{eqnarray}\label{Eq:4-pot}
A_{\mu}=\left(-\frac{Q}{r},0,0,0\right)\, .
\end{eqnarray}
Note that we consider the Hamiltonian as a constant, i.e. $H=-m^2/2$. For the Hamilton-Jacobi equation, the action $S$ can be then separated in the following form
\begin{eqnarray}\label{Eq:separation1}
S= \frac{1}{2}m^2\lambda-Et+L\varphi+S_{r}(r)+S_{\theta}(\theta)\ ,
\end{eqnarray}
where $E \equiv -\pi_t$ and $L \equiv \pi_{\varphi}$ are the constants of motion and respectively refer to the energy and angular momentum of the charged particle, while $S_{r}$ and $S_{\theta}$ are functions of $r$ and $\theta$. Let us then rewrite the Hamilton-Jacobi equation in the following form
\begin{equation}\label{Eq:separable}
-r^2\left(F\right)^{-1}\left(-E+\frac{qQ}{r}\right)^2
+ r^2F
\left(\frac{\partial S_{r}}{\partial r}\right)^2
+\left(\frac{\partial S_{\theta}}{\partial \theta}\right)^2
+
\frac{1}{\sin^2\theta}\left(L-qA_{\varphi}\right)^2+m^2r^2=0 \, . 
\end{equation}
\begin{figure}
\centering
 \includegraphics[width=0.45\textwidth]{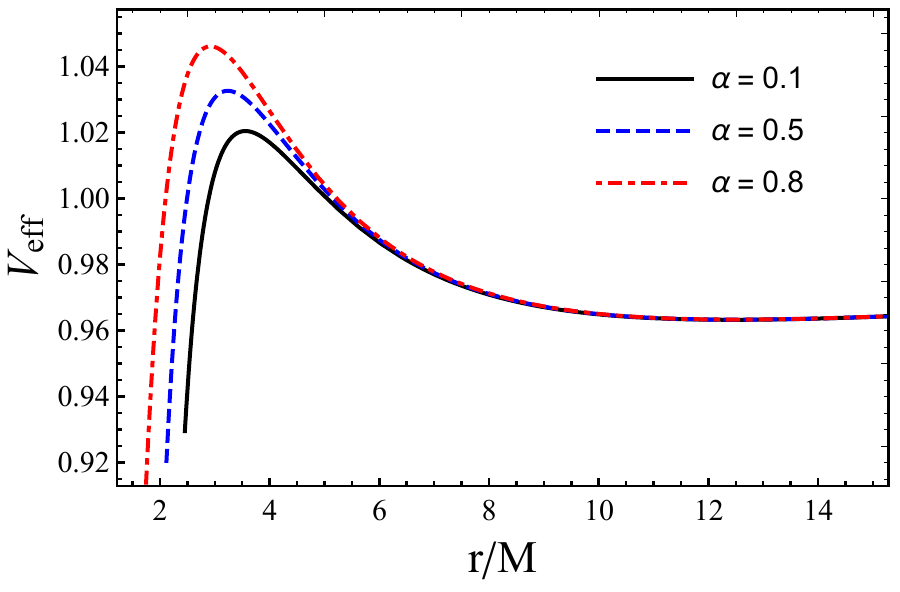}%
 \includegraphics[width=0.45\textwidth]{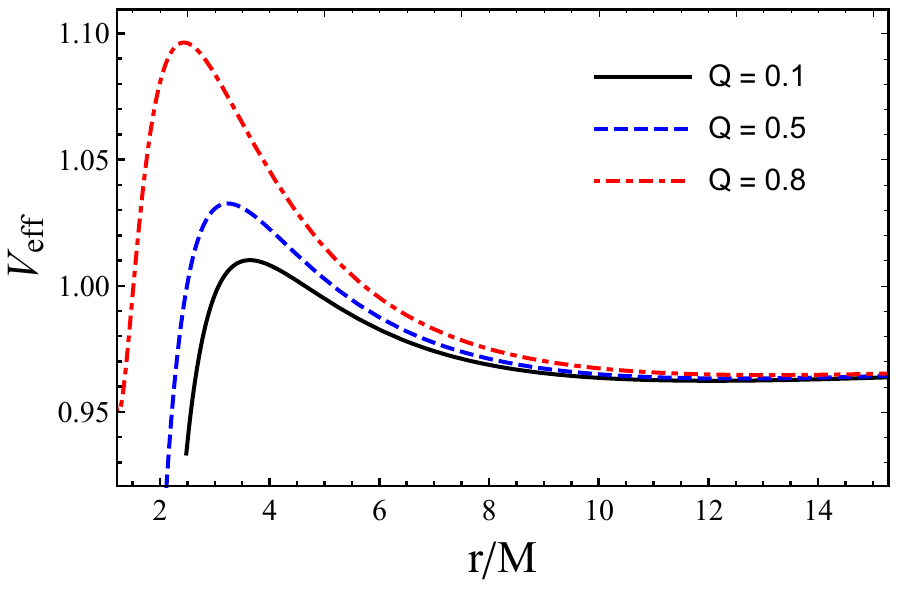}
 
 \includegraphics[width=0.45\textwidth]{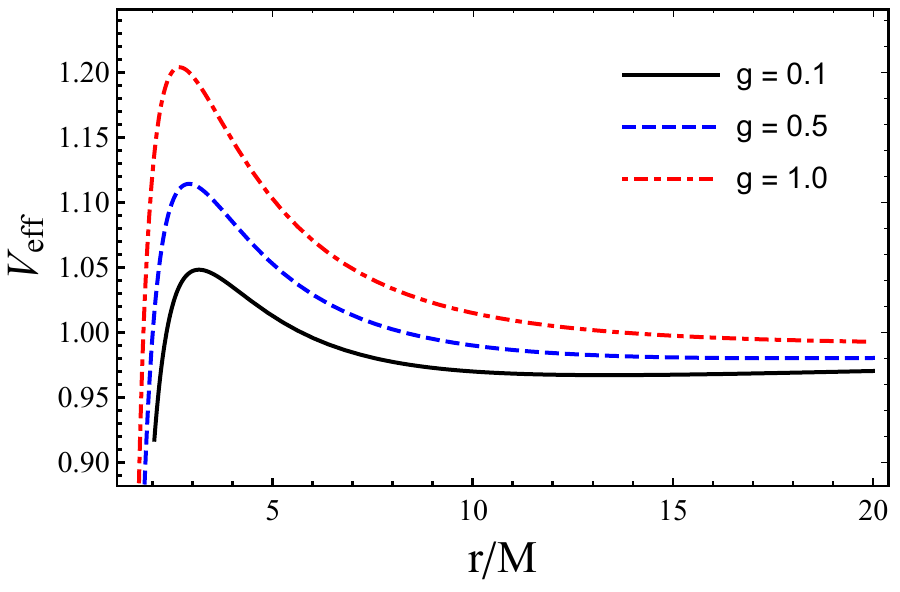}%
 \includegraphics[width=0.45\textwidth]{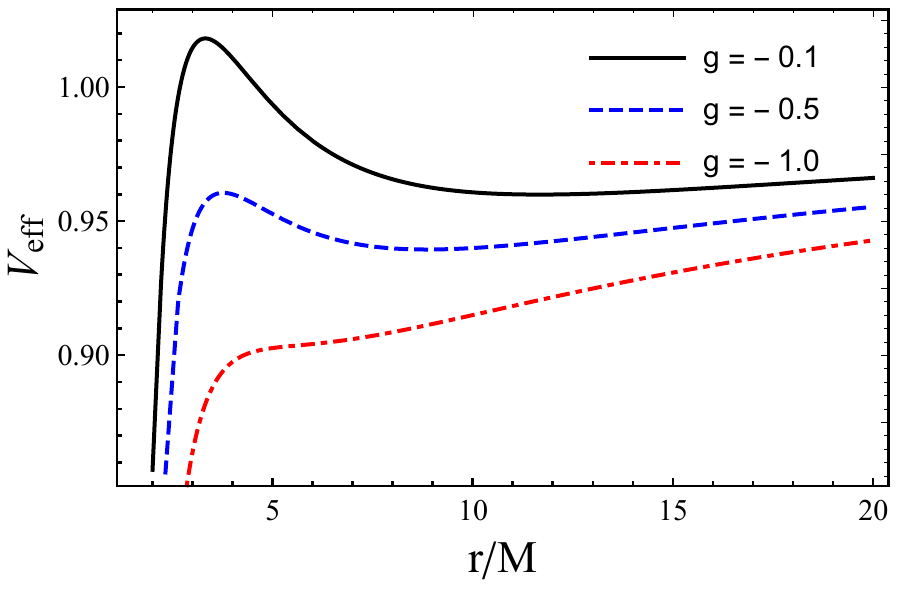}
 
\caption{\label{fig:Eff} Radial dependence of the effective potential for massive particles orbiting a $4D$ charged EGB BH for $g=0$ on the top and for $g\neq 0$ on the bottom. Top panels: $V_{eff}$ is plotted for different values of $\alpha$ for $Q=0.4$ (left panel) and for different values of $Q$ for $\alpha=0.3$ (right panel). Bottom panels: $V_{eff}$ is plotted for different values of charge parameter $\pm g$ in the case of $\alpha =0.5$ and $Q=0.5$.}
\end{figure}
\begin{table}[h]
\centering
\begin{tabular}{|c|c|c|c|c|c|c|}
\hline
&\multicolumn{5}{c}{$Q$}&  \\ \hline\hline
{$\alpha $} & $0.0$ & $0.1$ &  $0.2$
& $0.3$ & $0.5$ & $0.8$ \\
\hline
 0.0 & 6.00000 &5.98497 & 5.93957 & 5.86278 & 5.60664 & 4.89077 \\
\hline
0.1  &5.93782  &5.92248  &5.87608  &5.79755 &5.53491 &4.79238 \\
\hline
0.2  &5.87338  &5.85768 &5.81020 &5.72975 &5.45987 &4.68626 \\
\hline
0.3  &5.80644   &5.79035 &5.74167 &5.65910 &5.38111 &4.57055 \\
\hline
0.5 &5.66395   &5.64695 &5.59544 &5.50782 &5.21013 & $-$ \\
\hline
0.8 &5.42276   &5.40389 &5.34653 &5.24829 &$-$ &$-$ \\
\hline    
\end{tabular}
\caption{\label{1tab} The values of $r_{ISCO}$ are tabulated in the case of neutral particle orbiting $4D$ charged EGB BH for various values of GB coupling constant $\alpha$ and BH charge $Q$. }
\end{table}
\begin{table}[h]
\centering
\begin{tabular}{|c|c|c|c|c|c|c|}
\hline
&\multicolumn{5}{c}{$g$}& \\ \hline
\hline
{$\alpha;Q $} & $0.0$ & $0.01$ &  $0.05$
& $0.1$ & $0.2$ & $0.3$ \\
    &  & $-0.01$ &  $-0.05$ & $-0.1$ & $-0.2$ & $-0.3$ \\
\hline
0.1  &5.92248  &5.92244  &5.92230  &5.92215 &5.92192  &3.92179 \\
      &       &5.92251  &5.92267  &5.92290  &5.92341 &5.92402 \\
\hline
0.2  &5.81020  &5.81003 &5.80936 &5.80861 &5.80740 &5.80659 \\
&             &5.81038  &5.81113 &5.81215 &5.81443 &5.81703 \\
\hline
0.3 &5.65910   &5.65863 &5.65684 &5.65477 &5.65122 &5.64853 \\
      &     &5.65958 &5.66154 &5.66416 &5.66987 &5.67619 \\ 
\hline
0.5 & 5.21013   &5.20826 &5.20091 &5.19202 &5.17536 &5.16044 \\
      &     &5.21201 &5.21966 &5.22946 &5.24983 &5.27107 \\\hline
\end{tabular}
\caption{\label{2tab} The values of $r_{ISCO}$ are tabulated in the case of charged particle orbiting $4D$ charged EGB BH for various values of charge parameter $g$ in the case of fixed $\alpha$ and $Q$. Here we shall for simplicity consider that GB coupling constant and BH charge take the same values, i.e. $\alpha=Q$.}
\end{table}
\begin{figure}
\centering
 \includegraphics[scale=0.35]{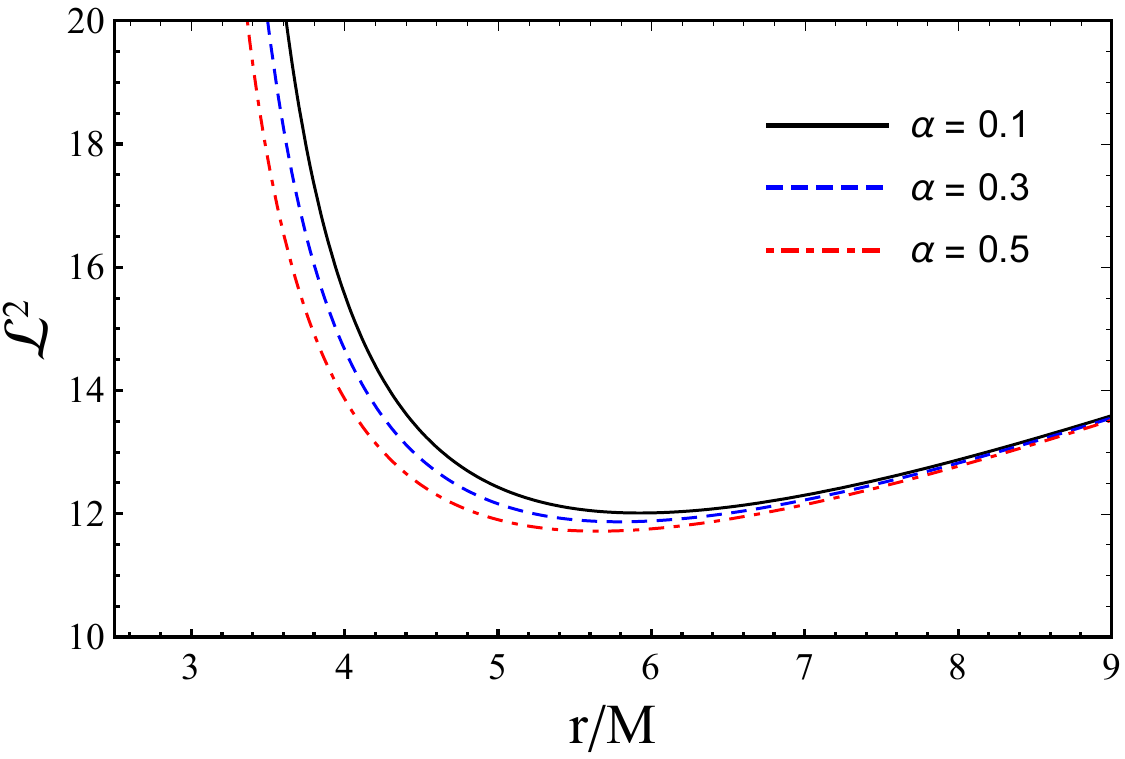}%
 \includegraphics[scale=0.35]{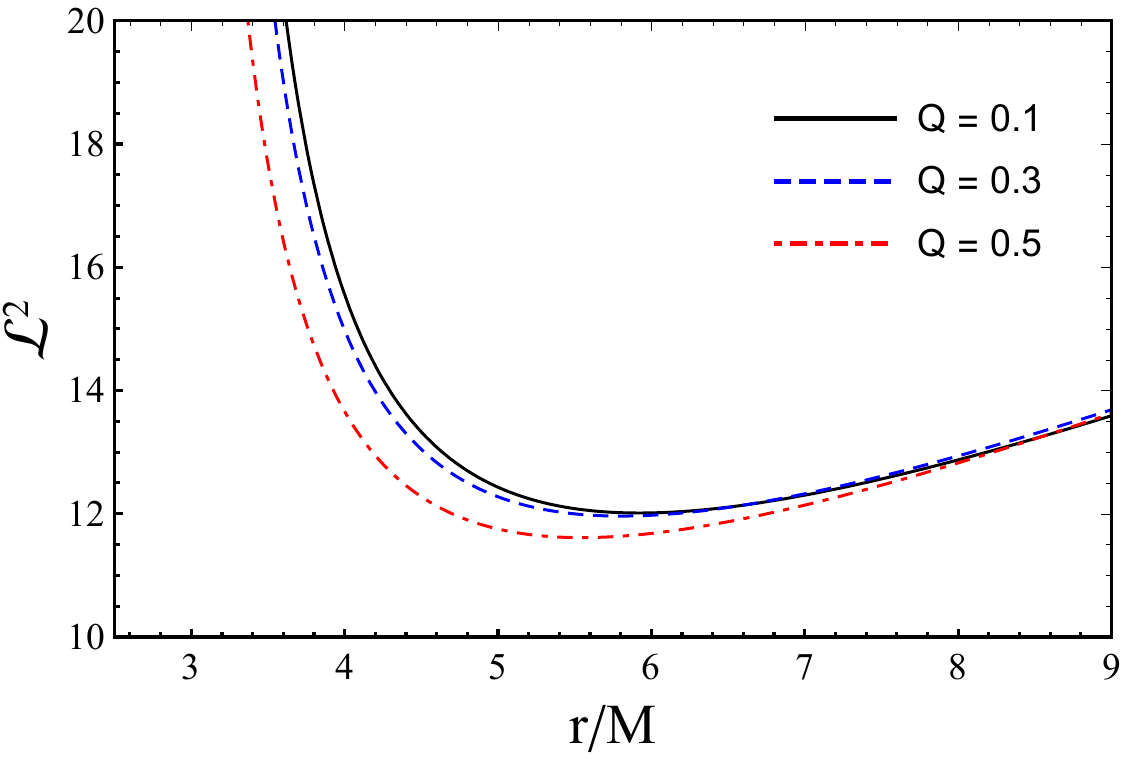}
\caption{\label{fig:an} Radial dependence of the angular momentum for charged test particles orbiting around $4D$ charged EGB BH in the case of $g=0.1$. Left panel: $\mathcal {L}^2$ is plotted for different values of GB coupling constant $\alpha$ for fixed $Q=0.1$. Right panel: $\mathcal {L}^2$ is plotted for different values of BH charge $Q$ for fixed $\alpha=0.1$.}
\end{figure}
Note that the system can be described by four independent constants of motion of which we have specified three ($E$, $L$, and $m^2$). We however restrict ourselves to the equatorial plane ($\theta=\pi/2$) so that we ignore the fourth constant of motion related to the latitudinal motion~\cite{Misner73}.

From Eq.~(\ref{Eq:separable}), the radial equation of motion for charged particle is written as
\begin{eqnarray}
\dot{r}^{2} = \Big(\mathcal{E} -
\mathcal{E_{+}}(r)\Big)\Big(\mathcal{E}
-\mathcal{E_{-}}(r)\Big)\, ,
\end{eqnarray}
where $\dot{r}^2$ must always be positive, and hence we have either $\mathcal{E}>\mathcal{E}_{+}(r)$ or $\mathcal{E}<\mathcal{E}_{-}(r)$. For further analysis, we choose the positive energy $\mathcal{E_{+}}(r)=V_{eff}(r)$ which is physically related to the effective potential. It is then straightforward to define the effective potential for the radial motion of particle in the field of $4D$ charged EGB BH. The effective potential then takes the form    
\begin{equation}\label{Eq:Veff}
V_{eff}(r)= \frac{gQ}{r}+\left(1+\frac{\mathcal{L}^2}{r^2}\right)^{1/2}\left[1+\frac{r^2}{2\alpha}\left(1-\sqrt{1+4\alpha\left(\frac{2 M}{r^3}-\frac{Q^2}{r^4}\right)}\right)\right]^{1/2} \, .
\end{equation}
Here $\mathcal{E}=E/m$ and $\mathcal{L}=L/m$ are the energy and angular momentum per unit mass and $g=q/m$ is the charge parameter per unit mass.

We then analyse the effective potential $V_{eff}(r)$ for the radial motion of both neutral and charged test particles. In Fig.~\ref{fig:Eff}, the top panels show the impact of both GB coupling constant $\alpha$ and BH charge $Q$ on the radial profiles of the effective potential for neutral particle, while the bottom panels show the impact of both negative and positive charge parameter $g$ for fixed values of $\alpha$ and $Q$. As seen from radial profile of $V_{eff}(r)$, the height of the effective potential increases with increasing both GB coupling constant $\alpha$ and BH charge $Q$ and the curves shift towards left to smaller $r$. Similarly, we notice that the height of the effective potential also grows for positive charge parameter $g>0$ while with increasing the negative $g<0$ this behaviour turns to opposite in the case of fixed $\alpha$ and $Q$. It is worth noting that the strength of effective potential becomes weaker only for $g<0$ as compered to the one for $g>0$. 

Now we turn to the study the circular orbits of charged test particles around $4D$ charged EGB BH characterized by GB coupling term $\alpha$, charge $Q$, and mass $M$.  For test particles to be on the circular orbits we need to solve simultaneously 
\begin{eqnarray}\label{Eq:cir}
V_{eff}(r)=0=V_{eff}^{\prime}(r)\, ,
\end{eqnarray}
where prime refers to a derivative with respect to $r$. The above required conditions give the radii of circular orbits for given values of $\mathcal{L}$. We then determine the angular momentum for charged test particles on the circular orbits 
%
\begin{equation}\label{Eq:L}
\mathcal{L}^2 = \frac{r^3 f'(r) \Big(2 f(r)-r f'(r)\Big)+2 Q^2 g^2 f(r)}{\Big(r f'(r)-2 f(r)\Big)^2}+ \frac{g \,Q\, f(r) \sqrt{Q^2 g^2+4 r^2 f(r)-2 r^3 f'(r)}}{\Big(r f'(r)-2 f(r)\Big)^2}
 \, .
\end{equation}

%
For neutral test particle, i.e. $g=0$, Eq.~(\ref{Eq:L}) takes the following form 
\begin{eqnarray}\label{Eq:min-ang}
\mathcal{L}^2=\frac{r^3 f'(r)}{2 f(r)-r f'(r)}\, .
\end{eqnarray}
The radial profiles of the angular momentum of the circular orbits for charged particles are shown in Fig.~\ref{fig:an}. As shown in Fig.~\ref{fig:an} one can easily see that the curves shift towards left to small radii as an increase in the value of both GB coupling constant $\alpha$ and BH charge $Q$ for fixed charge parameter $g$. 

Next, let us study the innermost stable circular orbit (ISCO) for both neutral and charged test particles orbiting around $4D$ charged EGB BH.  For determining radii of stable circular orbits we should solve the following required condition  
\begin{eqnarray}\label{Eq:isco}
V_{eff}^{\prime\prime}(r)\geq0\, .
\end{eqnarray}
However, for the ISCO radius one needs to solve $V_{eff}^{\prime\prime}(r)=0$. We explore the ISCO radius numerically by solving Eqs.~(\ref{Eq:cir}) and (\ref{Eq:isco}) simultaneously and provide results in  Tables~\ref{1tab} and \ref{2tab} for both neutral and charged particles. As shown in Table~\ref{1tab} the ISCO radius decreases as both $\alpha$ and $Q$ increase. However, the ISCO radius decreases in the case of negative charge parameter $g<0$ as shown in Table~\ref{2tab}.

\subsection{\label{massless}
massless particle (photon motion)}

\begin{figure}[t]
   \includegraphics[scale=0.4]{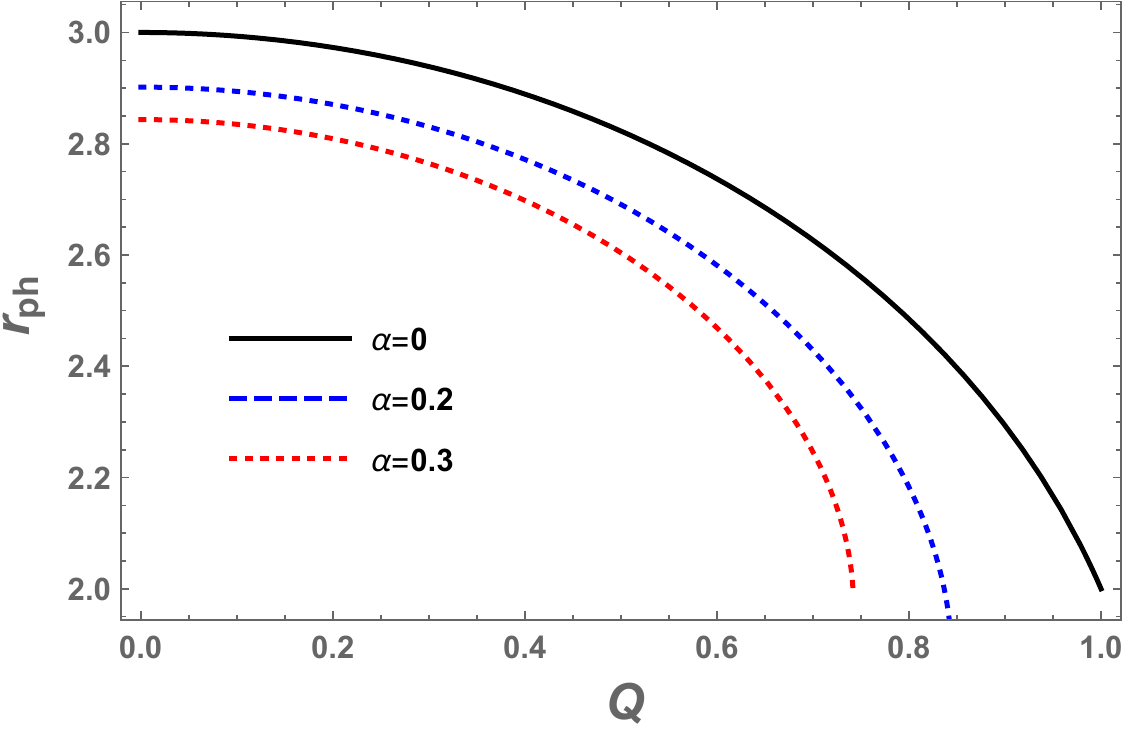}
   \includegraphics[scale=0.4]{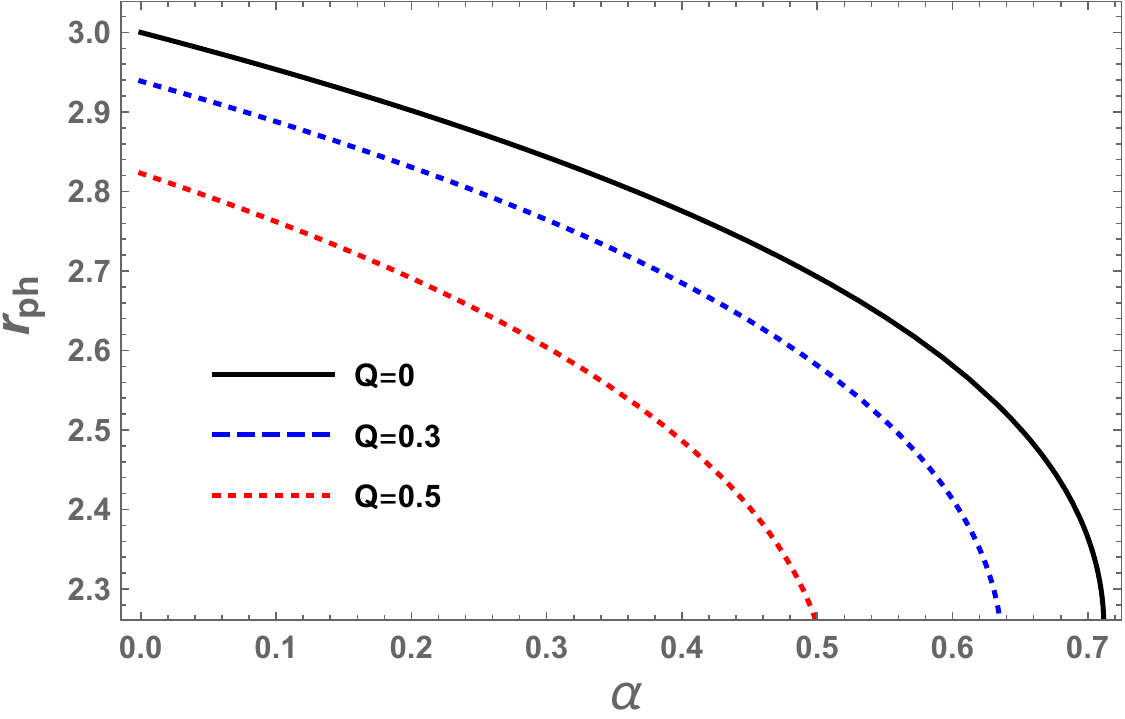}
\caption{The dependence of the photon radius $r_{ph}$ on the BH charge $Q$ for differnt values of $\alpha$ (left panel) and on the GB coupling constant $\alpha$ for different values of $Q$ (right panel). }
\label{fig:photon}
\end{figure}

Here we consider massless particle motion around $4D$ charged EGB BH for which the Hamilton-Jacobi equation is written in the following form
\begin{eqnarray}
 H  \equiv \frac{1}{2}g^{\mu\nu}\tilde{\pi}_\mu \tilde{\pi}_\nu =0\, ,
\label{Eq:Hphoton}
\end{eqnarray}
with $\tilde{\pi}_\mu$ and $\tilde{\pi}_\nu$ being conserved quantities which are the momentum of massless (photon) particles. It is then sufficient to use Eq.~(\ref{Eq:Hphoton}) to find the effective potential for the photon as follows  
\begin{eqnarray}\label{Eq:Veff_ph}
\tilde{V}_{eff}(r)=\tilde{E}^2-F(r)\frac{\tilde{L}^2}{r^2}\, .
\label{veffphoton}
\end{eqnarray}

where $\tilde{L}$ and $\tilde{E}$ are the conserved quantities, which are the angular momentum and energy of the photon, respectively. To obtain unstable circular orbits for photons we will need to solve following equations 
\begin{eqnarray}\label{Eq:con_ph}
\tilde{V}_{eff}(r)=0=\tilde{V'}_{eff}(r)\, .
\end{eqnarray}
In Fig.~\ref{fig:photon}, we show the radius of the photon orbit $r_{ph}$ for various values of $\alpha$ and $Q$ by using Eqs.~ (\ref{veffphoton}) and (\ref{Eq:con_ph}). As expected the photon radius decreases with increasing both GB coupling constant $\alpha$ and BH charge $Q$, see Fig~\ref{fig:photon}. 

\section{Weak-field lensing } \label{Sec:lensing}

In this section, we turn to the study optical properties, i.e. weak lensing around $4D$ charged EGB BH. Further we shall for simplicity consider a weak-field case for which Eq.~(\ref{Eq:lapse}) can be written as 
\begin{eqnarray}
F(r)=1-\frac{R_g}{r}+\frac{Q^2}{r^2}+\frac{\alpha R^2_{g}}{r^4}\, ,
\end{eqnarray}
with $R_g=2M$. For a weak-field approximation we have 
\begin{align}
g_{\alpha\beta}=\eta_{\alpha\beta}+h_{\alpha\beta}\, ,
\end{align}
where $\eta_{\alpha\beta}$ and $h_{\alpha\beta}$ respectively refer to the expression for Minkowski spacetime and perturbation gravity field describing GR part. Thus, we have 
\cite{Kogan10}
\begin{align}
\eta_{\alpha\beta}&={\rm diag} (-1, 1, 1, 1)\ ,\nonumber\\
 h_{\alpha\beta} &\ll 1, \quad  h_{\alpha\beta} \rightarrow 0 \quad {\rm under } \quad x^{\alpha}\rightarrow \infty\ , \nonumber \\
 g^{\alpha\beta}&=\eta^{\alpha\beta}-h^{\alpha\beta},\ \ \ h^{\alpha\beta}=h_{\alpha\beta}\ .
\end{align}
Following \cite{Kogan10}, we consider the weak-field approximation and plasma medium for the light propagation along $z$ axis~\cite{Kogan10,Babar2021a} to get the gravitational deflection equation as 
\begin{align}
\hat{\alpha}_i &= \frac{1}{2} \int_{-\infty}^\infty \bigg(h_{33_,i} +
\frac{ \omega^2}{\omega^2-\omega^2_{e}}h_{00_,i} -\frac{K_e }{\omega^2-\omega^2_{e}}N{_,i}\bigg) dz. \label{alphak}
\end{align}
Here, $\hat{\alpha}_i$ would be negative and positive respectively for the deflection of the light trajectory, depending on the light moving either towards or away from the central object. 
In the above equation, $\omega_e$ can be set $\omega(\infty)=\omega$ at infinity as limitations. For the weak field approach, the BH spacetime metric can be written in the following form~\cite{Babar2021a}
\begin{align}
ds^2 = ds^2_{0}+\left(\frac{R_{g}}{r}-\frac{Q^2}{r^2}-\frac{\alpha R^2_{g}}{r^4}\right)dt^2
+\left(\frac{R_{g}}{r}-\frac{Q^2}{r^2}-\frac{\alpha R^2_{g}}{r^4}\right) dr^2\, , \label{metrbb}
\end{align}
\\
with $ds^2_{0}=-dt^2+dr^2+r^2(d\theta^2+\sin^2\theta d\phi^2)$ corresponding to Minkowski metric. Further, $h_{\alpha\beta}$ can be written in the Cartesian coordinates system as follows
\begin{align}
 h_{00}=&\left(\frac{R_g}{r}-\frac{Q^2}{r^2}-\frac{\alpha R^2_{s}}{r^4}\right), \nonumber \\  h_{ik}=&\left(\frac{R_g}{r}-\frac{Q^2}{r^2}-\frac{\alpha R^2_{s}}{r^4}\right)n_{i}n_{k}\ ,\nonumber \\
 h_{33}=&\left(\frac{R_g}{r}-\frac{Q^2}{r^2}-\frac{\alpha R^2_{s}}{r^4}\right)\cos^2x\; ,\label{h}
\end{align}
with $\cos x=z/\sqrt{b^2+z^2}$ and $r=\sqrt{b^2+z^2}$ \cite{Babar2021a}.

Introducing new notation as $h_{\alpha\beta}$ in Eq.~(\ref{h}) one can rewrite the expression of the gravitational deflection angle in terms of the impact parameter $b$ for $4D$ charged EGB BH immersed in plasma medium. Thus, we have 
\begin{align}
\hat{\alpha}_{b}=\int_{-\infty}^\infty\frac{b}{2r}
\Bigg(\partial_r \bigg(\bigg(\frac{R_g}{r}-\frac{Q^2}{r^2}-\frac{\alpha R^2_{s}}{r^4}\bigg)\cos^2x\bigg)+
\partial_r\bigg(\frac{R_g}{r}-\frac{Q^2}{r^2}-\frac{\alpha R^2_{s}}{r^4}\bigg)\frac{\omega^2}{\omega^2-\omega^2_{e}}
-\frac{K_e}{\omega^2-\omega^2_{e}}\partial_r N\Bigg)dz\, .
\label{alfa}
\end{align}
%

Next, we will analyze the deflection angle for homogeneous and inhomogeneous cases in detail. 

\subsection{Uniform plasma}

\begin{figure}[t]
\begin{tabular}{c c}
   \includegraphics[scale=0.35]{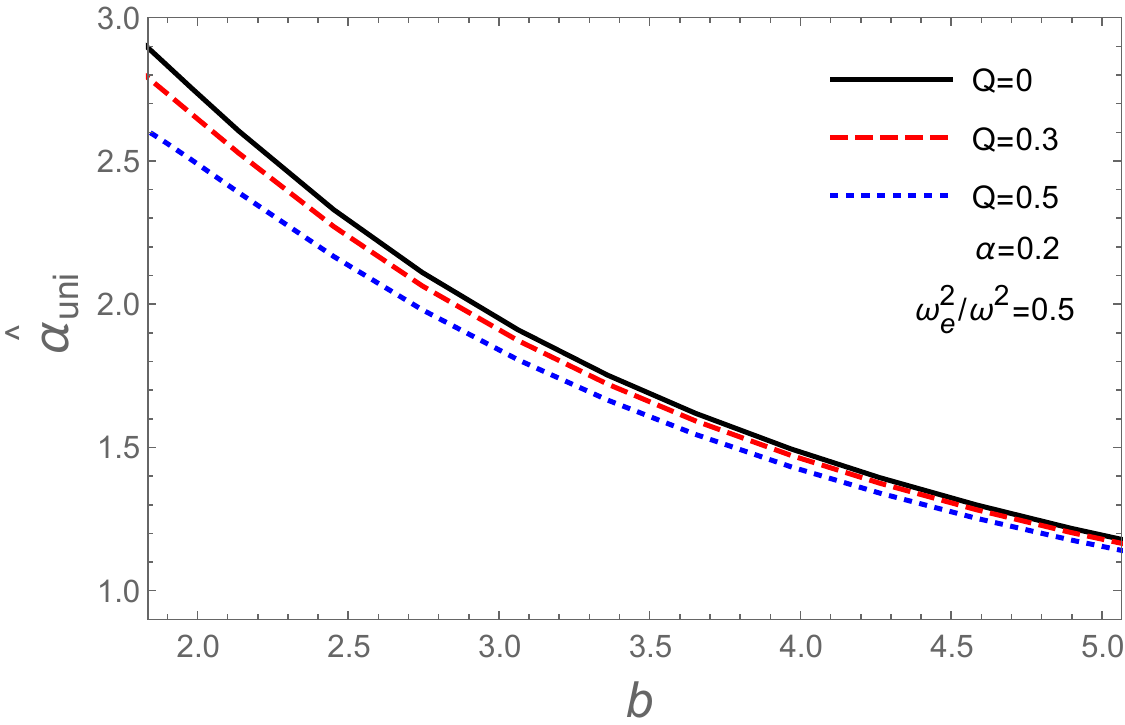}
 & \includegraphics[scale=0.35]{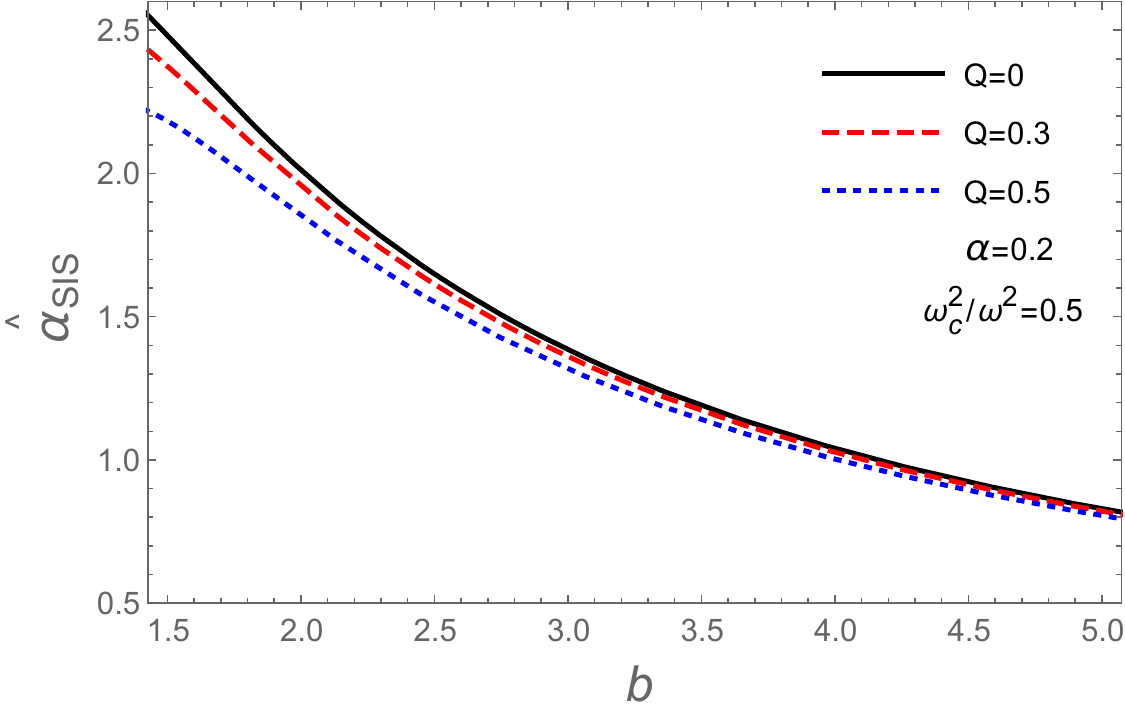}\\ 
   \includegraphics[scale=0.35]{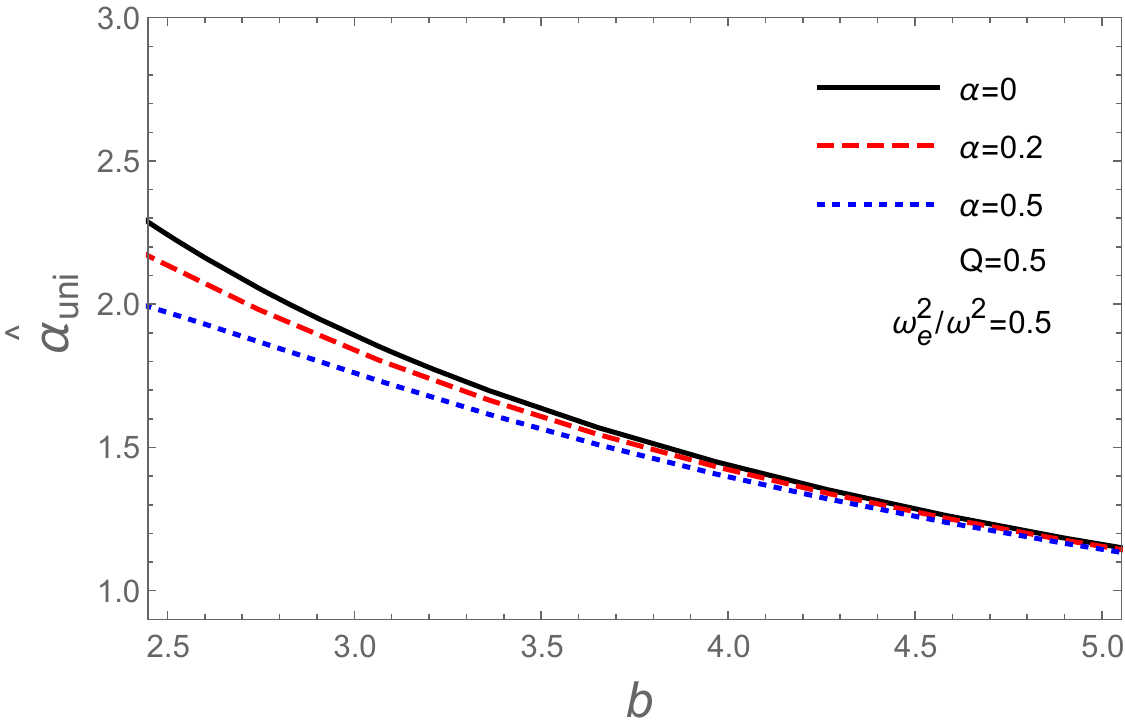}
 & \includegraphics[scale=0.35]{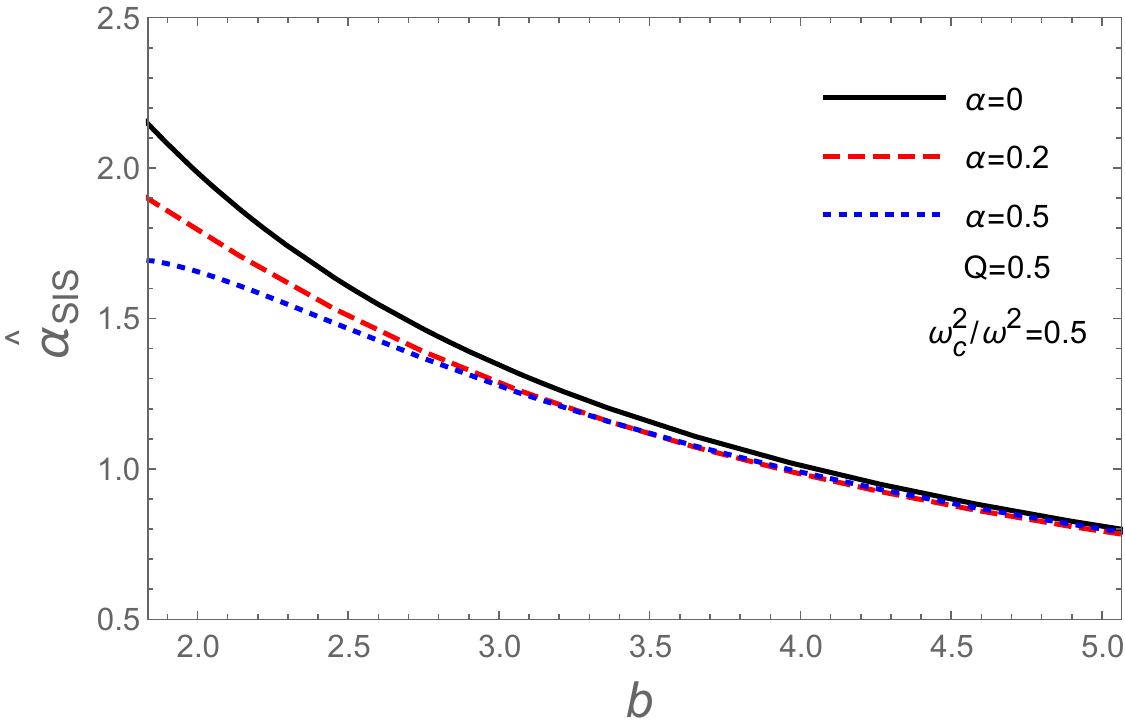}\\ 
   \includegraphics[scale=0.35]{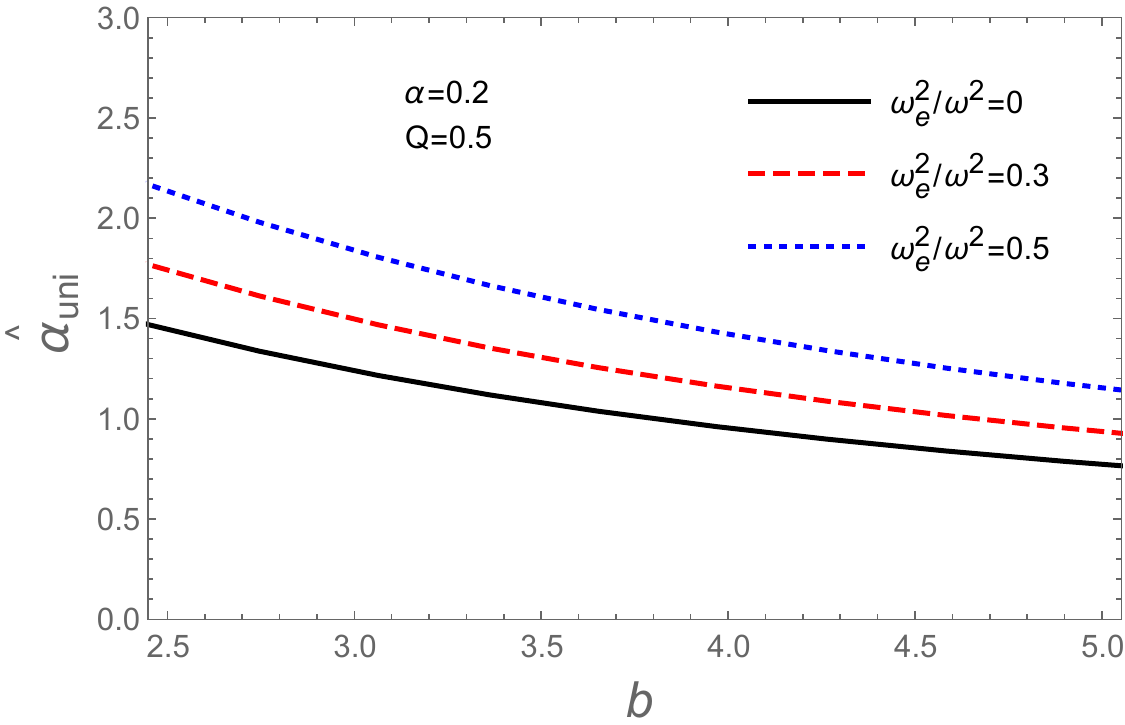}
 &  \includegraphics[scale=0.35]{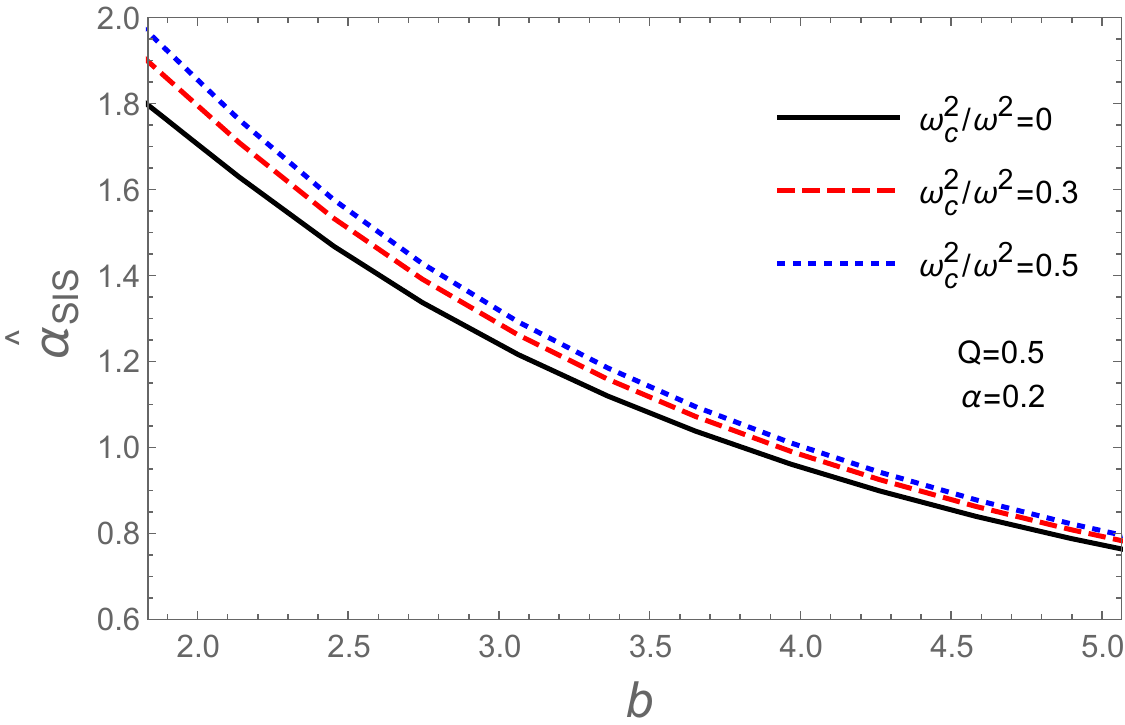} 
  \end{tabular}
\caption{\textit{Left column: }shows the variation of deflection angle $\hat{\alpha}_{b}$ as a function of the impact parameter $b$ in uniform plasma for different combinations of BH charge $Q$, GB coupling constant $\alpha$ and plasma parameter $\frac{\omega^2_{e}}{\omega^2}$. \textit{Right column: } shows the variation of deflection angle $\hat{\alpha}_{b}$ for the case corresponding cases in presence of non-uniform plasma.}
\label{deflectionunib}
\end{figure}


\begin{figure}[t]
\begin{tabular}{c c}
   \includegraphics[scale=0.35]{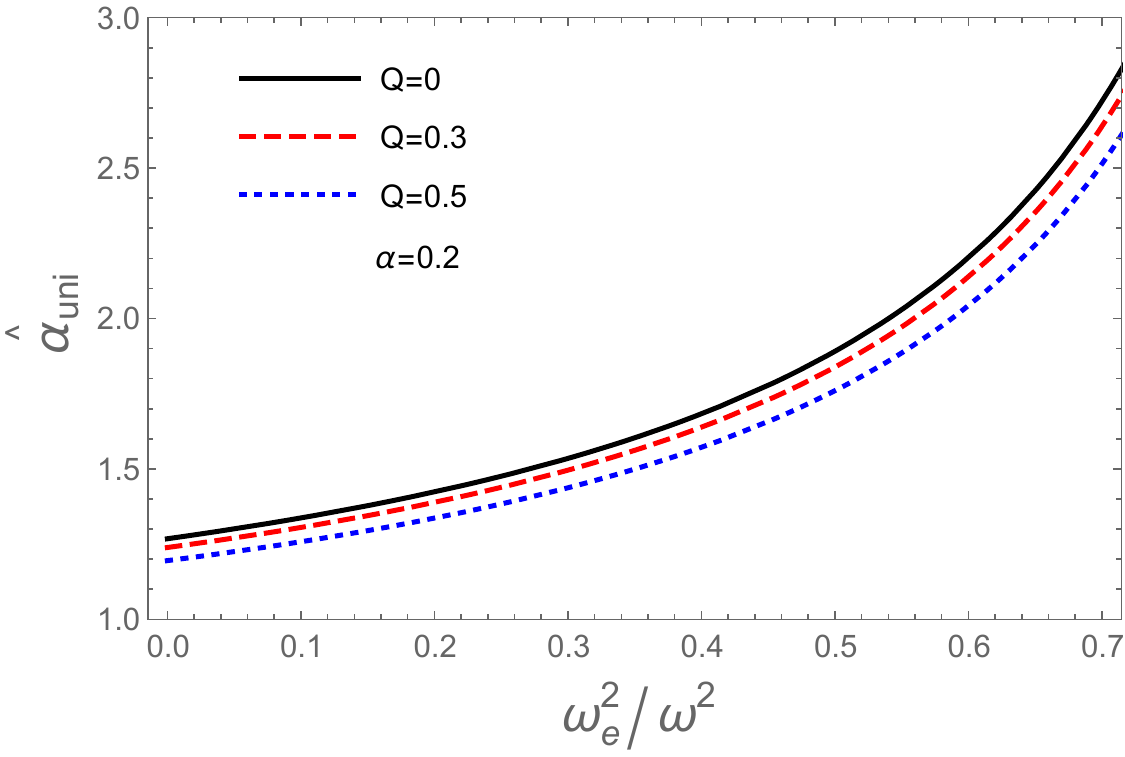}
&  \includegraphics[scale=0.35]{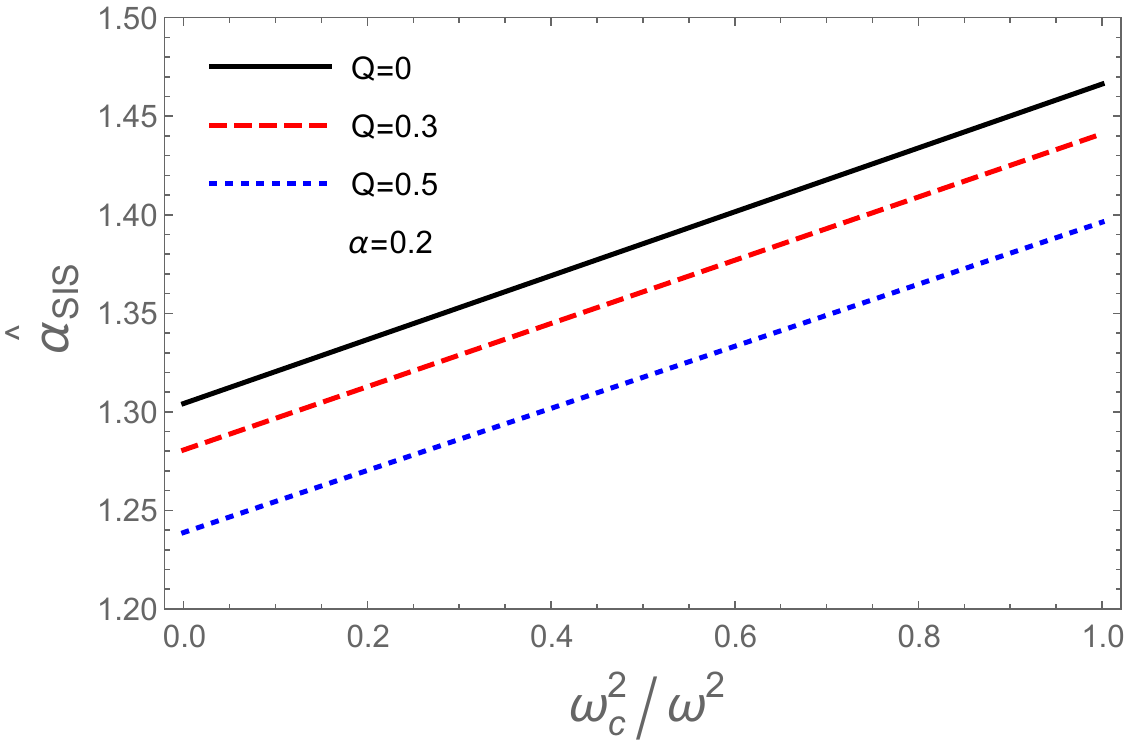}\\   
   \includegraphics[scale=0.35]{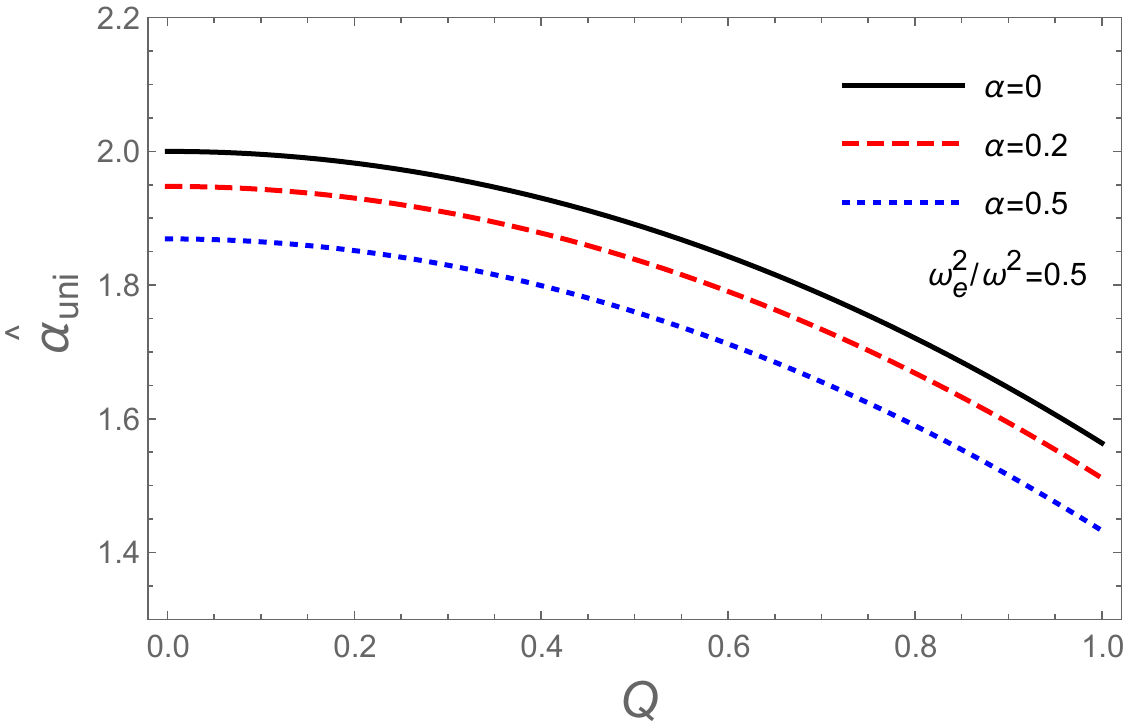}
&  \includegraphics[scale=0.35]{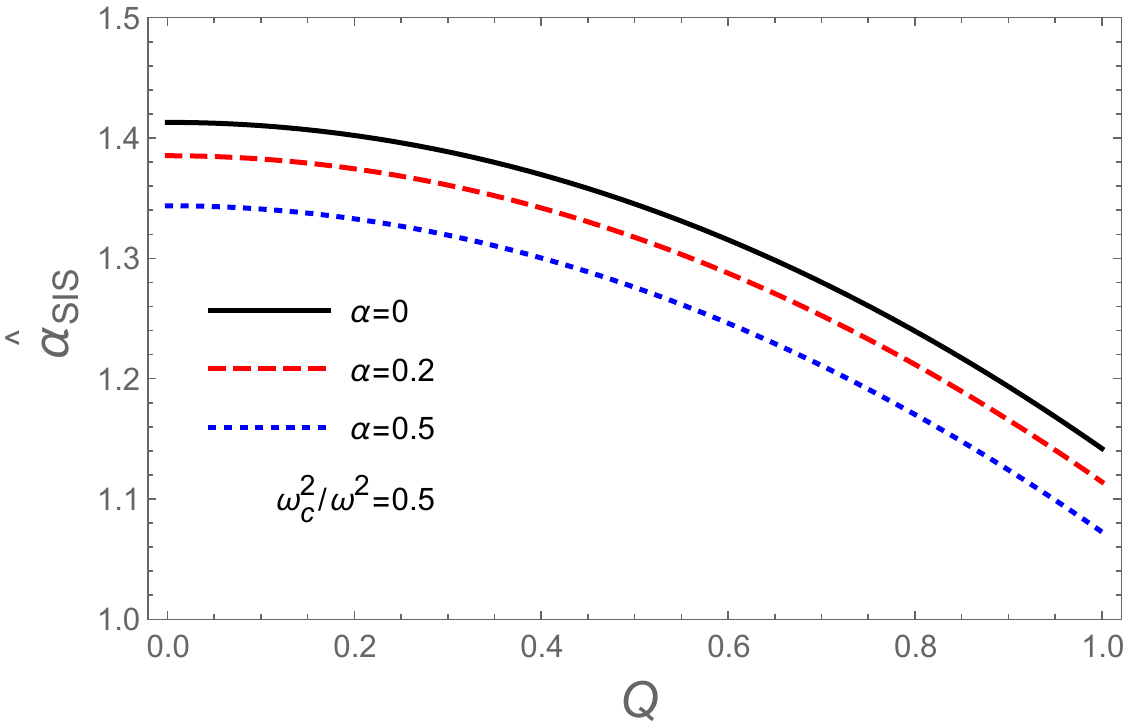}\\   
   \includegraphics[scale=0.35]{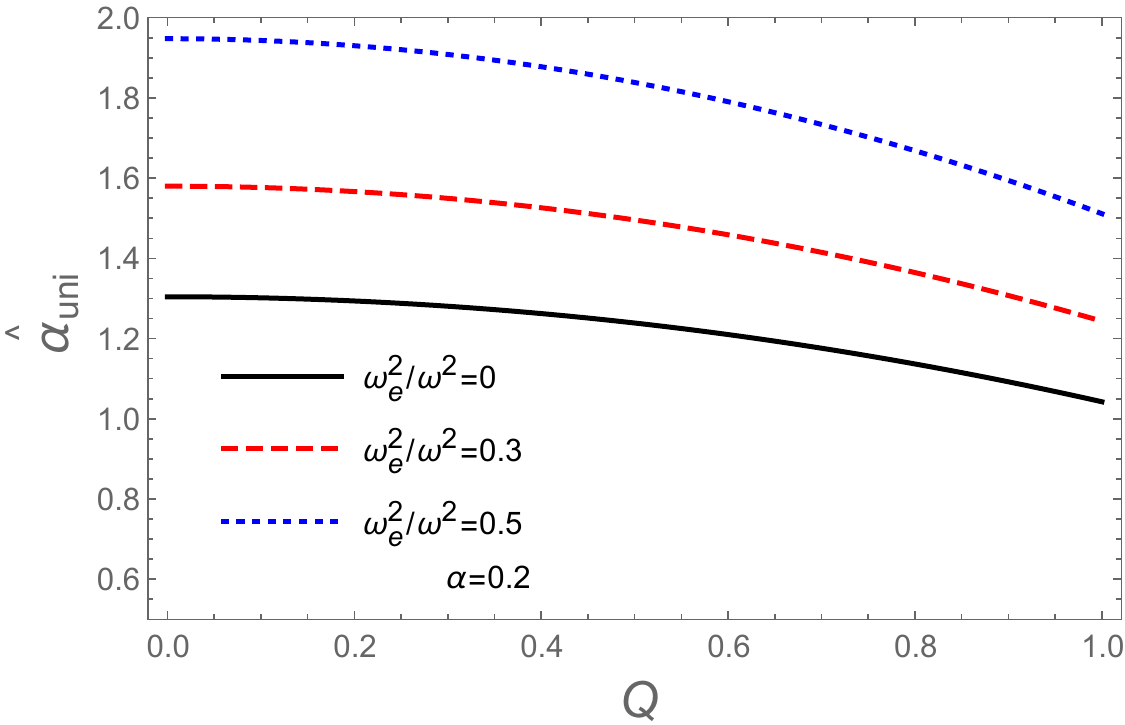}
&  \includegraphics[scale=0.35]{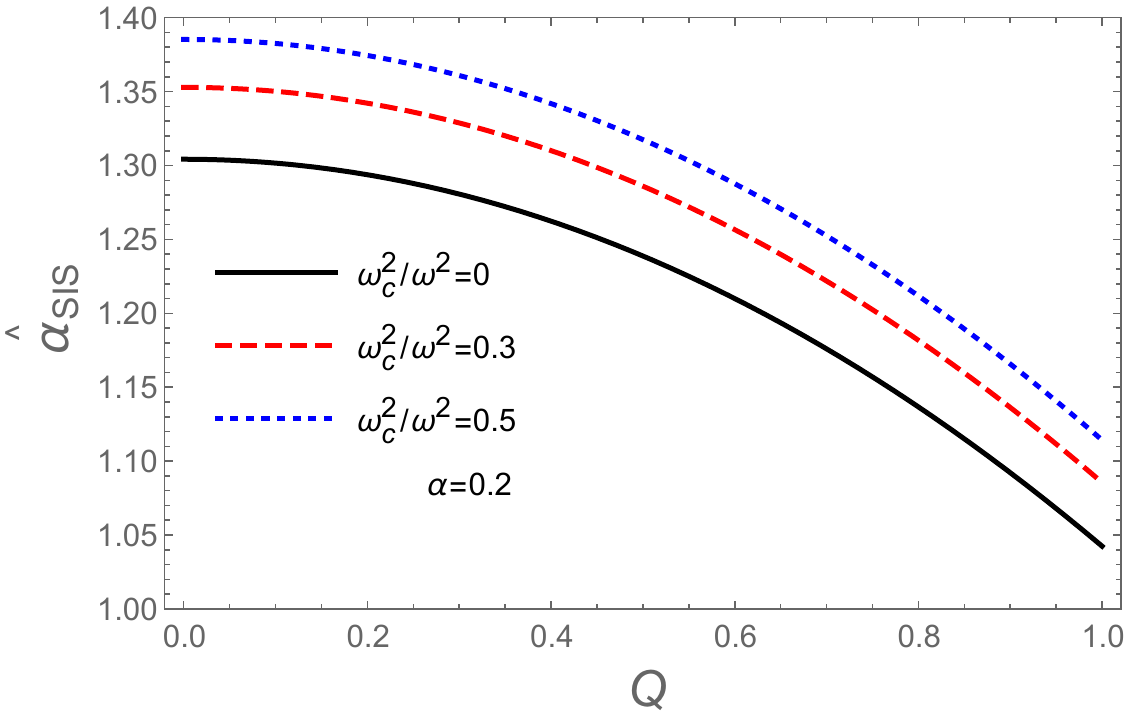} 
  \end{tabular}
\caption{\textit{Left column: }shows the variation of deflection angle $\hat{\alpha}_{b}$ in the presence of uniform-plasma as a function of plasma parameter (top panel) and charge $Q$ (middle and bottom panels) for different combinations of GB coupling constant $\alpha$. Similarly, \textit{right column: }shows the variation of deflection angle $\hat{\alpha}_{b}$ for the corresponding cases in the presence of non-uniform plasma. Here, for both the columns $b=3$.}
\label{deflectionsisb}
\end{figure}

%
We now analyse the light deflection with a uniform plasma medium.  From Eq.~(\ref{alfa}), one can easily obtain the following new expression 
\begin{align}
\hat{\alpha}_\mathrm{uni}=-\bigg(\frac{\omega^2}{\omega^2-\omega^2_{e}}\bigg)\bigg(\frac{\pi Q^2}{2b^2}-\frac{R_g}{b}+\frac{3\pi\alpha R^2_{s}}{4b^4}\bigg)
-\bigg(\frac{\pi Q^2}{4b^2}-\frac{R_g}{b}+\frac{3\pi\alpha R^2_{s}}{16b^4}\bigg)\, .
\label{uniequation}
\end{align}
For the uniform plasma $\partial_r N=0$ is always satisfied. 
We now provide plots representing the behaviour of deflection angle for various parameters such as the GB coupling constant $\alpha$, BH charge $Q$, and the plasma medium parameter $\frac{\omega^2_{e}}{\omega^2}$. Note that here we use Eq.~(\ref{uniequation}) to show the above mentioned plots in Figs.~\ref{deflectionunib} and \ref{deflectionsisb}. As could be seen from Figs.~\ref{deflectionunib} and \ref{deflectionsisb}  the deflection angle decreases with increasing the values of $\alpha$ and $Q$, while this behaviour is opposite with increasing the value of plasma medium parameter $\frac{\omega^2_{e}}{\omega^2}$. Thus, the circular orbits of light propagation can become close to the central object due to the impact of the GB coupling constant and BH charge, as shown in Figs.~\ref{deflectionunib} and \ref{deflectionsisb}. However, the deflection angle increases as that of the plasma medium parameter $\frac{\omega^2_{e}}{\omega^2}$ regardless of $\alpha$ and $Q$, which are regarded as gravitational repulsive charges. From an observational point of view, far away observer could observe an image larger than the object due to the plasma effect. 

\subsection{Non-uniform plasma}
Here we consider the gravitational deflection angle by the $4D$ charged EGB BH in presence of a singular isothermal sphere (SIS) plasma medium which would play an important role in studying lensing properties of galaxies and clusters~\cite{Kogan10}.

Following \cite{Kogan10,Babar2021a} the distribution of density and the plasma concentration will respectively read as follows: 
\begin{align}
\rho(r)=\frac{\sigma^2_{v}}{2\pi r^2},\label{rho}
\end{align}
with $\sigma^2_{v}$ being a 1D velocity dispersion and 
\begin{align}
N(r)=\frac{\rho(r)}{\kappa m_p}\, , \label{conelec}
\end{align}
with mass of proton $m_p$ and a 1D coefficient $\kappa$ related to the dark matter contribution. 

From Eq.~(\ref{alfa}) the gravitational deflection angle for the non-uniform plasma medium can be obtained as \cite{Kogan10},
\begin{align}
\hat{\alpha}_\mathrm{SIS}=\frac{\omega^2_{c}R^2_{g}}{\omega^2b^2}\bigg(\frac{1}{2}-\frac{3Q^2}{8b^2}
-\frac{2R{_g}}{3b\pi}+\frac{5\pi\alpha R^2_{s}}{8b^4}\bigg) 
-\bigg(\frac{3 \pi Q^2}{4b^2}-\frac{2R_g}{b}+\frac{15\pi\alpha R^2_{s}}{16b^4}\bigg)\, ,
\end{align}
where $\omega^2_{c}$ represents another plasma constant, and that is introduced as~\cite{Babar2021a,Far:2021a}
\begin{align}
\omega^2_{c}=\frac{\sigma^2_{v} K_e}{2\kappa m_p R^2_{g}}\, .
\label{sisequation}
\end{align}
Using Eq.~(\ref{sisequation}), we provide plots which reflect the role of non-uniform plasma, the GB coupling constant and BH charge for the gravitational lensing properties. The dependence of the deflection angle on impact parameter of orbits $b$ is shown in the right column of Fig.~\ref{deflectionunib}. For large values of the impact parameter $b$ the deflection angle approaches to the zero in the case when $Q$ and $\alpha$ and $\frac{\omega^2_{e}}{\omega^2}$ are taken into account, see right column of Fig.~\ref{deflectionunib}. The deflection angle as a function of BH charge $Q$ and plasma parameter $\frac{\omega^2_{e}}{\omega^2}$ is shown in right column of Fig.~\ref{deflectionsisb}. The top panel in the right column of Fig.~\ref{deflectionsisb}, reflects the deflection angle as a function of plasma parameter $\frac{\omega^2_{e}}{\omega^2}$ for different values of $Q$ for fixed $\alpha$, while the middle and bottom panels respectively reflect the deflection angle as a function of BH charge $Q$ for different values of $\alpha$ for fixed $\frac{\omega^2_{e}}{\omega^2}$ and for vise versa. From right column of Figs.~\ref{deflectionunib} and \ref{deflectionsisb} one can easily notice that the GB coupling constant and BH charge have similar effect that decreases the deflection angle. However, the deflection angle increases linearly in the presence of non-uniform plasma medium surrounding BH in comparison to uniform-plasma where it increases exponentially (see top row of Fig.~\ref{deflectionsisb}). It is worth noting here that the deflection angle in the presence of non-uniform plasma is smaller than in the presence of uniform plasma medium.

\section{Collision and Centre-of-mass energy}\label{Sec:energy}

\subsection{Centre-of-mass energy of two colliding spinless charged particles }\label{Sec:energy_charge_part}

 We now study the particle acceleration process of two charged particles colliding near the horizon of the $4D$ charged EGB BH. We shall consider two particles that have rest masses $m_{1}$ and $m_2$ at a distance far away from BH for simplicity. Let us write the four-momentum and the total momenta for colliding two particles $(i =1, 2)$  as 
\begin{eqnarray}\label{Eq:4-momentum}
p_{i}^{\alpha}&=&m_{i}u_{i}^{\alpha}, \nonumber\\
p_{t}^{\alpha}&=&p_{1}^{\alpha}+p_{2}^{\alpha}\,.
\end{eqnarray}
with the four velocity $u_{i}^{\mu}$ of particle $i$ and the four-momentum $\pi^{\alpha}$ which is given by
\begin{eqnarray}\label{Eq:canonical}
\pi^{\alpha}= g^{\mu\nu}\left(\pi_{\nu}-qA_{\nu}\right)\, .  
\end{eqnarray}
Following to \cite{Banados09} and Eq.~(\ref{Eq:4-momentum}) the extracted energy $E_{C.M.}$ by collision between two particles is defined by 
\begin{eqnarray}\label{Eq:cm1}
\frac{E_{C.M.}^{2}}{2m_1 m_2}=\frac{m_{1}^2+m_{2}^2}{2m_1
m_2}-g_{\alpha\beta}u^{\alpha}_{1}u^{\beta}_{2}\, .
\end{eqnarray}
The above the center of mass energy $E_{C.M.}$ allows to understand how the GB coupling constant $\alpha$ and BH charge $Q$ impact on the extracted energy by collision of two particles. In this respect, it would be important to study the extracted energy in the $4D$ charged BH in EGB gravity and compare to Einstein gravity. In other words this collision mechanism could play a crucial role in explaining high energy particle processes in the vicinity of BH candidates. Further for simplicity we assume two free falling particles collide at the near horizon. In doing so,  using Eqs.~(\ref{Eq:4-momentum}) and (\ref{Eq:cm1}) we obtain  %
\begin{figure*}
	\centering
	\includegraphics[width=0.45\textwidth]{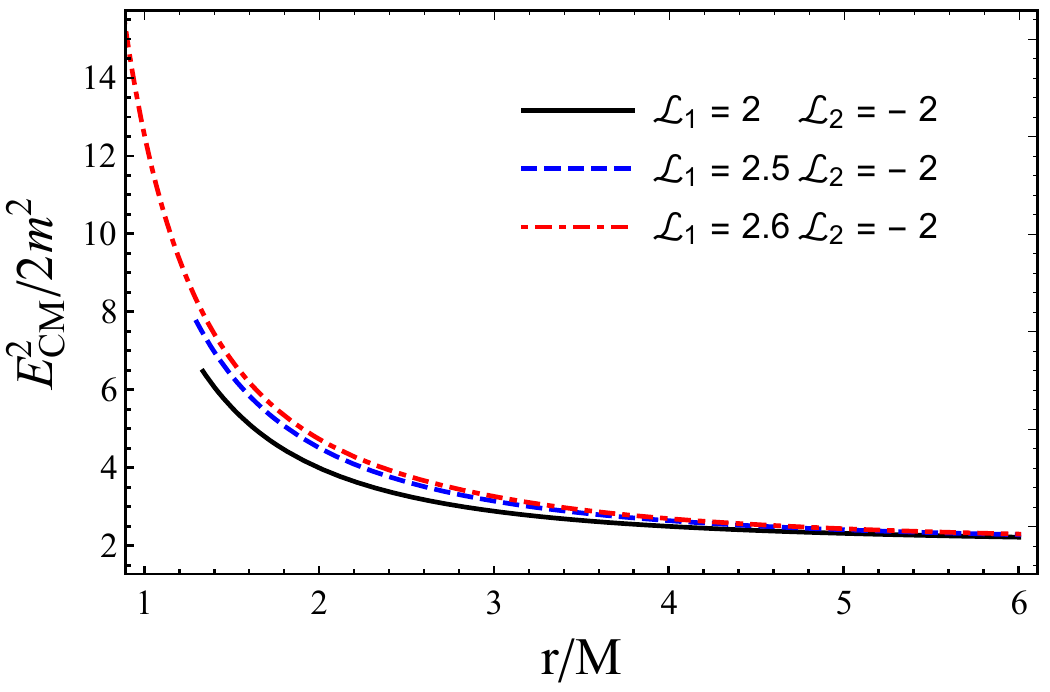}
	\includegraphics[width=0.45\textwidth]{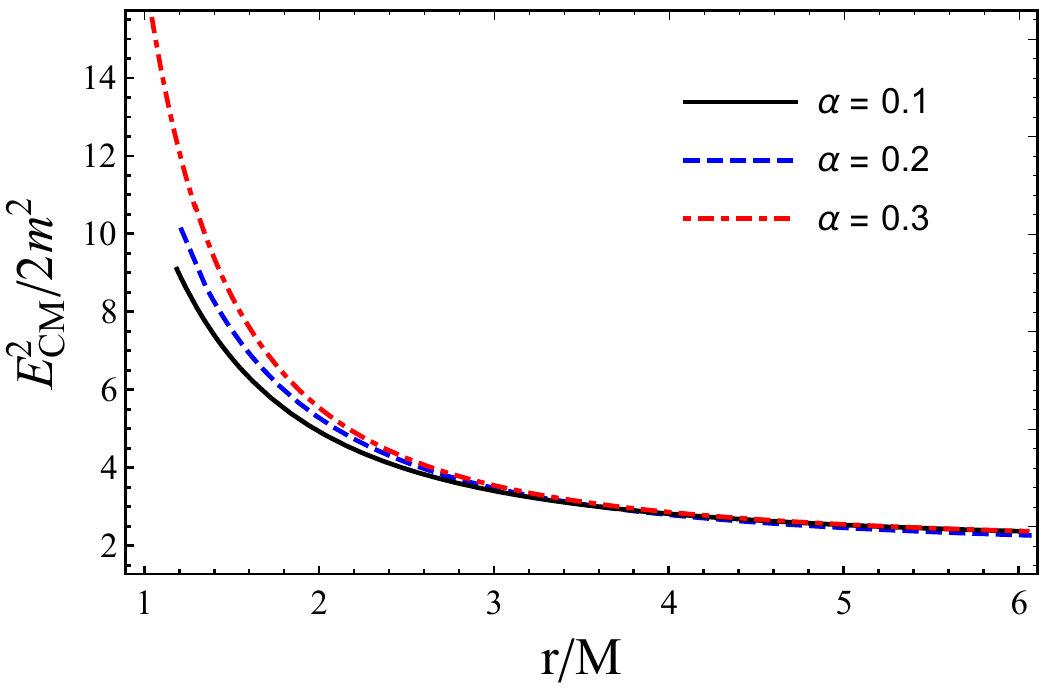}
	
	\caption{\label{fig:cm1} 
	The behaviour of $E_{C.M}$ vs $r$ for two non-spinning particles colliding near the horizon of a 4D charged EGB BH is presented. For the \textit{left panel}, we select the extreme BH case (i.e. $r_{h}/M=1$) and set the values of the parameters $\alpha$ and $Q$ appropriately for various combinations of $\mathcal{L}_1$ and $\mathcal{L}_2$. For the \textit{right panel}, the parameters $Q$, $\mathcal{L}_1$, and $\mathcal{L}_2$ are set to $0.8$, $2.6$, and $-2$, respectively, and the parameter $\alpha$ is varied in 0.1-step increments from 0.1 to 0.3. Here, the value energy for both particles is set to unity.
}
\end{figure*}
\begin{align}\label{Eq:cm2} 
\frac{E^{2}_{C.M.}}{2m_{1}m_{2}} & = 1+\frac{\left(m_{1}-m_{2}\right)^2}{2m_1
m_2}+\frac{\tilde{\mathcal{E}}_1 \tilde{\mathcal{E}}_2}{F(r)}-\frac{\mathcal{L}_1\mathcal{L}_2}{r^2}-\frac{\sqrt{\tilde{\mathcal{E}}_1^2-F(r) \left(1+\frac{\mathcal{L}_1^2}{r^2}\right)}\sqrt{\tilde{\mathcal{E}}_2^2-F(r) \left(1+\frac{\mathcal{L}_2^2}{r^2}\right)}}{F(r)}\, ,
\end{align}

where we have defined $\tilde{\mathcal{E}}_{i}=\mathcal{E}_{i}-q_{i}Q/r$ and we have set $M=1$. The extracted energy is given by Eq.~(\ref{Eq:cm2}) for the collision of two particles with energies $\mathcal E_{1,2}$ and specific angular momenta $\mathcal L_{1,2}$. Let us then evaluate the center-of-mass energy extracted by the freely falling charged particles colliding near the  BH horizon. For evaluating $E_{C.M.}$ we shall for simplicity assume free falling particles $m_{1,2} = m$ and these two particles have different angular momenta $\mathcal{L}_1=\mathcal{L}_{max}$ and $\mathcal{L}_2=\mathcal{L}_{min}$ (where $\mathcal{L}_{min}$ and $\mathcal{L}_{max}$ are angular momenta of the two charged particles falling into the black hole with opposite directions). In Fig.~\ref{fig:cm1}, we show the radial dependence of $E_{C.M.}$ energy. As can bee seen from Fig.~\ref{fig:cm1} the minimum energy occurs when two charged particles have opposite angular momentum. However, the extracted energy goes up with increasing $\mathcal{L}_1=\mathcal{L}_{max}$. We note that  $E_{C.M.}$ reaches its maximum value as a consequence of the increase in value for the coupling constant $\alpha$ for fixed value of $Q$, thus resulting in increasing the value for the extracted energy in contrast to the RN BH.      

We now consider the limiting case $r\to r_{h}$ so as to find the highest values of the center of mass energy. Considering $E_1/m_1=E_2/m_2=1$ for free falling particles we then have the limiting value of $E_{C.M.}$ at the horizon
\begin{align}\label{Eq:cm3}
 \frac{E^{2}_{C.M.}(r\rightarrow r_{h})}{m^2}
\approx \Big[4+\left(1-6\sqrt{1-Q^2-\alpha}\right) \Big(\mathcal{L}_1-\mathcal{L}_2\Big)^2 \Big]\, .
\end{align}
As seen from Eq.~(\ref{Eq:cm3}) this clearly shows that the limiting value of center of mass energy extracted by collision of two particles at the horizon increases due to the impact of the GB coupling constant and BH charge. That is, the horizon radius can become arbitrarily close to the BH as the coupling constant and BH charge increase.  

\subsection{Centre-of-mass energy in the context of spinning massive test particles}\label{subsec:Ecm}

It is shown that centre of mass energy ($E_{C.M.}$) for two colliding equal-mass non-(spinning \cite{Sheoran:2020kmn}) \cite{Baushev:2008yz} particles becomes maximum. However, the maximum $E_{C.M.}$ for non-spinning particles case has an upper bound \cite{Banados09} while there is no upper bound if the colliding particles are spinning \cite{Armaza:2015eha,Zaslavskii:2016dfh}. For rotating BHs, the maximum $E_{C.M.}$ increases in comparison with their non-rotating counterparts and becomes infinite when the rotating BH satisfies an extremal condition. In this subsection, our main goal is to bring out the effect of particle's spin on the $E_{C.M.}$.

The $E_{C.M.}$ of two colliding equal mass spinning test particles falling from rest at infinity is defined as
\begin{align}
    E_{C.M.}^{2} &\equiv -(P_{1\mu}+P_{2\mu})(P_{1}^{\mu}+P_{2}^{\mu}),\nonumber\\
    &= -\left(P_{1\mu}P_{1}^{\mu}+P_{2\mu}P_{2}^{\mu}+P_{1\mu}P_{2}^{\mu}+P_{2\mu}P_{1}^{\mu}\right),\nonumber\\
    &=-2\left(-m^{2}+P_{1\mu}P_{2}^{\mu}\right)=2m^{2}\left(1-\frac{P_{1\mu}P_{2}^{\mu}}{m^{2}}\right).\label{Ecm_gen}
\end{align}
Now, by using Eqs. \eqref{Pt}, \eqref{Pph} and \eqref{Pr} in the above Eq. \eqref{Ecm_gen}, the $E_{C.M.}$ per unit mass reads as
\begin{align}
    \left(\frac{E_{C.M.}}{m}\right)^2 &=2\left[1+\left(1+\frac{\left(r^2-\mathcal{B}\right)}{2\alpha}\right)^{-1}\mathcal{K}_{1}\mathcal{K}_{2}-\frac{1}{r^{2}}\mathcal{Z}_{1}\mathcal{Z}_{2}- \left(1+\frac{\left(r^2-\mathcal{B}\right)}{2\alpha}\right)^{-1}\right.\nonumber
    \\
    & \left. \times \sqrt{\mathcal{K}_{1}^{2}-\left(1+\frac{\left(r^2-\mathcal{B}\right)}{2\alpha}\right)\left[1+\frac{\mathcal{Z}_{1}^{2}}{r^{2}}\right]}\, .\label{Ecm1}
    \sqrt{\mathcal{K}_{2}^{2}-\left(1+\frac{\left(r^2-\mathcal{B}\right)}{2\alpha}\right)\left[1+\frac{\mathcal{Z}_{2}^{2}}{r^{2}}\right]}
    \right] 
\end{align}
From straightforward calculations, the above equation yields
\begin{align}
    \left(\frac{E_{C.M.}}{m}\right)^2 &= \frac{2}{\Delta \Pi_{1}\Pi_{2}}\Bigg[\left(2\alpha r \mathcal{B}E_{1}-\tilde{S}_{1}J_{1}\mathcal{C}\right)\left(2\alpha r \mathcal{B}E_{2}-\tilde{S}_{2}J_{2}\mathcal{C}\right)+\Delta\Big[\Pi_{1}\Pi_{2}-4\alpha^{2}\mathcal{B}^{2}\nonumber
    \\
   &  \times \left(J_{1}-E_{1}S_{1}\right)\left(J_{2}-E_{2}S_{2}\right)\Big]-\sqrt{\left(2\alpha r \mathcal{B}E_{1}-\tilde{S}_{1}J_{1}\mathcal{C}\right)^{2}-\Delta\Big[\Pi_{1}^{2}+4\alpha^{2}\mathcal{B}^{2}\left(J_{1}-E_{1}S_{1}\right)^{2}\Big]}\nonumber
   \\
   & \times \sqrt{\left(2\alpha r \mathcal{B}E_{2}-\tilde{S}_{2}J_{2}\mathcal{C}\right)^{2}-\Delta\Big[\Pi_{2}^{2}+4\alpha^{2}\mathcal{B}^{2}\left(J_{2}-E_{2}S_{2}\right)^{2}\Big]}\Bigg],\label{Ecm2}
\end{align}
where $\Delta\equiv F(r) \equiv \left[1+(1/2\alpha)\left(r^2-\mathcal{B}\right)\right]$, $\mathcal{C}\equiv 2\alpha r \mathcal{AB}$ and $\Pi_{1,2}\equiv (2\alpha r \mathcal{B}-\tilde{S}_{1,2}^{2}\mathcal{C})$. From Eq. \eqref{Ecm2}, one can easily deduce that $E_{C.M.}$ could possibly become infinitely high not only at horizon of $4D$ charged EGB BH $\Delta=0$, but also at either $\Pi_{1}=0$ or $\Pi_{2}=0$. However, for the former case, it is already a well established fact that a static spherically symmetric BH $E_{C.M.}$ will not diverge at the horizon. Also, for the later case, the potential divergence of $E_{C.M.}$ will occurs at some radial distance, but this is not straightforward in nature due to the same reason, as mentioned in \cite{Armaza:2015eha} for Schwarzschild BH. In the present subsection, we will consider another interesting scenario where the collision occurs near the horizon but outside it (i.e., $r_{0}<r_{h}=M+\sqrt{M^{2}-Q^{2}-a}$). Therefore, the only way to diverge for $E_{C.M.}$ extracted by the collision of two spinning particles is to make $\Delta$ small enough.

\subsubsection{Classification of spinning particle and their different combinations}\label{classification of spinning particle}
Now, in order to discuss this case in more detail, we need to classify the spinning particle into three types, i.e. \textit{usual particle, critical particle} and \textit{near-critical particle}, as mentioned in \cite{Zaslavskii:2016dfh}. A particle is known as \textit{usual} if:
\begin{align}
    \mathcal{K}_{h}\equiv \frac{E-\tilde{S}J\mathcal{A}_{h}}{1-\tilde{S}^{2}A_{h}}\neq 0\, .\label{mathcalK}
\end{align}
Here and hereafter ``$h$" in the subscript means that the value of corresponding quantity is taken at the horizon $r_{h}$. On the other hand, a particle is considered as \textit{critical} if $\mathcal{K}_{h}=0$ always
\begin{align}
    E-\tilde{S}J\mathcal{A}_{h}=0\, .\label{critical_con}     
\end{align}
Then, $\mathcal{K}$ near the horizon reads as
\begin{align}
   \mathcal{K} \approx \left[\frac{J \tilde{S} \left(3 M r_{h} \left(r_{h}^3-4 \alpha  M\right)-4 Q^2 \left(r_{h}^3-\alpha  M\right)\right)}{r_{h}^2 \left(8 \alpha  M r_{h}-4 \alpha  Q^2+r_{h}^4\right)^{3/2}\left(1-\tilde{S}^{2}A_{h}\right)}\right](r-r_{h})\, .
\end{align}
Thus, at the colliding radius $r=r_{c}$ where $\mathcal{K}_{c}=O(\Delta_{c})$, the second term dominates Eq. \eqref{Pr} so that
$(P^{r}/m)_{c}$ becomes imaginary, and hence such a spinning particle cannot reach the $r_{h}$ of $4D$ charged EGB BH (hereafter ``$c$" in the subscript means that the value of corresponding quantity is taken at the point of collision $r_{c}$).
And, a colliding spinning particle is known as \textit{near-critical} particle if and only if $\mathcal{K}_{h}=\sqrt{r_{c}-r_{h}}$ is satisfied 
\begin{align}
   E-\tilde{S}J\mathcal{A}_{h} = O\left(\sqrt{r_{c}-r_{h}}\right)\, .
\end{align}
Let us then consider different combinations of colliding particles depending upon the classification done till now and check, for which combination of two colliding spinning particles the divergence of $E_{C.M.}$ can occur as follows:
\begin{itemize}
    \item \textit{Case I:} When both colliding spinning particles are usual (i.e. $P_1 = P_2 = usual$) it is easy to check from Eq. \eqref{Ecm2} that $E_{C.M.}$ is finite and hence the BSW effect is absent. The same has already been stated earlier in this section that if the collision of two spinning particles will take place at the $r_{h}$, $E_{C.M.}$ then becomes finite.
    \item \textit{Case II:} When one of the two colliding particles is critical and the other one is usual (i.e. $P_1 = critical$ and $P_2 = usual$), then $P_1$ cannot reach the horizon. Hence, no collision will take place near $r_{h}$ and $E_{C.M.}$ becomes finite, thereby indicating that the BSW effect is absent.
    \item \textit{Case III:} When both the particles are critical (i.e. $P_1 = P_2 = critical$), then again both the particles cannot reach the horizon. Hence no collision take place near $r_{h}$, and thus the BSW effect is absent.
    \item \textit{Case IV}: In the case in which particle $P_{1}$ is $near-critical$ and particle $P_2$ is $usual$: let us suppose that for $P_{1}$
    \begin{align}
        \mathcal{K}_{h}=\mathcal{C}\sqrt{\Delta_{c}}+O(\Delta_{c}),\label{near_critic_con}
    \end{align}
    where $\mathcal{C}$ is some finite nonvanishing coefficient corresponding to $P_{1}$. Let us now consider that at the point of collision $\mathcal{K}_{c} \approx \mathcal{K}_{h}+O(\Delta_{c})$ and therefore the second part of Eq. \eqref{Ecm_gen} (i.e. $-P_{1\mu}P_{2}^{\mu}/m^{2}$) becomes
    \begin{align}
       -\frac{P_{1\mu}P_{2}^{\mu}}{m^{2}}&\approx \frac{1}{m^{2}} \left(\frac{P_{1c}^{t}P_{2h}^{t}-P_{1c}^{r}P_{2h}^{r}}{\Delta_{c}}\right)\nonumber\\
       &= \frac{\mathcal{K}_{2h}}{\Delta_{c}}\left[\mathcal{K}_{1c}-\left(\sqrt{\mathcal{K}^{2}-\Delta\left(1+\frac{\mathcal{Z}^{2}}{r^{2}}\right)}\right)_{1c}\right],
       \label{P1P2}
    \end{align}
   where $\mathcal{Z}=(J-E\tilde{S})/(1-\tilde{S}^{2}\mathcal{A})$. It is worth to mention here that in Eq. \eqref{P1P2}, we neglect the expression for ($P_{1r}P_{2}^{r}$) as it remains finite and does not contribute to $E_{C.M.}$  to let it diverge. Using Eqs. \eqref{critical_con} and \eqref{near_critic_con}, Eq. \eqref{P1P2} takes the form
    \begin{align}
       -\frac{P_{1\mu}P_{2}^{\mu}}{m^{2}}&\approx  \frac{\mathcal{K}_{2h}}{\sqrt{\Delta_{c}}}\left[\mathcal{C}_{1}-\left(\sqrt{\mathcal{C}^{2}-1-\frac{E^{2}}{r_{h}^{2}\tilde{S}^{2}\mathcal{A}_{h}^{2}}}\right)_{1}\right]\, .\label{Final_P1P1}
    \end{align}
    It is easy to check that Eq. \eqref{Final_P1P1} diverges in the limit when $\Delta_{c}\to 0$. Therefore, we have successfully found the case where $E_{C.M.}$ will grow illimitably. Further, in the limit when $\alpha \to 0$ and $Q=0$, the expression for $A_{h}$ reduces to $M/r_{h}^{3}$ and Eq. \eqref{Final_P1P1} takes the form
    \begin{align}
      -\frac{P_{1\mu}P_{2}^{\mu}}{m^{2}}&\approx \frac{\mathcal{K}_{2h}}{\sqrt{\Delta_{c}}}\left[\mathcal{C}_{1}-\left(\sqrt{\mathcal{C}^{2}-1-\frac{16M^{2}E^{2}}{\tilde{S}^{2}}}\right)_{1}\right]\, ,
    \end{align}
     which is the same as found in \cite{Zaslavskii:2016dfh} for Schwarzschild BH case.
\item \textit{Case V:} When particle $P_{1}$ is \textit{near-critical} and particle $P_{2}$ is \textit{critical}, then again $P_{2}$ cannot reach the horizon. Hence, no collision will take place near the horizon and $E_{C.M.}$ will not grow infinitely.
\item \textit{Case VI:} When both $P_{1}$ and $P_{2}$ are \textit{near-critical} particles, $-P_{1\mu}P_{2}^{\mu}/m^{2}$ comes out as
\begin{align}
    -\frac{P_{1\mu}P_{2}^{\mu}}{m^{2}} &\approx 1+\frac{E^{2}}{\left(r_{h}\tilde{S}\mathcal{A}_{h}\right)^{2}}\, ,
    \end{align}
    which is a finite quantity. Hence, $E_{C.M.}$ is finite and thus again BSW effect is absent.
\end{itemize}
It is important to note here that out of all six possible cases of two spinning particles only for the \textit{case IV} (i.e. combination of \textit{near-critical} and \textit{usual} particles) unbound growth of $E_{C.M.}$ is obtained if and only if the term under square root in \eqref{Final_P1P1} is positive. Thus, any $near-critical$ particle having the same mass as \textit{usual} particle and satisfying the following condition:
\begin{align}
    \mathcal{C}^{2}&> 1+\frac{E^{2}}{r_{h}^{2}\tilde{S}^{2}\mathcal{A}_{h}^{2}}\, ,
\end{align}
which is required for obtaining the unbounded $E_{C.M.}$

\subsubsection{Bypassing the superluminal motion of spinning particle}
\begin{figure}[H]
\begin{tabular}{c c c}
  \includegraphics[width=0.35\textwidth]{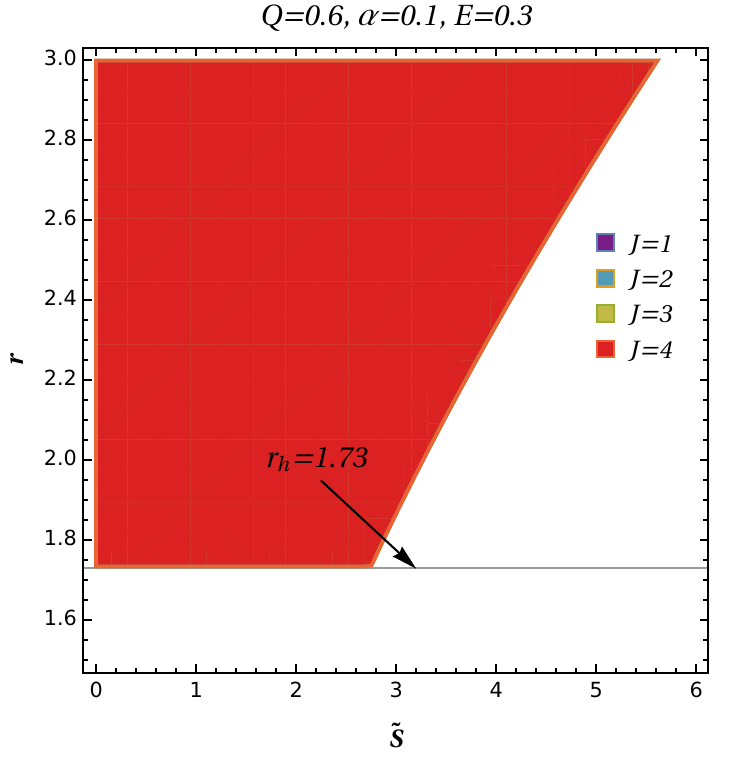}\hspace{-0.5cm}
  &  \includegraphics[width=0.35\textwidth]{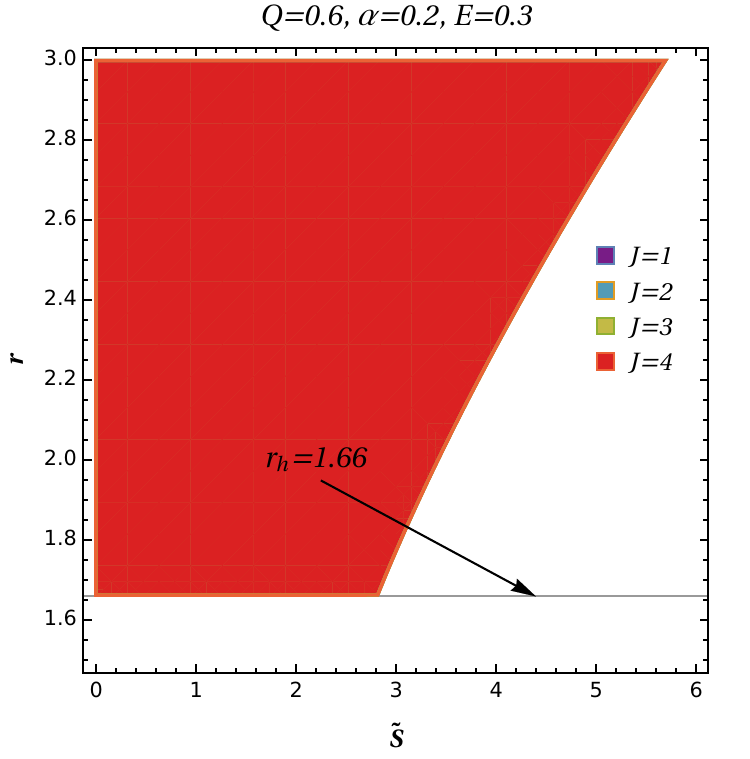}\hspace{-0.5cm}
  &  \includegraphics[width=0.35\textwidth]{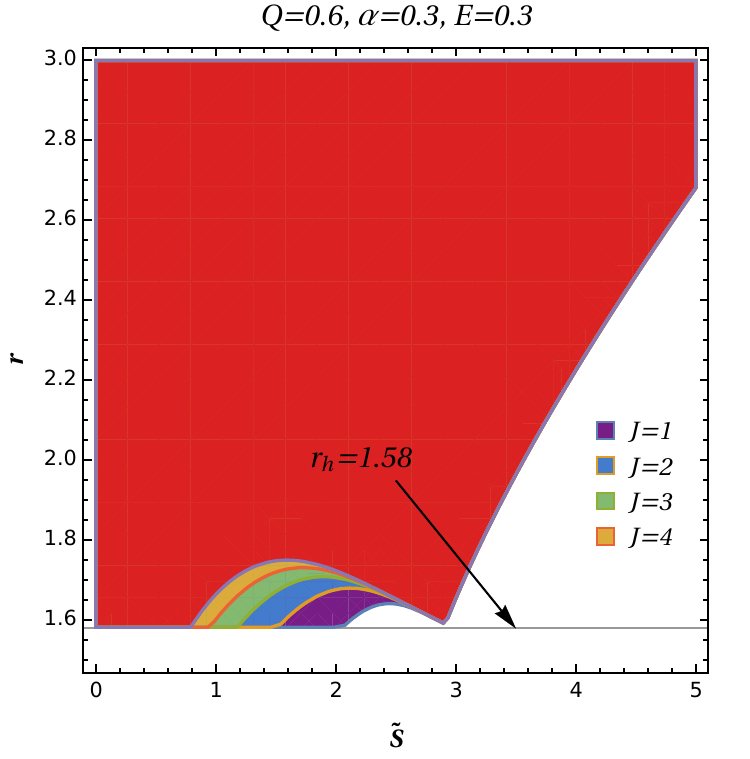}\\
  \includegraphics[width=0.35\textwidth]{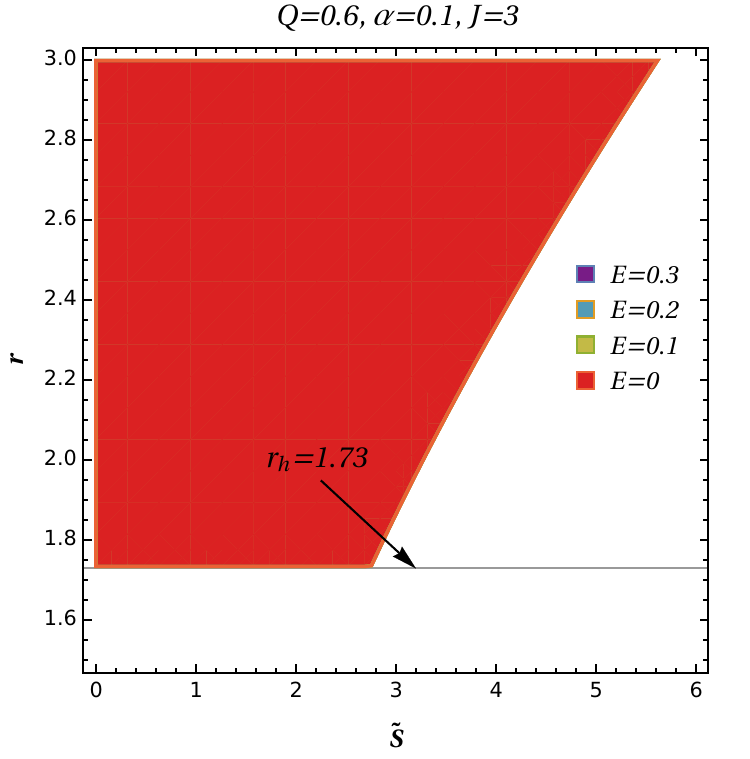}\hspace{-0.5cm}
  &  \includegraphics[width=0.35\textwidth]{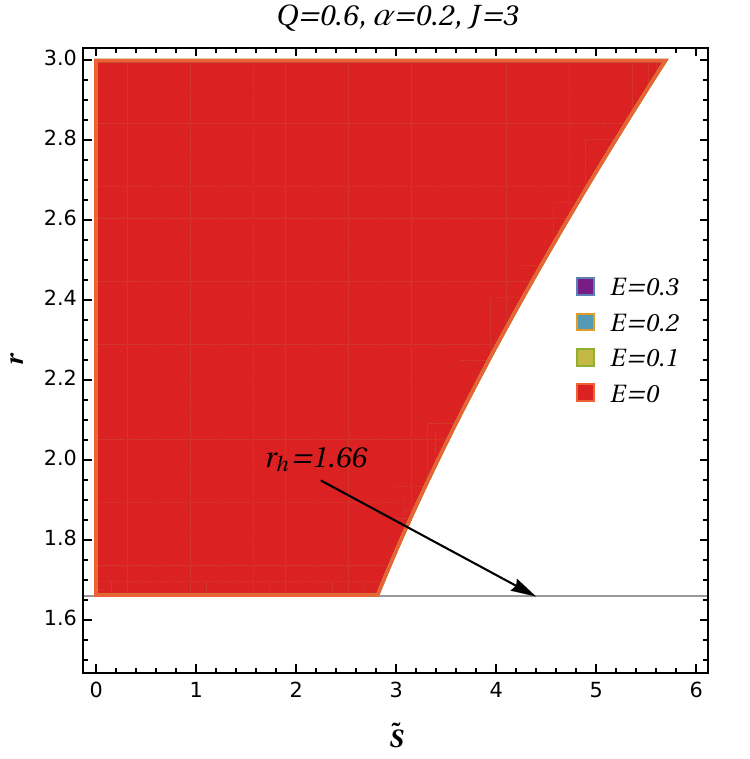}\hspace{-0.5cm}
  &  \includegraphics[width=0.35\textwidth]{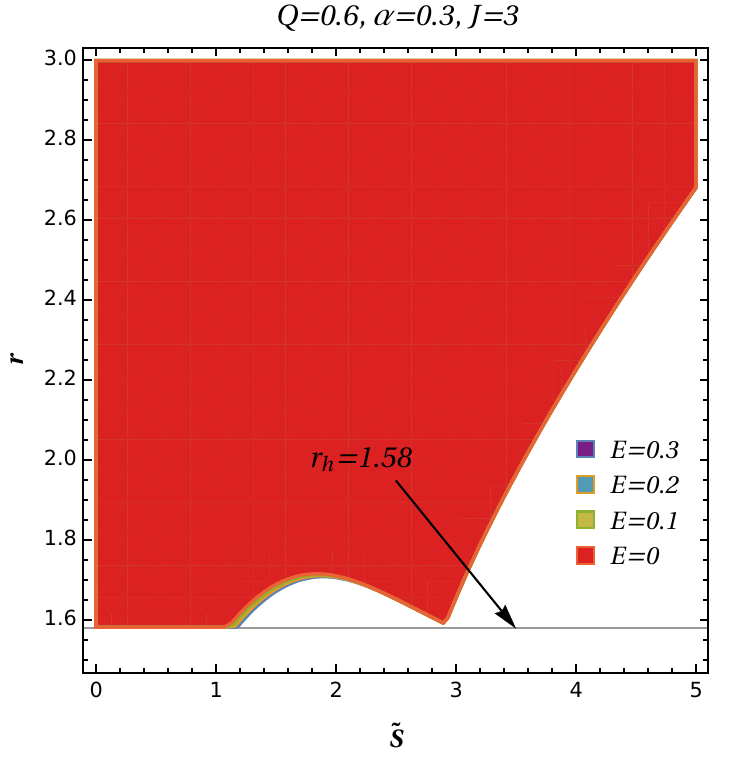}\\
  \includegraphics[width=0.35\textwidth]{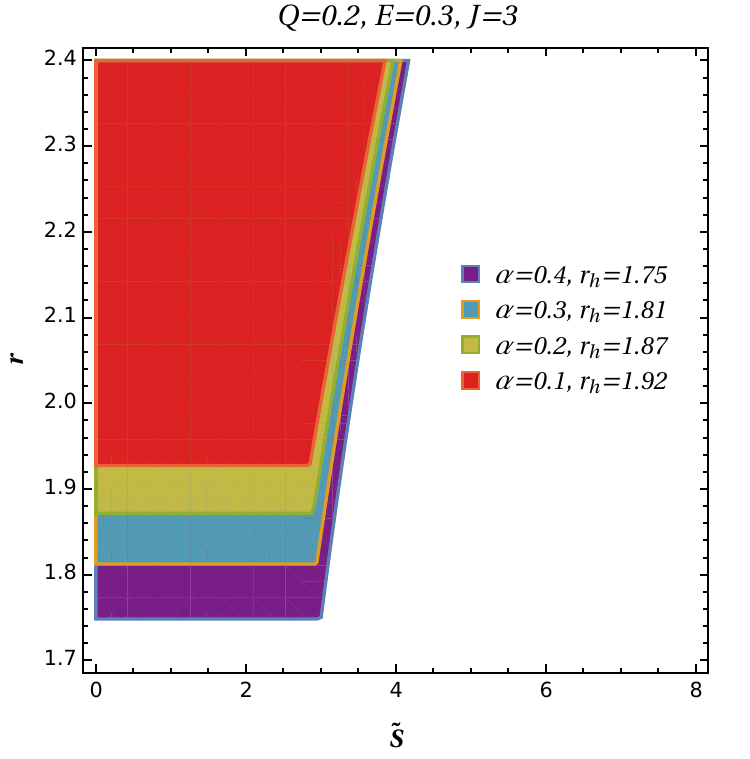}\hspace{-0.5cm}
  &  \includegraphics[width=0.35\textwidth]{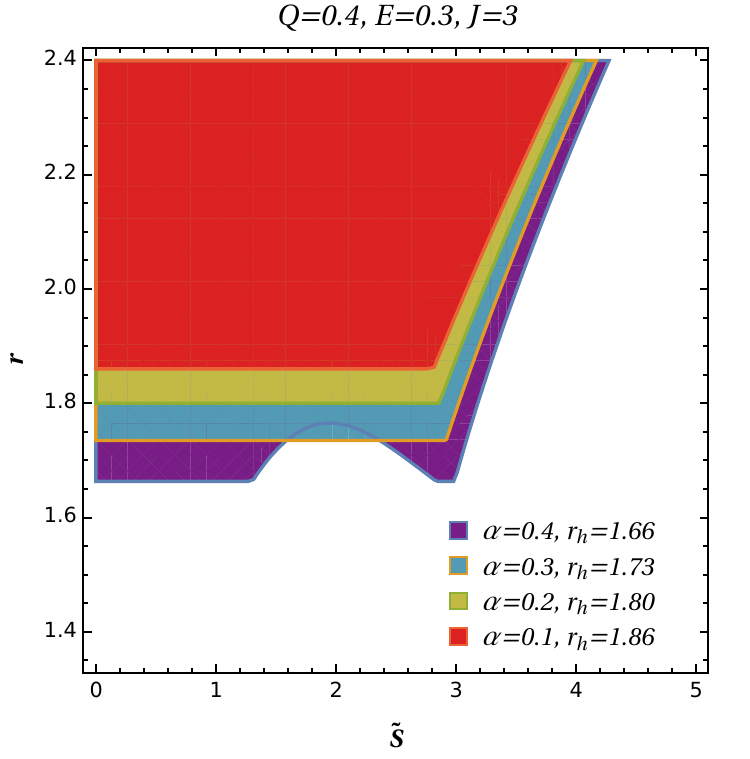}\hspace{-0.5cm}
  &  \includegraphics[width=0.35\textwidth]{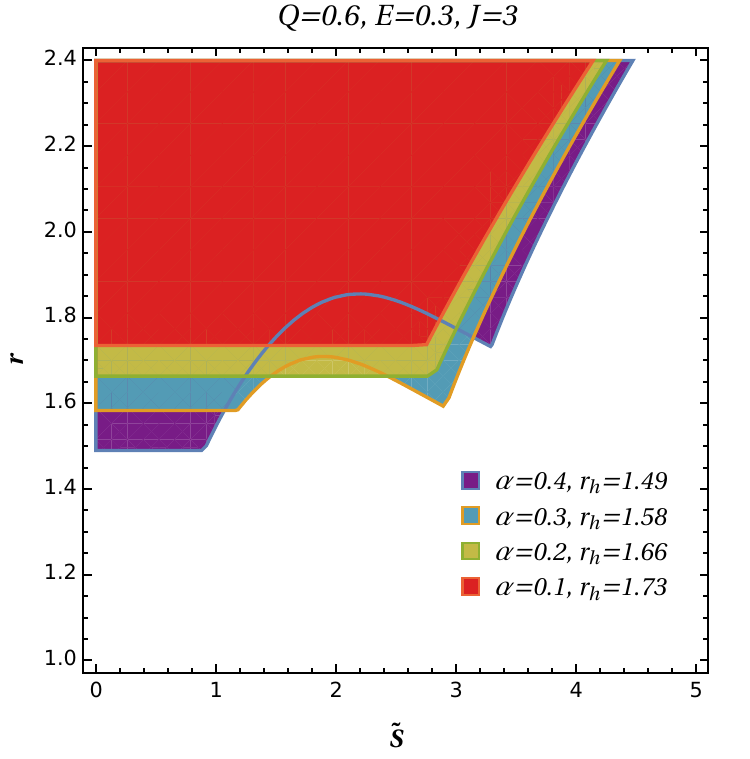}
\end{tabular}
	\caption{\label{fig;SR1} Plots show the allowed region between spin parameter $\Tilde{S}$ and $r$ for which $U^{2}$ is subluminal (i.e. less than zero). In \textit{first row:} variation of parameter space between $\tilde{S}$ and $r$ for different values of $J$ is shown. Here, $Q$ and $E$ are fixed to $0.6$ and $0.3$, respectively while parameter $\alpha$ varies from 0.1 (left) to 0.3 (right). 
	In \textit{second row:} variation of parameter space between $\tilde{S}$ and $r$ for different values of $E$ is shown. Here, $Q$ and $J$ are fixed to $0.6$ and $3$, respectively while parameter $\alpha$ again varies from 0.1 (left) to 0.3 (right).
	In \textit{third row:} variation of parameter space between $\tilde{S}$ and $r$ for different values of $\alpha$ is shown. Here, $E$ and $J$ are fixed to $0.3$ and $3$, respectively while parameter $Q$ varies from 0.2 (left) to 0.6 (right).	
	Here, mass parameter of $4D$ EGB BH set to unity.}
\end{figure}

For spinning particles it is a well known fact that the four-velocity $u^{\mu}$ is not a conserved quantity and hence their motion can be timelike (subluminal) and spacelike (superluminal). Using the relation between $u^{\mu}$ and four-momentum $P^{\mu}$ from Eqs. \eqref{dotr} and \eqref{dotph}, the ratio of the square of the four velocity ($u_{\mu}u^{\mu}$) to the square of time component $\left(u^{t}\right)^{2}$ of $u^{\mu}$ reads as
\begin{align}
   U^{2}\equiv \frac{u_{\mu}u^{\mu}}{\left(u^{t}\right)^{2}} &= -\Delta^{2}\left(\frac{1-\tilde{S}^2\mathcal{A}}{E-\tilde{S}J\mathcal{A}}\right)^{2}\left(1-\Psi\right)\, ,\label{uu}
\end{align}
where
\begin{align}
    \Psi&=\tilde{S}^{2}\left(\frac{J-E\tilde{S}}{r\left(1-\tilde{S}^{2}\mathcal{A}\right)^{2}}\right)^{2}\left[\mathcal{A}\left(2-\tilde{S}^{2}\mathcal{A}\right)+\Delta''\left(1-\frac{\tilde{S}^{2}\Delta''}{4}\right)\right]\label{psi}
\end{align}
Here, ``$\;'\;$'' denotes the derivative with respect to radial coordinate $r$. The variation of parameter space between $r$ and $\tilde{S}$ where $U^{2}$ is subluminal (i.e., less than zero) is shown in Fig. \ref{fig;SR1} for different combinations of $Q, \alpha, E$ and $J$. As was mentioned previously \ref{subsec:Ecm} there appears a potential divergence of $E_{C.M.}$ (via Eq. \eqref{Ecm2}) near the point $\Pi(r_{0})=0 \Rightarrow 1-\tilde{S}^{2}\mathcal{A}=0$ which leads to the problems such as superluminal motion and causality, as Eq. \eqref{uu} changes its sign from negative to positive at this location $r_{0}$. However, we have shown in previous subsection \ref{classification of spinning particle} that if we consider the collision of \textit{near-critical} and \textit{usual} particles near the horizon but outside it, then also there is a possiblity of unbounded $E_{C.M.}$. Hence, we are only interested in the region $r>r_{h}$ and want to have square of the four velocity less than zero in this region in order to bypass the superluminal motion of spinning particles. This necessity of bypassing of superlumial region leads $\Psi<1$. Taking into account the condition $\mathcal{K}\geq 0$, from Eq. \eqref{mathcalK} we have:
\begin{align}
    \varkappa\equiv\tilde{S}^{2}\mathcal{A}_{h}<1\, ,
    \label{varkappa}
\end{align}
and from Eq. \eqref{psi}, one can infer that $\Psi$ is a monotonically decreasing function of $r$. Therefore, it is ample to show that
\begin{align}
    \Psi(r_{h})<1\, ,\label{cond_psi}
\end{align}
because for any radial distance $r>r_{h}$, one will have $\Psi(r)<\Psi(r_{h})<1$. As was shown in the previous subsection that unbound $E_{C.M.}$ is possible only for the \textit{case IV} (i.e. $P_{1}=$\textit{near-critical} and $P_{2}=$\textit{usual}). Therefore, as far as particle $p_{2}$ is concerned, it is enough to take a $\tilde{S}=0$ particle. This condition leads to $\Psi=0$ and hence condition \eqref{cond_psi} is satisfied automatically. Also, for $\tilde{S}\neq 0$ but finite, the condition \eqref{cond_psi} holds well because one can easily choose $\left|J-E\tilde{S}\right|$ small enough as $J$ and $E$ are independent quantities. Hence, we assume that for \textit{usual} particle the condition \eqref{cond_psi} holds well.

Now for the \textit{near-critical} particle, the conserved energy $E$ and the total angular momentum $J$ are related via Eq. \eqref{critical_con}. Using Eqs. \eqref{critical_con}, \eqref{psi}  and \eqref{varkappa}, we obtain the condition
\begin{align}
    E<\frac{\mathcal{A}_{h} r_{h}(1-\varkappa)}{\sqrt{\mathcal{A}_{h}(2-\varkappa)+\Delta''_{h}\left(1-\frac{\tilde{S}^2\Delta''_{h}}{4}\right)}}\, .\label{Con_E}
\end{align}

\begin{figure}[H]
\begin{tabular}{cc}
  \includegraphics[width=0.45\textwidth]{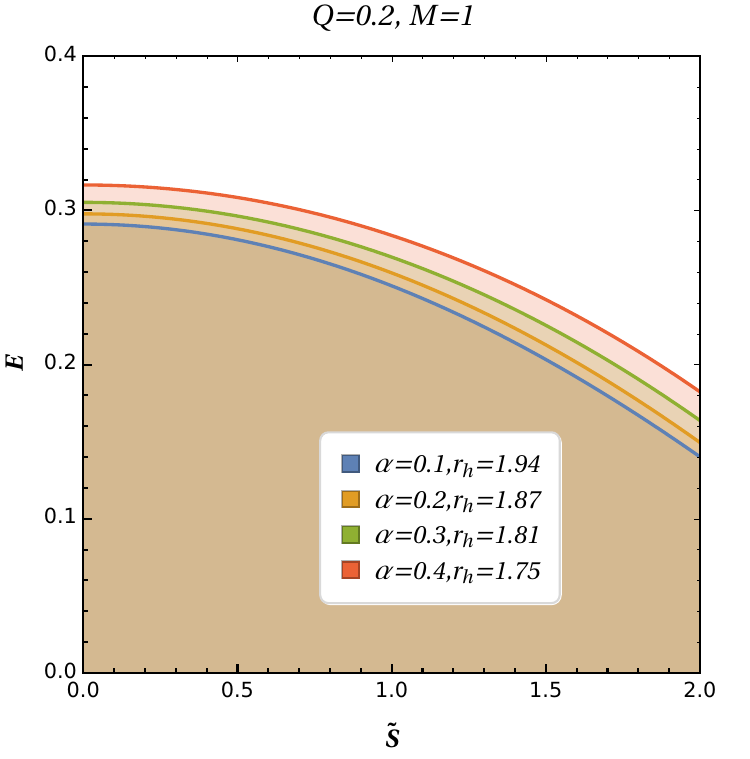}
 & \includegraphics[width=0.45\textwidth]{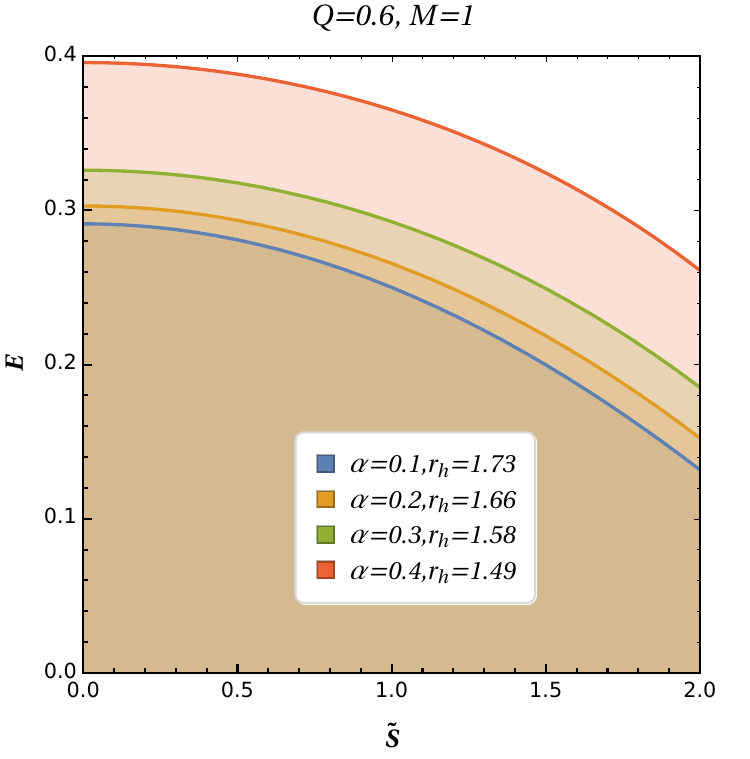}
\end{tabular}
	\caption{\label{fig;E1} Plots show the variation of allowed (shaded) region for $E$ as parameter $\tilde{S}$ increases for different combinations of $Q$ and $\alpha$.}\label{plot_E}
\end{figure}

Inequality \eqref{Con_E} is multi-variable in nature and hence it is difficult to analyse it analytically. Therefore, we numerically find the region (see Fig. \ref{plot_E}) where this inequality is satisfied. From Fig. \ref{plot_E}, it is easy to conclude that as the parameter $Q$ increases the allowed (shaded) region for $E$ increases. However, this increment is small for smaller values of GB coupling constant $\alpha$. Similarly, the allowed region increases as we increase the parameter $\alpha$, keeping parameter $Q$ constant. On contrary to $Q$ and $\alpha$, the allowed region decreases as parameter $\tilde{S}$ increases.

Using Eq. \eqref{critical_con} in condition \eqref{Con_E}, the inequality in $J$ becomes
\begin{align}
    J<\frac{\tilde{S}\mathcal{A}_{h} r_{h}(1-\varkappa)}{\varkappa\sqrt{\mathcal{A}_{h}(2-\varkappa)+\Delta''_{h}\left(1-\frac{\tilde{S}^2\Delta''_{h}}{4}\right)}}\label{Con_J}
\end{align}

\begin{figure}[H]
\begin{tabular}{cc}
  \includegraphics[width=0.45\textwidth]{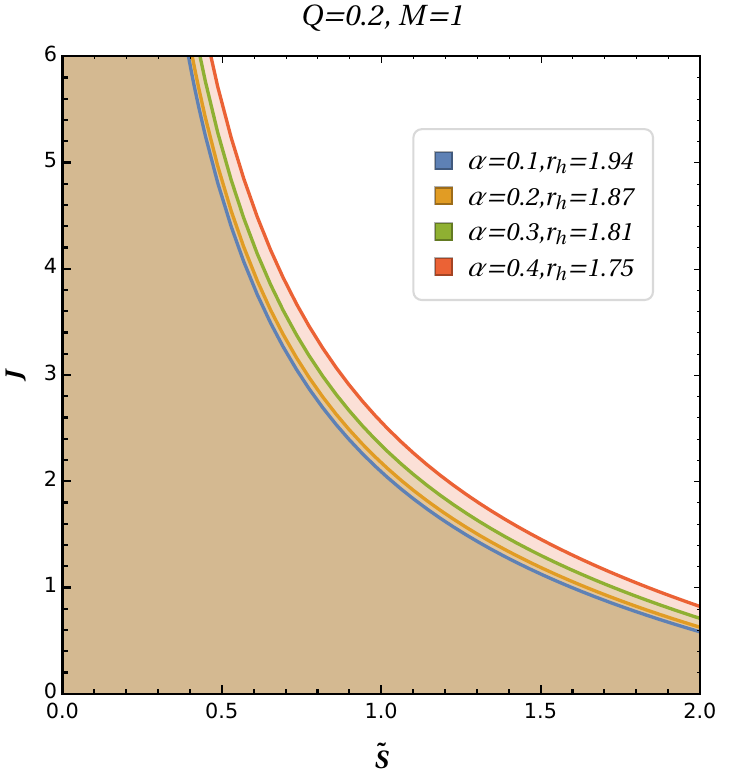}
 & \includegraphics[width=0.45\textwidth]{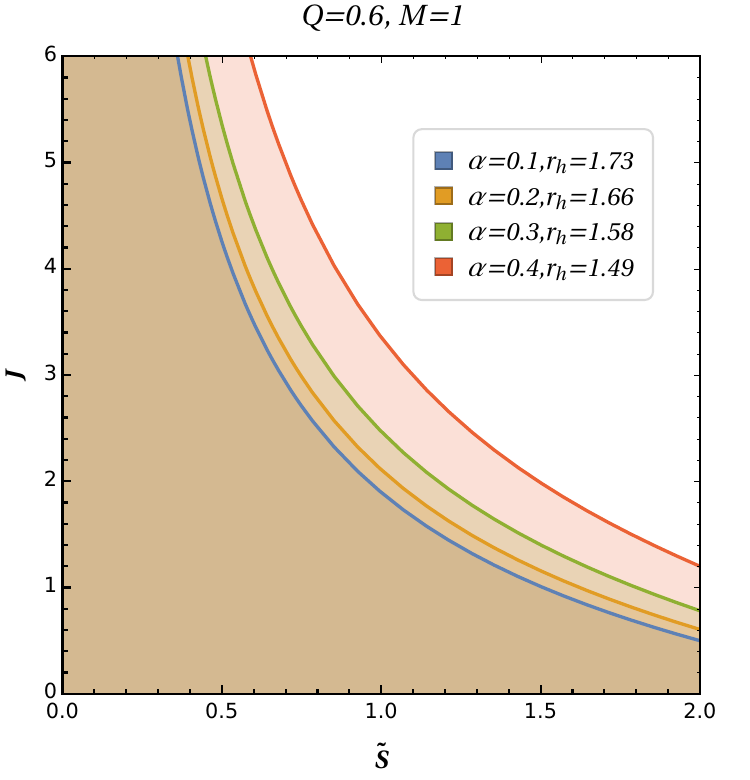}
\end{tabular}
	\caption{\label{fig;J1} Plots show the variation of allowed (shaded) region for $J$ as parameter $\tilde{S}$ increases for different combinations of $Q$ and $\alpha$.}\label{plot_J}
\end{figure}
In Fig. \ref{plot_J}, we show the variation of total angular momentum $J$ as a function of particle spin $\tilde{S}$ for different combinations of $Q$ and $\alpha$, using the inequality relation \eqref{Con_J}. From Fig. \ref{plot_J} one can observe that as the parameter $Q$ increases the allowed (shaded) region for $J$ increases only for the fixed $\alpha \geq 0.3$. However, there is no visible increment for smaller values $\alpha$. Additionally, the allowed region increases as we increase the parameter $\alpha$ for the fixed $Q$. In contrary to $Q$ and $\alpha$, the allowed region decreases as parameter $\tilde{S}$ increases. It is interesting to note from Fig. \ref{plot_J} that for smaller values of $\tilde{S}$ there exists no bound on the value of $J$.

Using Eq. \eqref{critical_con} in Eq. \eqref{Pph} for \textit{near critical} particle, we have
\begin{align}
     \left(\frac{J-E\tilde{S}}{1-\tilde{S}^2\mathcal{A}}\right)\approx J = \frac{E}{\tilde{S}\mathcal{A}_{h}}>0\, ,
\end{align}
which indicates that only prograde orbits can lead to infinitely high $E_{C.M.}$ values.

From condition \eqref{Con_E} and Fig. \ref{plot_J}, we conclude that if a \textit{near-critical} particle is coming from infinity (i.e., $E\geq1$) then infinitely high $E_{C.M.}$ cannot be observed until and unless a multiple scattering scenario is achieved rather than the direct collision. This infers that a particle coming from rest at infinity collides near the horizon of BH and achieves the bounding values to satisfy the \textit{near-critical} conditions for infinitely $E_{C.M.}$ to be produced in the next collision. However, infinitely high $E_{C.M.}$ could possibly be observed if and only if the particle is regarded as a \textit{near-critical} particle which starts from a distance $r>r_{h}$, where condition \eqref{Con_E} is always satisfied. 

\section{Conclusions and discussions}
\label{Sec:conclusion}
In the this paper, we have carried out the calculations for the weak gravitational lensing in the context of both  uniform and non-uniform plasma, whereas the BSW mechanism (i.e. BH as a particle accelerator) is studied for non-spinning and spinning particles around the charged $4D$ EGB BH spacetime, and examined the effect of Gauss-Bonnet coupling paramter $\alpha$ and charge parameter $Q$ on these phenomena. We also compared the results with those of RN (i.e., $\alpha \rightarrow 0$) and $4D$ EGB BHs (i.e., $Q \rightarrow 0$). Although the measurement of weak gravitational lensing effects with enough accuracy are improbable at present but in near future with the advancement of technology and high precision instruments, if it occurs, it would be extremely useful in precisely determining the location of the lensing object as well as the properties of the surrounding medium.

The main findings of our study are as follows:
\begin{itemize}
    \item It is found that the parameter space between horizon radius $r_{h}$ and charge $Q$ as well as between horizon radius $r_{h}$ and Gauss-Bonnet coupling parameter $\alpha$ decreases with the increase in the value of parameters $\alpha$ and $Q$ respectively. This in turns mean that the size of event horizon decreases as we deviate from the RN and $4D$ EGB BH spacetimes (see Fig. \ref{fig:horizon}).
    \item We have also found that the maximum of effective potential $V_{eff}$ (Fig.\ref{fig:Eff}) increases with the increase in parameters $\alpha, Q$ and $g$ .
    \item In tables \ref{1tab} and \ref{2tab}, the numerical value ISCO radius of both neutral and charged test particles is provided.The combined impacts of the GB coupling constant and the BH charge reduce the values of the ISCO radius and the photon sphere radii, as shown in the table \ref{1tab} and Fig. \ref{fig:photon}. The combined impacts of the GB coupling constant and the BH charge reduce the values of the ISCO radius and the photon sphere radii, as shown in the table ref1tab (see Fig. \ref{fig:photon}). The ISCO radius for a charged particle decreases (increases) with increasing the value of positively (negatively) charged particle as a result of the repulsive ($qQ>0$) and attractive ($qQ<0$) Coulomb forces, as seen in table \ref{2tab}.
    \item Intriguingly, in Fig.\ref{fig:an}, we discovered that both the parameters $\alpha$ and $Q$ have a similar effect, reducing the angular moment of test particles to be in circular orbits when increased.
    \item We noted that for the increase in the value of parameters $\alpha$ and $Q$, the deflection angle $\hat{\alpha}$ decreases and it does however increases by the impact of plasma medium parameter. It's also worth noting that in the case of non-uniform plasma, the angle of deflection is lower than in the case of uniform plasma (see Figs. \ref{deflectionunib} and \ref{deflectionsisb}).
    \item When a collision of two non-spinning particles is considered, the $E_{C.M.}$ stays finite, regardless of the value of the parameters $\alpha, Q, \mathcal{L}_{1,2}$ at the horizon of the $4D$ charged EGB BH as seen from Fig. \ref{fig:cm1}. This conclusion serves as a consistency check for the well-known fact in the literature that the $E_{C.M.}$ of two colliding non-spinning particles in the presence of a static spherically symmetric BH will always stay finite. On the contrary, it was acknowledged that infinitely high $E_{C.M.}$ is possible for spinning particles if the following conditions are fulfilled: (\textit{i}) the two spinning particles must collide near the horizon of charged $4D$ EGB BH, (\textit{ii}) the two colliding particles must only be \textit{near-critical} particle and \textit{usual} particle. Aside from these criteria, it's worth noting that the divergence of a \textit{near-critical} particle from the \text{critical} one at the site of contact is of the order $\sqrt{\Delta_c}$. We also found that if one of the colliding particles is a \text{near-critical} particle and falls from infinity, endlessly high $E_{C.M.}$ is not feasible. However, It makes no difference whether the particle is spinning or non-spinning until it collides numerous times, at which point an arbitrarily high $E_{C.M}$ is achieved, similar to what Grib and Pavlov proposed \cite{Grib:2010bs} for non-spinning particles.
    \item Furthermore, we investigated the superluminal (i.e., when a spinning particle's velocity is greater than unity) motion of spinning particles, which is yet another important aspect when dealing with spinning bodies, and the conditions for the particle's energy $E$ (see Eq. \ref{Con_E}) and total orbital angular momentum $J$ (see Eq. \ref{Con_J}) were obtained in order to avoid it. The region between the parameters $r$ and $S$ for which the motion of a spinning particle is subluminal (i.e. when the velocity of the spinning particle is less than unity) is displayed in Fig. \ref{fig;SR1} to highlight the influence of the GB coupling constant $\alpha$ and parameter $Q$. As the parameters $\alpha$ and $Q$ rise, the allowable parameter space between $r$ and $S$ grows. This indicates that as we depart further from the geometry of RN and $4D$ EGB BHs, the permitted domain of spinning particles expands, increasing the potential of charged $4D$ EGB BH acting as a particle accelerator in contrast to their counterparts. When Eqs. \ref{Con_E} and \ref{Con_J} are quantitatively examined using Figs. \ref{fig;E1} and \ref{fig;J1}, similar effects are seen.
\end{itemize}

\vspace{0.4cm}
\section*{Acknowledgments}
We would like to thank the referee for useful
comments. F.A. acknowledges the support of INHA University in Tashkent. S.S. acknowledges the support of Uzbekistan Ministry for Innovative Development. P.S. acknowledges support under Council of Scientific and Industrial Research (CSIR)-RA scheme (Govt. of India). 
\appendix

\section{General equations for Spinning massive test particles in curved spacetime}\label{EOM_spin_particles}
The study of spinning massive particle in curved spacetime started with the pioneering work of Mathisson \cite{Mathisson:1937zz} and Papapetrou \cite{Papapetrou:1951pa,Corinaldesi:1951pb}, and was further extended by Tulczyjew \cite{Tulczyjew:1959a}, Taub \cite{doi:10.1063/1.1704055}, and Dixon \cite{Dixon:1964aa}. Mathisson and Papapetrou (MP) in their work considered the spinning particle of finite length much shorter than the characteristic length of the spacetime, and hence posses a dipole moment besides the monopole moment one defined by the nonspinning test particles. Due to this dipole moment there occurs a spin-gravity coupling, resulting in the non-geodesic equations of motion (EOM) for a spinning massive test particle.

On the other hand, the EOM for spinning massive test particles can also be derived from a Lagrangian theory, as shown in \cite{Hanson:1974qy,Hojmanphdthesis:1975,Hojman:1976kn}. The Lagrangian approach provides two main advantages over the MP approach: (\textit{i}) The EOM are automatically reparametrization-covariant. (\textit{ii}) We can properly define the four-momentum of the spinning particle rather using its ad-hoc or hindered definition. 
As the four velocity and the four-momentum are not parallel to each other, the latter advantage of Lagrangian theory approach becomes important. Hence, this motivates us to follow this approach for the study of spinning massive test particles moving in the vicinity of $4D$ charged EGB BH. 

The EOM that come out from Lagrangian theory of a spinning massive test particle read as
\begin{align}
  &\;\;\;\;  u^{\mu} \equiv \frac{dx^{\mu}}{d\lambda},\label{ua}\\
   &  \frac{DP^{\mu}}{D\lambda} = -\frac{1}{2}R^{\mu}_{\nu\alpha\beta}u^{\nu}S^{\alpha\beta},\label{Pa}\\
    \text{and}&\nonumber\\
  &  \frac{DS^{\mu\nu}}{D\lambda} = S^{\mu\beta}\sigma_{beta}^{\nu}-\sigma^{\mu\beta}S_{\beta}^{\nu}=P^{\mu}u^{\nu}-u^{\mu}P^{\nu},\label{Sab}
\end{align}
where $\lambda, u^{\mu}, \sigma^{\mu\nu}, P^{\mu}, S^{\mu\nu}, R^{\mu}_{\nu\alpha\beta}$, and $D/D\lambda \equiv u^{\mu}\nabla_{\mu}$ are the affine parameter, four-velocity, angular velocity, four-momentum, spin-tensor (anti-symmetric), Riemann tensor and covariant derivative along four-velocity $u^{\mu}$, respectively.

For the case of spinning massive test particle, one can define the mass ($m$) and the spin ($S$) as
\begin{align}
 m^{2}&= -P_{\mu}P^{\mu},\label{cons_m}\\
 S^{2} &=\frac{1}{2}S_{\mu\nu}S^{\mu\nu}\, . \label{cons_S}
\end{align}
Here, one can easily check that the particle spin $S^{2}$ is a conserved quantity and the conservation is coming from contracting Eq. \eqref{Sab} with $S_{\mu\nu}$. It is also important to note here that the conservation of spin comes naturally in Lagrangian theory, unlike \cite{Dixon:1964aa}, where an additional assumption is needed.

Finally, due to symmetries of the spacetime we have more conserved quantities found by using
\begin{align}
    C_{\xi} = P^{\mu}\xi_{\mu}-\frac{1}{2}S^{\mu\nu}\xi_{\mu;\nu},\label{cosnstant_motion}
\end{align}
where $\xi_{\mu}$ is any Killing vector associated to the metric satisfying the following relation
\begin{align}
    \xi_{\mu;\nu}+\xi_{\nu;\mu}=0\label{Killing_eq}\, .
\end{align}
By examining carefully the EOM for spinning massive test particles one can conclude that there appear a number of unknown parameters than the number of equations. Hence, the system is incomplete and we need extra conditions known as the \textit{spin supplementary conditions} in order to close the system.

For the study of spinning particle, the spin supplementary condition we are choosing is the Tulczyjew condition, which reads as
\begin{align}
S^{\mu\nu}P_{\nu}=0. \label{Tulc_cond}
\end{align}
The main purpose of choosing the Tulczyjew spin supplementary condition is that $m^{2}$ defined via Eq. \eqref{cons_m} now refers to a constant of motion as well besides $S^{2}$. Even though, the invariant $u_{\mu}u^{\mu}$ is not, in general, a constant of motion. While using the condition \eqref{Tulc_cond}, we arbitrarily set three of the six components of $S^{\mu\nu}$ equal to zero, i.e. $S^{0i}$. This freedom to choose any of the three out of the six non-zero components of $S^{\mu\nu}$ comes from the arbitrariness of how one fixes the trajectory of the spinning particle \cite{Costa:2014nta}. 

\section{Components of four-momentum and four velocity of spinning particle}\label{four_p_u}
 Let us consider the motion of spinning massive test particle in static spherically symmetric spacetime e.g. (\ref{solution}). We will be interested in motion in the equatorial plane ($\theta=\pi/2$) since the spin stays perpendicular to the plane of motion for equatorial trajectories. Hence, for simplicity we consider these trajectories, and the non-zero components found with the help of the equations mentioned in the appendix \ref{EOM_spin_particles} are then given by
\begin{align}
    \frac{P^{t}}{m}&=\left[1+\frac{1}{2\alpha}\left(r^2-\mathcal{B}\right)\right]^{-1} \mathcal{K},\label{Pt}\\
    \frac{P^{\phi}}{m}&=\frac{ \mathcal{Z}}{r^2},\label{Pph}\\
    \left(\frac{P^{r}}{m}\right)^{2}&=\mathcal{K}^{2}-\left(1+\frac{r^2-\mathcal{B}}{2\alpha}\right)\left[1+\frac{\mathcal{Z}^{2}}{r^2}\right],\label{Pr}
\end{align}
where
\begin{align}
    \mathcal{K}&=\frac{E-\tilde{S}J\mathcal{A}}{1-\tilde{S}^2\mathcal{A}},\\
    \mathcal{Z}&=\frac{J-E\tilde{S}}{1-\tilde{S}^2\mathcal{A}},\\
    \mathcal{A}&= \frac{r\left(\mathcal{B}-r^2\right)-2M\alpha}{2\alpha r \mathcal{B}},\\
    \mathcal{B}&=\sqrt{r^4-4\alpha\left(Q^2-2Mr\right)}.
  \end{align}
The quantities $E$, $J$ and $\tilde{S}=\pm S$ respectively refer to the conserved energy, total angular momentum and spin per unit mass. Positive (negative) sign of spin signifies the alignment of the spin parallel (anti-parallel) to the $J$. One can also find by using the equations from the previous section that the $S_{z}$ component of the spin perpendicular to $\pi/2$ plane is
\begin{align}
    S_{z} = r S^{r\phi} = m\tilde{S}\left(\frac{E-\tilde{S}J\mathcal{A}}{1-\tilde{S}^2\mathcal{A}}\right),
\end{align}
such that in the limit ($r\rightarrow \infty$) far away from the event horizon of BH the $S_{z}/m=E\tilde{S}$ which is same as that for Schwazschild BH \cite{Armaza:2015eha}. Eventually, the coordinate velocities comes out to be
\begin{align}
    \dot{r}\equiv \frac{u^{r}}{u^{t}} &= \frac{P^{r}}{P^{t}},\label{dotr}\\
     \dot{\phi}\equiv \frac{u^{\phi}}{u^{t}} &= \left(\frac{F(r)'-\zeta r F(r)''}{F(r)'(1-\zeta)}\right)\frac{P^{\phi}}{P^{t}},\label{dotph}
\end{align}
where, 
\begin{align}
    \zeta = \left(\frac{\tilde{S}^2}{2r}\right)\left(\frac{r\left(\mathcal{B}-r^2\right)-2M \alpha}{\alpha\mathcal{B}}\right).
\end{align}
It is important to note here that the ``$\cdot$" in Eqs. \eqref{dotr} and \eqref{dotph} represents the derivative with respect to coordinate time. However, the non-zero components of four velocity ($u^{t}, u^{r}$ and $u^{\phi}$) are obtained by fixing the affine parameter $\lambda$ via some external condition. For our study of relativistic invariants no such choice is required all through.


\bibliographystyle{JHEP}
\bibliography{gravreferences,Lensing,Spinning}

\providecommand{\href}[2]{#2}\begingroup\raggedright\begin{thebibliography}{100}

\bibitem{Abbott16a}
B.P.~{Abbott} and et~al. {(Virgo and LIGO Scientific Collaborations)},
  \emph{{Observation of Gravitational Waves from a Binary Black Hole Merger}},
  \href{https://doi.org/10.1103/PhysRevLett.116.061102}{\emph{Phys. Rev. Lett.}
  {\bfseries 116} (2016) 061102}
  [\href{https://arxiv.org/abs/1602.03837}{{\ttfamily 1602.03837}}].

\bibitem{Abbott16b}
B.P.~{Abbott} and et~al. {(Virgo and LIGO Scientific Collaborations)},
  \emph{{Properties of the Binary Black Hole Merger GW150914}},
  \href{https://doi.org/10.1103/PhysRevLett.116.241102}{\emph{Phys. Rev. Lett.}
  {\bfseries 116} (2016) 241102}
  [\href{https://arxiv.org/abs/1602.03840}{{\ttfamily 1602.03840}}].

\bibitem{Akiyama19L1}
K.~{Akiyama} and et~al. {(Event Horizon Telescope Collaboration)}, \emph{{First
  M87 Event Horizon Telescope Results. I. The Shadow of the Supermassive Black
  Hole}}, \href{https://doi.org/10.3847/2041-8213/ab0ec7}{\emph{Astrophys. J.}
  {\bfseries 875} (2019) L1}
  [\href{https://arxiv.org/abs/1906.11238}{{\ttfamily 1906.11238}}].

\bibitem{Akiyama19L6}
K.~{Akiyama} and et~al. {(Event Horizon Telescope Collaboration)}, \emph{{First
  M87 Event Horizon Telescope Results. VI. The Shadow and Mass of the Central
  Black Hole}},
  \href{https://doi.org/10.3847/2041-8213/ab1141}{\emph{Astrophys. J.}
  {\bfseries 875} (2019) L6}
  [\href{https://arxiv.org/abs/1906.11243}{{\ttfamily 1906.11243}}].

\bibitem{Dadhich12c}
N.~{Dadhich}, S.G.~{Ghosh} and S.~{Jhingan}, \emph{{The Lovelock gravity in the
  critical spacetime dimension}},
  \href{https://doi.org/10.1016/j.physletb.2012.03.084}{\emph{Phys. Lett. B}
  {\bfseries 711} (2012) 196}
  [\href{https://arxiv.org/abs/1202.4575}{{\ttfamily 1202.4575}}].

\bibitem{Lovelock1971}
D.~{Lovelock}, \emph{{The Einstein Tensor and Its Generalizations}},
  \href{https://doi.org/10.1063/1.1665613}{\emph{J. Math. Phys.} {\bfseries 12}
  (1971) 498}.

\bibitem{Glavan20prl}
D.~{Glavan} and C.~{Lin}, \emph{{Einstein-Gauss-Bonnet Gravity in
  Four-Dimensional Spacetime}},
  \href{https://doi.org/10.1103/PhysRevLett.124.081301}{\emph{Phys. Rev. Lett.}
  {\bfseries 124} (2020) 081301}
  [\href{https://arxiv.org/abs/1905.03601}{{\ttfamily 1905.03601}}].

\bibitem{Lu:2020iav}
H.~Lu and Y.~Pang, \emph{{Horndeski gravity as $D \rightarrow 4$ limit of
  Gauss-Bonnet}},
  \href{https://doi.org/10.1016/j.physletb.2020.135717}{\emph{Phys. Lett. B}
  {\bfseries 809} (2020) 135717}
  [\href{https://arxiv.org/abs/2003.11552}{{\ttfamily 2003.11552}}].

\bibitem{Bonifacio:2020vbk}
J.~Bonifacio, K.~Hinterbichler and L.A.~Johnson, \emph{{Amplitudes and 4D
  Gauss-Bonnet Theory}},
  \href{https://doi.org/10.1103/PhysRevD.102.024029}{\emph{Phys. Rev. D}
  {\bfseries 102} (2020) 024029}
  [\href{https://arxiv.org/abs/2004.10716}{{\ttfamily 2004.10716}}].

\bibitem{Fernandes:2020nbq}
P.G.S.~Fernandes, P.~Carrilho, T.~Clifton and D.J.~Mulryne, \emph{{Derivation
  of Regularized Field Equations for the Einstein-Gauss-Bonnet Theory in Four
  Dimensions}}, \href{https://doi.org/10.1103/PhysRevD.102.024025}{\emph{Phys.
  Rev. D} {\bfseries 102} (2020) 024025}
  [\href{https://arxiv.org/abs/2004.08362}{{\ttfamily 2004.08362}}].

\bibitem{Kobayashi:2020wqy}
T.~Kobayashi, \emph{{Effective scalar-tensor description of regularized
  Lovelock gravity in four dimensions}},
  \href{https://doi.org/10.1088/1475-7516/2020/07/013}{\emph{JCAP} {\bfseries
  07} (2020) 013} [\href{https://arxiv.org/abs/2003.12771}{{\ttfamily
  2003.12771}}].

\bibitem{Hennigar:2020lsl}
R.A.~Hennigar, D.~Kubiz\v{n}\'ak, R.B.~Mann and C.~Pollack, \emph{{On taking
  the D \textrightarrow{} 4 limit of Gauss-Bonnet gravity: theory and
  solutions}}, \href{https://doi.org/10.1007/JHEP07(2020)027}{\emph{JHEP}
  {\bfseries 07} (2020) 027}
  [\href{https://arxiv.org/abs/2004.09472}{{\ttfamily 2004.09472}}].

\bibitem{Aoki20egb}
K.~{Aoki}, M.A.~{Gorji} and S.~{Mukohyama}, \emph{{A consistent theory of D
  {\textrightarrow} 4 Einstein-Gauss-Bonnet gravity}},
  \href{https://doi.org/10.1016/j.physletb.2020.135843}{\emph{Phys. Lett. B}
  {\bfseries 810} (2020) 135843}
  [\href{https://arxiv.org/abs/2005.03859}{{\ttfamily 2005.03859}}].

\bibitem{Dadhich20egb}
N.~{Dadhich}, \emph{{On causal structure of 4D-Einstein-Gauss-Bonnet black
  hole}}, \href{https://doi.org/10.1140/epjc/s10052-020-8422-8}{\emph{Eur.
  Phys. J. C} {\bfseries 80} (2020) 832}
  [\href{https://arxiv.org/abs/2005.05757}{{\ttfamily 2005.05757}}].

\bibitem{Hennigar20egb}
R.A.~{Hennigar}, D.~{Kubiz{\v{n}}{\'a}k}, R.B.~{Mann} and C.~{Pollack},
  \emph{{On taking the D {\textrightarrow} 4 limit of Gauss-Bonnet gravity:
  theory and solutions}},
  \href{https://doi.org/10.1007/JHEP07(2020)027}{\emph{JHEP} {\bfseries 2020}
  (2020) 27} [\href{https://arxiv.org/abs/2004.09472}{{\ttfamily 2004.09472}}].

\bibitem{Arrechea20egb}
J.~{Arrechea}, A.~{Delhom} and A.~{Jim{\'e}nez-Cano}, \emph{{Comment on
  ``Einstein-Gauss-Bonnet Gravity in Four-Dimensional Spacetime''}},
  \href{https://doi.org/10.1103/PhysRevLett.125.149002}{\emph{Phys. Rev. Lett.}
  {\bfseries 125} (2020) 149002}
  [\href{https://arxiv.org/abs/2009.10715}{{\ttfamily 2009.10715}}].

\bibitem{Gurses20egb}
M.~{G{\"u}rses}, T.{\c{c}}.~{{\c{S}}i{\textcommabelow s}man} and B.~{Tekin},
  \emph{{Is there a novel Einstein-Gauss-Bonnet theory in four dimensions?}},
  \href{https://doi.org/10.1140/epjc/s10052-020-8200-7}{\emph{Eur. Phys. J. C}
  {\bfseries 80} (2020) 647}
  [\href{https://arxiv.org/abs/2004.03390}{{\ttfamily 2004.03390}}].

\bibitem{Mahapatra20egb}
S.~{Mahapatra}, \emph{{A note on the total action of 4D Gauss-Bonnet theory}},
  \href{https://doi.org/10.1140/epjc/s10052-020-08568-6}{\emph{Eur. Phys. J. C}
  {\bfseries 80} (2020) 992}
  [\href{https://arxiv.org/abs/2004.09214}{{\ttfamily 2004.09214}}].

\bibitem{Liu20egb}
C.~{Liu}, T.~{Zhu} and Q.~{Wu}, \emph{{Thin accretion disk around a
  four-dimensional Einstein-Gauss-Bonnet black hole}},
  \href{https://doi.org/10.1088/1674-1137/abc16c}{\emph{Chin. Phys. C}
  {\bfseries 45} (2021) 015105}
  [\href{https://arxiv.org/abs/2004.01662}{{\ttfamily 2004.01662}}].

\bibitem{Guo20egb}
M.~{Guo} and P.-C.~{Li}, \emph{{Innermost stable circular orbit and shadow of
  the 4D Einstein-Gauss-Bonnet black hole}},
  \href{https://doi.org/10.1140/epjc/s10052-020-8164-7}{\emph{Eur. Phys. J. C}
  {\bfseries 80} (2020) 588}
  [\href{https://arxiv.org/abs/2003.02523}{{\ttfamily 2003.02523}}].

\bibitem{Wei20egb}
S.-W.~{Wei} and Y.-X.~{Liu}, \emph{{Testing the nature of Gauss-Bonnet gravity
  by four-dimensional rotating black hole shadow}}, {\emph{arXiv e-prints}
  (2020) } [\href{https://arxiv.org/abs/2003.07769}{{\ttfamily 2003.07769}}].

\bibitem{Kumar20egb}
R.~{Kumar} and S.G.~{Ghosh}, \emph{{Rotating black holes in 4D
  Einstein-Gauss-Bonnet gravity and its shadow}},
  \href{https://doi.org/10.1088/1475-7516/2020/07/053}{\emph{JCAP} {\bfseries
  2020} (2020) 053} [\href{https://arxiv.org/abs/2003.08927}{{\ttfamily
  2003.08927}}].

\bibitem{Konoplya20egb}
R.A.~{Konoplya} and A.F.~{Zinhailo}, \emph{{Quasinormal modes, stability and
  shadows of a black hole in the 4D Einstein-Gauss-Bonnet gravity}},
  \href{https://doi.org/10.1140/epjc/s10052-020-08639-8}{\emph{Eur. Phys. J. C}
  {\bfseries 80} (2020) 1049}
  [\href{https://arxiv.org/abs/2003.01188}{{\ttfamily 2003.01188}}].

\bibitem{Churilova20egb}
M.S.~{Churilova}, \emph{{Quasinormal modes of the Dirac field in the novel 4D
  Einstein-Gauss-Bonnet gravity}}, {\emph{arXiv e-prints} (2020) }
  [\href{https://arxiv.org/abs/2004.00513}{{\ttfamily 2004.00513}}].

\bibitem{Malafarina20egb}
D.~{Malafarina}, B.~{Toshmatov} and N.~{Dadhich}, \emph{{Dust collapse in 4D
  Einstein-Gauss-Bonnet gravity}},
  \href{https://doi.org/10.1016/j.dark.2020.100598}{\emph{Phys. Dark Universe}
  {\bfseries 30} (2020) 100598}
  [\href{https://arxiv.org/abs/2004.07089}{{\ttfamily 2004.07089}}].

\bibitem{Aragon20egb}
A.~{Arag{\'o}n}, R.~{B{\'e}car}, P.A.~{Gonz{\'a}lez} and Y.~{V{\'a}squez},
  \emph{{Perturbative and nonperturbative quasinormal modes of 4D
  Einstein-Gauss-Bonnet black holes}},
  \href{https://doi.org/10.1140/epjc/s10052-020-8298-7}{\emph{Eur. Phys. J. C}
  {\bfseries 80} (2020) 773}
  [\href{https://arxiv.org/abs/2004.05632}{{\ttfamily 2004.05632}}].

\bibitem{Mansoori20egb}
S.A.H.~{Mansoori}, \emph{{Thermodynamic geometry of the novel 4-D Gauss Bonnet
  AdS Black Hole}}, {\emph{arXiv e-prints} (2020) }
  [\href{https://arxiv.org/abs/2003.13382}{{\ttfamily 2003.13382}}].

\bibitem{Ge20egb}
X.-H.~{Ge} and S.-J.~{Sin}, \emph{{Causality of black holes in 4-dimensional
  Einstein-Gauss-Bonnet-Maxwell theory}},
  \href{https://doi.org/10.1140/epjc/s10052-020-8288-9}{\emph{Eur. Phys. J. C}
  {\bfseries 80} (2020) 695}
  [\href{https://arxiv.org/abs/2004.12191}{{\ttfamily 2004.12191}}].

\bibitem{Rayimbaev20egb}
J.~{Rayimbaev}, A.~{Abdujabbarov}, B.~{Turimov} and F.~{Atamurotov},
  \emph{{Magnetized particle motion around 4-D Einstein-Gauss-Bonnet Black
  Hole}}, {\emph{arXiv e-prints} (2020) }
  [\href{https://arxiv.org/abs/2004.10031}{{\ttfamily 2004.10031}}].

\bibitem{Chakraborty20egb}
S.~{Chakraborty} and N.~{Dadhich}, \emph{{Limits on stellar structures in
  Lovelock theories of gravity}},
  \href{https://doi.org/10.1016/j.dark.2020.100658}{\emph{Phys. Dark Universe}
  {\bfseries 30} (2020) 100658}
  [\href{https://arxiv.org/abs/2005.07504}{{\ttfamily 2005.07504}}].

\bibitem{Odintsov20egb}
S.D.~{Odintsov}, V.K.~{Oikonomou} and F.P.~{Fronimos}, \emph{{Rectifying
  Einstein-Gauss-Bonnet inflation in view of GW170817}},
  \href{https://doi.org/10.1016/j.nuclphysb.2020.115135}{\emph{Nucl. Phys. B.}
  {\bfseries 958} (2020) 115135}
  [\href{https://arxiv.org/abs/2003.13724}{{\ttfamily 2003.13724}}].

\bibitem{Odintsov20plb}
S.D.~{Odintsov} and V.K.~{Oikonomou}, \emph{{Swampland implications of
  GW170817-compatible Einstein-Gauss-Bonnet gravity}},
  \href{https://doi.org/10.1016/j.physletb.2020.135437}{\emph{Phys. Lett. B}
  {\bfseries 805} (2020) 135437}
  [\href{https://arxiv.org/abs/2004.00479}{{\ttfamily 2004.00479}}].

\bibitem{Lin20egb}
Z.-C.~{Lin}, K.~{Yang}, S.-W.~{Wei}, Y.-Q.~{Wang} and Y.-X.~{Liu},
  \emph{{Equivalence of solutions between the four-dimensional novel and
  regularized EGB theories in a cylindrically symmetric spacetime}},
  \href{https://doi.org/10.1140/epjc/s10052-020-08612-5}{\emph{Eur. Phys. J. C}
  {\bfseries 80} (2020) 1033}
  [\href{https://arxiv.org/abs/2006.07913}{{\ttfamily 2006.07913}}].

\bibitem{Shaymatov20egb}
S.~{Shaymatov}, J.~{Vrba}, D.~{Malafarina}, B.~{Ahmedov} and
  Z.~{Stuchl{\'\i}k}, \emph{{Charged particle and epicyclic motions around 4 D
  Einstein-Gauss-Bonnet black hole immersed in an external magnetic field}},
  \href{https://doi.org/10.1016/j.dark.2020.100648}{\emph{Phys. Dark Universe}
  {\bfseries 30} (2020) 100648}
  [\href{https://arxiv.org/abs/2005.12410}{{\ttfamily 2005.12410}}].

\bibitem{Islam20egb}
S.U.~{Islam}, R.~{Kumar} and S.G.~{Ghosh}, \emph{{Gravitational lensing by
  black holes in the 4D Einstein-Gauss-Bonnet gravity}},
  \href{https://doi.org/10.1088/1475-7516/2020/09/030}{\emph{JCAP} {\bfseries
  2020} (2020) 030} [\href{https://arxiv.org/abs/2004.01038}{{\ttfamily
  2004.01038}}].

\bibitem{Singh20-egb}
D.V.~{Singh} and S.~{Siwach}, \emph{{Thermodynamics and P-v criticality of
  Bardeen-AdS black hole in 4D Einstein-Gauss-Bonnet gravity}},
  \href{https://doi.org/10.1016/j.physletb.2020.135658}{\emph{Phys. Lett. B}
  {\bfseries 808} (2020) 135658}.

\bibitem{EslamPanah:2020hoj}
B.~Eslam~Panah, K.~Jafarzade and S.H.~Hendi, \emph{{Charged 4D
  Einstein-Gauss-Bonnet-AdS black holes: Shadow, energy emission, deflection
  angle and heat engine}},
  \href{https://doi.org/10.1016/j.nuclphysb.2020.115269}{\emph{Nucl. Phys. B}
  {\bfseries 961} (2020) 115269}
  [\href{https://arxiv.org/abs/2004.04058}{{\ttfamily 2004.04058}}].

\bibitem{Zhang20aegb}
C.-Y.~{Zhang}, S.-J.~{Zhang}, P.-C.~{Li} and M.~{Guo}, \emph{{Superradiance and
  stability of the novel 4D charged Einstein-Gauss-Bonnet black hole}},
  {\emph{arXiv e-prints} (2020) arXiv:2004.03141}
  [\href{https://arxiv.org/abs/2004.03141}{{\ttfamily 2004.03141}}].

\bibitem{Zhang20egb}
Y.-P.~{Zhang}, S.-W.~{Wei} and Y.-X.~{Liu}, \emph{{Spinning Test Particle in
  Four-Dimensional Einstein-Gauss-Bonnet Black Holes}},
  \href{https://doi.org/10.3390/universe6080103}{\emph{Universe} {\bfseries 6}
  (2020) 103} [\href{https://arxiv.org/abs/2003.10960}{{\ttfamily
  2003.10960}}].

\bibitem{Mishra:2020gce}
A.K.~Mishra, \emph{{Quasinormal modes and strong cosmic censorship in the
  regularised 4D Einstein\textendash{}Gauss\textendash{}Bonnet gravity}},
  \href{https://doi.org/10.1007/s10714-020-02763-2}{\emph{Gen. Rel. Grav.}
  {\bfseries 52} (2020) 106}
  [\href{https://arxiv.org/abs/2004.01243}{{\ttfamily 2004.01243}}].

\bibitem{Donmez2021}
O.~{Donmez}, \emph{{Bondi-Hoyle accretion around the non-rotating black hole in
  4D Einstein-Gauss-Bonnet gravity}},
  \href{https://doi.org/10.1140/epjc/s10052-021-08923-1}{\emph{European
  Physical Journal C} {\bfseries 81} (2021) 113}
  [\href{https://arxiv.org/abs/2011.04399}{{\ttfamily 2011.04399}}].

\bibitem{Fernandes20plb}
P.G.S.~{Fernandes}, \emph{{Charged black holes in AdS spaces in 4D Einstein
  Gauss-Bonnet gravity}},
  \href{https://doi.org/10.1016/j.physletb.2020.135468}{\emph{Phys. Lett. B}
  {\bfseries 805} (2020) 135468}
  [\href{https://arxiv.org/abs/2003.05491}{{\ttfamily 2003.05491}}].

\bibitem{Abdujabbarov15a}
A.~{Abdujabbarov}, F.~{Atamurotov}, N.~{Dadhich}, B.~{Ahmedov} and
  Z.~{Stuchl{\'{\i}}k}, \emph{{Energetics and optical properties of
  6-dimensional rotating black hole in pure Gauss-Bonnet gravity}},
  \href{https://doi.org/10.1140/epjc/s10052-015-3604-5}{\emph{Eur. Phys. J. C}
  {\bfseries 75} (2015) 399}
  [\href{https://arxiv.org/abs/1508.00331}{{\ttfamily 1508.00331}}].

\bibitem{Aguilar19}
G.~{Aguilar-P{\'e}rez}, M.~{Cruz}, S.~{Lepe} and I.~{Moran-Rivera},
  \emph{{Hairy black hole stability under odd parity perturbations in the
  Einstein-Gauss-Bonnet model}}, {\emph{arXiv e-prints} (2019)
  arXiv:1907.06168} [\href{https://arxiv.org/abs/1907.06168}{{\ttfamily
  1907.06168}}].

\bibitem{Shaymatov20-pl}
S.~{Shaymatov} and N.~{Dadhich}, \emph{{Weak cosmic censorship conjecture in
  the pure Lovelock gravity}}, {\emph{arXiv e-prints} (2020) }
  [\href{https://arxiv.org/abs/2008.04092}{{\ttfamily 2008.04092}}].

\bibitem{Dadhich21}
N.~{Dadhich} and S.~{Shaymatov}, \emph{{Circular orbits around higher
  dimensional Einstein and pure Gauss-Bonnet rotating black holes}},
  {\emph{arXiv e-prints} (2021) arXiv:2104.00427}
  [\href{https://arxiv.org/abs/2104.00427}{{\ttfamily 2104.00427}}].

\bibitem{Wu:2021zyl}
C.-H.~Wu, Y.-P.~Hu and H.~Xu, \emph{{Hawking evaporation of
  Einstein\textendash{}Gauss\textendash{}Bonnet AdS black holes in $D\geqslant
  4$ dimensions}},
  \href{https://doi.org/10.1140/epjc/s10052-021-09140-6}{\emph{Eur. Phys. J. C}
  {\bfseries 81} (2021) 351}
  [\href{https://arxiv.org/abs/2103.00257}{{\ttfamily 2103.00257}}].

\bibitem{Hioki09}
K.~{Hioki} and K.-I.~{Maeda}, \emph{{Measurement of the Kerr spin parameter by
  observation of a compact object's shadow}},
  \href{https://doi.org/10.1103/PhysRevD.80.024042}{\emph{Phys. Rev. D}
  {\bfseries 80} (2009) 024042}.

\bibitem{Atamurotov13}
F.~{Atamurotov}, A.~{Abdujabbarov} and B.~{Ahmedov}, \emph{{Shadow of rotating
  Ho{\v r}ava-Lifshitz black hole}},
  \href{https://doi.org/10.1007/s10509-013-1548-5}{\emph{Astrophys Space Sci}
  {\bfseries 348} (2013) 179}.

\bibitem{Atamurotov13b}
F.~{Atamurotov}, A.~{Abdujabbarov} and B.~{Ahmedov}, \emph{{Shadow of rotating
  non-Kerr black hole}},
  \href{https://doi.org/10.1103/PhysRevD.88.064004}{\emph{Phys. Rev. D}
  {\bfseries 88} (2013) 064004}.

\bibitem{Abdujabbarov13aa}
A.~{Abdujabbarov}, F.~{Atamurotov}, Y.~{Kucukakca}, B.~{Ahmedov} and
  U.~{Camci}, \emph{{Shadow of Kerr-Taub-NUT black hole}},
  \href{https://doi.org/10.1007/s10509-012-1337-6}{\emph{Astrophys. Space.
  Sci.} {\bfseries 344} (2013) 429}
  [\href{https://arxiv.org/abs/1212.4949}{{\ttfamily 1212.4949}}].

\bibitem{Atamurotov2015a}
F.~{Atamurotov}, B.~{Ahmedov} and A.~{Abdujabbarov}, \emph{{Optical properties
  of black holes in the presence of a plasma: The shadow}},
  \href{https://doi.org/10.1103/PhysRevD.92.084005}{\emph{Phys. Rev. D}
  {\bfseries 92} (2015) 084005}
  [\href{https://arxiv.org/abs/1507.08131}{{\ttfamily 1507.08131}}].

\bibitem{Atamurotov2016a}
F.~{Atamurotov}, S.G.~{Ghosh} and B.~{Ahmedov}, \emph{{Horizon structure of
  rotating Einstein-Born-Infeld black holes and shadow}},
  \href{https://doi.org/10.1140/epjc/s10052-016-4122-9}{\emph{European Physical
  Journal C} {\bfseries 76} (2016) 273}
  [\href{https://arxiv.org/abs/1506.03690}{{\ttfamily 1506.03690}}].

\bibitem{Papnoi2015}
U.~{Papnoi}, F.~{Atamurotov}, S.G.~{Ghosh} and B.~{Ahmedov}, \emph{{Shadow of
  five-dimensional rotating Myers-Perry black hole}},
  \href{https://doi.org/10.1103/PhysRevD.90.024073}{\emph{Phys. Rev. D}
  {\bfseries 90} (2014) 024073}
  [\href{https://arxiv.org/abs/1407.0834}{{\ttfamily 1407.0834}}].

\bibitem{Babar:2020a}
G.Z.~Babar, A.Z.~Babar and F.~Atamurotov, \emph{{`Optical properties of
  Kerr–Newman spacetime in the presence of plasma'}},
  \href{https://doi.org/10.1140/epjc/s10052-020-8346-3}{\emph{Eur.~Phys.~J.~C.}
  {\bfseries 80} (2020) 761}.

\bibitem{Rahul:2020a}
R.~Kumar and S.G.~Ghosh, \emph{{`Rotating black holes in 4D
  Einstein-Gauss-Bonnet gravity and its shadow'}},
  \href{https://doi.org/10.1088/1475-7516/2020/07/053}{\emph{J.~Cosmol.~A.~P}
  {\bfseries 2020} (2020) 053}.

\bibitem{Cunha20a}
P.V.P.~{Cunha}, N.A.~{Eir{\'o}}, C.A.R.~{Herdeiro} and J.P.S.~{Lemos},
  \emph{{Lensing and shadow of a black hole surrounded by a heavy accretion
  disk}},
  \href{https://doi.org/10.1088/1475-7516/2020/03/035}{\emph{J.~Cosmol.~A.~P}
  {\bfseries 2020} (2020) 035}
  [\href{https://arxiv.org/abs/1912.08833}{{\ttfamily 1912.08833}}].

\bibitem{Cunha17a}
P.V.P.~{Cunha}, C.A.R.~{Herdeiro}, B.~{Kleihaus}, J.~{Kunz} and E.~{Radu},
  \emph{{Shadows of Einstein-dilaton-Gauss-Bonnet black holes}},
  \href{https://doi.org/10.1016/j.physletb.2017.03.020}{\emph{Physics Letters
  B} {\bfseries 768} (2017) 373}
  [\href{https://arxiv.org/abs/1701.00079}{{\ttfamily 1701.00079}}].

\bibitem{Atamurotov21b}
F.~{Atamurotov} and U.~{Papnoi}, \emph{{Shadow of charged rotating black hole
  surrounded by perfect fluid dark matter}}, {\emph{arXiv e-prints} (2021)
  arXiv:2104.14898} [\href{https://arxiv.org/abs/2104.14898}{{\ttfamily
  2104.14898}}].

\bibitem{Virbha:2000a}
K.S.~Virbhadra and G.F.R.~Ellis, \emph{{`Schwarzschild black hole lensing'}},
  \href{https://doi.org/10.1103/PhysRevD.62.084003}{\emph{Phys.~Rev.~D.}
  {\bfseries 62} (2000) 084003}.

\bibitem{Bozza:2001a}
â.~Bozza, S.~Capozziello, G.~Iovane and G.~Scarpetta, \emph{{`Strong field
  limit of black hole gravitational lensing'}},
  \href{https://doi.org/10.1023/A:1012292927358}{\emph{Gen.~Rel.~Grav.}
  {\bfseries 33} (2001) 1535}.

\bibitem{Bozza:2002b}
â.~Bozza, \emph{{`Gravitational lensing in the strong field limit'}},
  \href{https://doi.org/10.1103/PhysRevD.66.103001}{\emph{Phys.~Rev.~D.}
  {\bfseries 66} (2002) 103001}.

\bibitem{Zhao:2017a}
S.-S.~Zhao and Y.~Xie, \emph{{`Strong deflection lensing by a Lee–Wick black
  hole'}},
  \href{https://doi.org/10.1016/j.physletb.2017.09.090}{\emph{Phys.~Lett.~B.}
  {\bfseries 774} (2017) 357}.

\bibitem{Vazquez04}
S.E.~{V{\'a}zquez} and E.P.~{Esteban}, \emph{{Strong-field gravitational
  lensing by a Kerr black hole}},
  \href{https://doi.org/10.1393/ncb/i2004-10121-y}{\emph{Nuovo Cim. B}
  {\bfseries 119} (2004) 489}.

\bibitem{Eiroa:2002b}
E.F.~Eiroa, G.E.~Romero and D.F.~Torres, \emph{{`Reissner-Nordstr$\ddot{o}$m
  black hole lensing'}},
  \href{https://doi.org/10.1103/PhysRevD.66.024010}{\emph{Phys.~Rev.~D.}
  {\bfseries 66} (2002) 024010}.

\bibitem{Eiroa:2004a}
E.F.~Eiroa and D.F.~Torres, \emph{{`Strong field limit analysis of
  gravitational retro lensing'}},
  \href{https://doi.org/10.1103/PhysRevD.69.063004}{\emph{Phys.~Rev.~D.}
  {\bfseries 69} (2004) 063004}.

\bibitem{Chak:2017a}
S.~{Chakraborty} and S.~Soumitra, \emph{{`Strong gravitational lensing---A
  probe for extra dimensions and Kalb-Ramond field'}},
  \href{https://doi.org/10.1088/1475-7516/2017/07/045}{\emph{J.~Cosmol.~A.~P.}
  {\bfseries 07} (2017) 045}.

\bibitem{Perlick04}
V.~{Perlick}, \emph{{Gravitational Lensing from a Spacetime Perspective}},
  \href{https://doi.org/10.12942/lrr-2004-9}{\emph{Living Reviews in
  Relativity} {\bfseries 7} (2004) 9}.

\bibitem{Babar2021c}
G.~{Zaman Babar}, F.~{Atamurotov} and A.~{Zaman Babar}, \emph{{Retrolensing by
  a spherically symmetric naked singularity}}, {\emph{arXiv e-prints} (2021)
  arXiv:2104.01340} [\href{https://arxiv.org/abs/2104.01340}{{\ttfamily
  2104.01340}}].

\bibitem{Abu:2017a}
A.~Abdujabbarov, B.~Ahmedov, N.~Dadhich and F.~Atamurotov, \emph{{`Optical
  properties of a braneworld black hole: Gravitational lensing and
  retrolensing'}},
  \href{https://doi.org/10.1103/PhysRevD.96.084017}{\emph{Phys.~Rev.~D.}
  {\bfseries 96} (2017) 084017}.

\bibitem{Virbha:2002a}
K.S.~Virbhadra and G.F.R.~Ellis, \emph{{`Gravitational lensing by naked
  singularities'}},
  \href{https://doi.org/10.1103/PhysRevD.65.103004}{\emph{Phys.~Rev.~D.}
  {\bfseries 65} (2002) 103004}.

\bibitem{Kumar2020a}
R.~{Kumar}, S.U.~{Islam} and S.G.~{Ghosh}, \emph{{Gravitational lensing by
  charged black hole in regularized 4D Einstein-Gauss-Bonnet gravity}},
  \href{https://doi.org/10.1140/epjc/s10052-020-08606-3}{\emph{European
  Physical Journal C} {\bfseries 80} (2020) 1128}
  [\href{https://arxiv.org/abs/2004.12970}{{\ttfamily 2004.12970}}].

\bibitem{Kogan10}
G.S.~{Bisnovatyi-Kogan} and O.Y.~{Tsupko}, \emph{{Gravitational lensing in a
  non-uniform plasma}},
  \href{https://doi.org/10.1111/j.1365-2966.2010.16290.x}{\emph{Monthly Notices
  of the Royal Astronomical Society} {\bfseries 404} (2010) 1790}
  [\href{https://arxiv.org/abs/1006.2321}{{\ttfamily 1006.2321}}].

\bibitem{Tsupko12}
O.Y.~{Tsupko} and G.S.~{Bisnovatyi-Kogan}, \emph{{On gravitational lensing in
  the presence of a plasma}},
  \href{https://doi.org/10.1134/S0202289312020120}{\emph{Gravitation and
  Cosmology} {\bfseries 18} (2012) 117}.

\bibitem{Bisnovatyi15}
G.S.~{Bisnovatyi-Kogan} and O.Y.~{Tsupko}, \emph{{Gravitational lensing in
  plasmic medium}},
  \href{https://doi.org/10.1134/S1063780X15070016}{\emph{Plasma Physics
  Reports} {\bfseries 41} (2015) 562}
  [\href{https://arxiv.org/abs/1507.08545}{{\ttfamily 1507.08545}}].

\bibitem{Babar2021a}
G.Z.~{Babar}, F.~{Atamurotov} and A.Z.~{Babar}, \emph{{Gravitational lensing in
  4-D Einstein-Gauss-Bonnet gravity in the presence of plasma}},
  \href{https://doi.org/10.1016/j.dark.2021.100798}{\emph{Physics of the Dark
  Universe} {\bfseries 32} (2021) 100798}.

\bibitem{Synge:1960b}
J.L.~Synge{\emph{Relativity: The General Theory. North-Holland, Amsterdam,
  1960} }.

\bibitem{Hakimov2016a}
A.~Hakimov and F.~Atamurotov, \emph{{`Gravitational lensing by a
  non-Schwarzschild black hole in a plasma'}},
  \href{https://doi.org/10.1007/s10509-016-2702-7}{\emph{Astrophys.~Space.~Sci.}
  {\bfseries 361} (2016) 112}.

\bibitem{Rog:2015a}
A.~Rogers, \emph{{`Frequency-dependent effects of gravitational lensing within
  plasma'}},
  \href{https://doi.org/10.1093/mnras/stv903}{\emph{Mon.~Not.~R.~Astron.~Soc.}
  {\bfseries 451} (2015) 17}.

\bibitem{Turi:2019a}
B.~{Turimov}, B.~{Ahmedov}, A.~{Abdujabbarov} and C.~{Bambi},
  \emph{{Gravitational lensing by a magnetized compact object in the presence
  of plasma}},
  \href{https://doi.org/10.1142/S0218271820400131}{\emph{Int.~J.~Mod.~Phys.~D.}
  {\bfseries 28} (2019) 2040013}.

\bibitem{Far:2021a}
F.~{Atamurotov}, A.~{Abdujabbarov} and J.~{Rayimbaev}, \emph{{`Weak
  gravitational lensing Schwarzschild-MOG black hole in plasma'}},
  \href{https://doi.org/10.1140/epjc/s10052-021-08919-x}{\emph{Eur.~Phys.~J.~C.}
  {\bfseries 81} (2021) 118}.

\bibitem{Car:2018a}
C.~Benavides-Gallego, A.~Abdujabbarov and Bambi, \emph{{`Gravitational lensing
  for a boosted Kerr black hole in the presence of plasma'}},
  \href{https://doi.org/10.1140/epjc/s10052-018-6170-97}{\emph{Eur.~Phys.~J.~C.}
  {\bfseries 78} (2018) 694}.

\bibitem{Chak:2018a}
H.~{Chakrabarty}, A.B.~{Abdikamalov}, A.A.~{Abdujabbarov} and C.~{Bambi},
  \emph{{Weak gravitational lensing: A compact object with arbitrary quadrupole
  moment immersed in plasma}},
  \href{https://doi.org/10.1103/PhysRevD.98.024022}{\emph{Phys.~Rev.~D.}
  {\bfseries 98} (2018) 024022}.

\bibitem{Abu:2017aa}
A.~Abdujabbarov, B.~Toshmatov, J.~Schee, Z.~Stuchl{\'\i}k and B.~Ahmedov,
  \emph{{`Gravitational lensing by regular black holes surrounded by plasma'}},
  \href{https://doi.org/10.1142/S0218271817410115}{\emph{Int.~J.~Mod.~Phys.~D.}
  {\bfseries 26} (2017) 1741011}.

\bibitem{Li2020}
Z.~{Li}, G.~{Zhang} and A.~{{\"O}vg{\"u}n}, \emph{{Circular orbit of a particle
  and weak gravitational lensing}},
  \href{https://doi.org/10.1103/PhysRevD.101.124058}{\emph{Phys. Rev. D}
  {\bfseries 101} (2020) 124058}
  [\href{https://arxiv.org/abs/2006.13047}{{\ttfamily 2006.13047}}].

\bibitem{nla.cat-vn2032203}
L.D.~Landau and E.M.~Lifshitz, \emph{Electrodynamics of continuous media / by
  L.D. Landau and E.M. Lifshitz ; translated from the Russian by J.B. Skyes and
  J.S. Bell}, Pergamon Press ; Addison-Wesley Oxford : Reading, Mass (1960).

\bibitem{PhysRevLett_24_1377}
D.O.~Muhleman, R.D.~Ekers and E.B.~Fomalont, \emph{Radio interferometric test
  of the general relativistic light bending near the sun},
  \href{https://doi.org/10.1103/PhysRevLett.24.1377}{\emph{Phys. Rev. Lett.}
  {\bfseries 24} (1970) 1377}.

\bibitem{1975pbrg.book.....L}
A.P.~{Lightman}, W.H.~{Press}, R.H.~{Price} and S.A.~{Teukolsky},
  \emph{{Problem Book in Relativity and Gravitation}}, Princeton University
  Press; Princeton, NJ (1975).

\bibitem{B-Minakov}
P.V.~Bliokh and A.A.~Minakov, \emph{Gravitational lenses}, {\emph{Naukova
  Dumka, Kiev (in Russian)} (1989) }.

\bibitem{Bisnovatyi-Kogan:2010flt}
G.S.~Bisnovatyi-Kogan and O.Y.~Tsupko, \emph{{Gravitational lensing in a
  non-uniform plasma}},
  \href{https://doi.org/10.1111/j.1365-2966.2010.16290.x}{\emph{Mon. Not. Roy.
  Astron. Soc.} {\bfseries 404} (2010) 1790}
  [\href{https://arxiv.org/abs/1006.2321}{{\ttfamily 1006.2321}}].

\bibitem{1986A&A...166...36K}
R.~{Kayser}, S.~{Refsdal} and R.~{Stabell}, \emph{{Astrophysical applications
  of gravitational micro-lensing.}}, {\emph{Astron. Astrophys.} {\bfseries 166}
  (1986) 36}.

\bibitem{1991AJ....102..864W}
J.~{Wambsganss} and B.~{Paczynski}, \emph{{Expected Color Variations of the
  Gravitationally Microlensed QSO 2237+0305}},
  \href{https://doi.org/10.1086/115916}{\emph{Astron. J.} {\bfseries 102}
  (1991) 864}.

\bibitem{Frolov10}
V.P.~{Frolov} and A.A.~{Shoom}, \emph{{Motion of charged particles near a
  weakly magnetized Schwarzschild black hole}},
  \href{https://doi.org/10.1103/PhysRevD.82.084034}{\emph{Phys. Rev. D}
  {\bfseries 82} (2010) 084034}
  [\href{https://arxiv.org/abs/1008.2985}{{\ttfamily 1008.2985}}].

\bibitem{Aliev02}
A.N.~{Aliev} and N.~{{\"O}zdemir}, \emph{{Motion of charged particles around a
  rotating black hole in a magnetic field}},
  \href{https://doi.org/10.1046/j.1365-8711.2002.05727.x}{\emph{Mon. Not. R.
  Astron. Soc.} {\bfseries 336} (2002) 241}
  [\href{https://arxiv.org/abs/gr-qc/0208025}{{\ttfamily gr-qc/0208025}}].

\bibitem{Abdujabbarov10}
A.~{Abdujabbarov} and B.~{Ahmedov}, \emph{{Test particle motion around a black
  hole in a braneworld}},
  \href{https://doi.org/10.1103/PhysRevD.81.044022}{\emph{Phys. Rev. D}
  {\bfseries 81} (2010) 044022}
  [\href{https://arxiv.org/abs/0905.2730}{{\ttfamily 0905.2730}}].

\bibitem{Shaymatov14}
S.~{Shaymatov}, F.~{Atamurotov} and B.~{Ahmedov}, \emph{{Isofrequency pairing
  of circular orbits in Schwarzschild spacetime in the presence of magnetic
  field}}, \href{https://doi.org/10.1007/s10509-013-1752-3}{\emph{Astrophys
  Space Sci} {\bfseries 350} (2014) 413}.

\bibitem{Toshmatov15d}
B.~{Toshmatov}, A.~{Abdujabbarov}, B.~{Ahmedov} and Z.~{Stuchl{\'{\i}}k},
  \emph{{Motion and high energy collision of magnetized particles around a
  Ho{\v r}ava-Lifshitz black hole}},
  \href{https://doi.org/10.1007/s10509-015-2533-y}{\emph{Astrophys Space Sci}
  {\bfseries 360} (2015) 19}.

\bibitem{Shaymatov15}
S.~{Shaymatov}, M.~{Patil}, B.~{Ahmedov} and P.S.~{Joshi}, \emph{{Destroying a
  near-extremal Kerr black hole with a charged particle: Can a test magnetic
  field serve as a cosmic censor?}},
  \href{https://doi.org/10.1103/PhysRevD.91.064025}{\emph{Phys. Rev. D}
  {\bfseries 91} (2015) 064025}
  [\href{https://arxiv.org/abs/1409.3018}{{\ttfamily 1409.3018}}].

\bibitem{Pavlovic19}
P.~{Pavlovi{\'c}}, A.~{Saveliev} and M.~{Sossich}, \emph{{Influence of the
  vacuum polarization effect on the motion of charged particles in the magnetic
  field around a Schwarzschild black hole}},
  \href{https://doi.org/10.1103/PhysRevD.100.084033}{\emph{Phys. Rev. D}
  {\bfseries 100} (2019) 084033}
  [\href{https://arxiv.org/abs/1908.01888}{{\ttfamily 1908.01888}}].

\bibitem{Shaymatov19b}
S.~{Shaymatov}, \emph{{Magnetized Reissner–Nordström black hole restores
  cosmic censorship conjecture}},
  \href{https://doi.org/10.1142/S2010194519600206}{\emph{Int. J. Mod. Phys.
  Conf. Ser.} {\bfseries 49} (2019) 1960020}.

\bibitem{Jamil15}
M.~{Jamil}, S.~{Hussain} and B.~{Majeed}, \emph{{Dynamics of particles around a
  Schwarzschild-like black hole in the presence of quintessence and magnetic
  field}}, \href{https://doi.org/10.1140/epjc/s10052-014-3230-7}{\emph{Eur.
  Phys. J. C} {\bfseries 75} (2015) 24}
  [\href{https://arxiv.org/abs/1404.7123}{{\ttfamily 1404.7123}}].

\bibitem{Hussain17}
{Hussain, S}, {Hussain, I} and {Jamil, M.}, \emph{Dynamics of a charged
  particle around a slowly rotating kerr black hole immersed in magnetic
  field}, \href{https://doi.org/10.1140/epjc/s10052-014-3210-y}{\emph{Eur.
  Phys. J. C} {\bfseries 74} (2014) 210}.

\bibitem{Shaymatov20a}
S.~{Shaymatov}, N.~{Dadhich} and B.~{Ahmedov}, \emph{{Six-dimensional
  Myers-Perry rotating black hole cannot be overspun}},
  \href{https://doi.org/10.1103/PhysRevD.101.044028}{\emph{Phys. Rev. D}
  {\bfseries 101} (2020) 044028}
  [\href{https://arxiv.org/abs/1908.07799}{{\ttfamily 1908.07799}}].

\bibitem{Toshmatov19d}
B.~{Toshmatov} and D.~{Malafarina}, \emph{{Spinning test particles in the
  {\ensuremath{\gamma}} spacetime}},
  \href{https://doi.org/10.1103/PhysRevD.100.104052}{\emph{Phys. Rev. D}
  {\bfseries 100} (2019) 104052}
  [\href{https://arxiv.org/abs/1910.11565}{{\ttfamily 1910.11565}}].

\bibitem{Rayimbaev20c}
J.~{Rayimbaev}, M.~{Figueroa}, Z.~{Stuchl{\'\i}k} and B.~{Juraev}, \emph{{Test
  particle orbits around regular black holes in general relativity combined
  with nonlinear electrodynamics}},
  \href{https://doi.org/10.1103/PhysRevD.101.104045}{\emph{Phys. Rev. D}
  {\bfseries 101} (2020) 104045}.

\bibitem{Shaymatov20b}
S.~{Shaymatov}, D.~{Malafarina} and B.~{Ahmedov}, \emph{{Effect of perfect
  fluid dark matter on particle motion around a static black hole immersed in
  an external magnetic field}}, {\emph{arXiv e-prints} (2020) arXiv:2004.06811}
  [\href{https://arxiv.org/abs/2004.06811}{{\ttfamily 2004.06811}}].

\bibitem{Narzilloev20a}
B.~{Narzilloev}, J.~{Rayimbaev}, S.~{Shaymatov}, A.~{Abdujabbarov},
  B.~{Ahmedov} and C.~{Bambi}, \emph{{Can the dynamics of test particles around
  charged stringy black holes mimic the spin of Kerr black holes?}},
  \href{https://doi.org/10.1103/PhysRevD.102.044013}{\emph{Phys. Rev. D}
  {\bfseries 102} (2020) 044013}
  [\href{https://arxiv.org/abs/2007.12462}{{\ttfamily 2007.12462}}].

\bibitem{Narzilloev20b}
B.~{Narzilloev}, J.~{Rayimbaev}, S.~{Shaymatov}, A.~{Abdujabbarov},
  B.~{Ahmedov} and C.~{Bambi}, \emph{{Dynamics of test particles around a
  Bardeen black hole surrounded by perfect fluid dark matter}},
  \href{https://doi.org/10.1103/PhysRevD.102.104062}{\emph{Phys. Rev. D}
  {\bfseries 102} (2020) 104062}
  [\href{https://arxiv.org/abs/2011.06148}{{\ttfamily 2011.06148}}].

\bibitem{Stuchlik20}
Z.~{Stuchl{\'\i}k}, M.~{Kolo{\v{s}}}, J.~{Kov{\'a}{\v{r}}}, P.~{Slan{\'y}} and
  A.~{Tursunov}, \emph{{Influence of Cosmic Repulsion and Magnetic Fields on
  Accretion Disks Rotating around Kerr Black Holes}},
  \href{https://doi.org/10.3390/universe6020026}{\emph{Universe} {\bfseries 6}
  (2020) 26}.

\bibitem{Shaymatov21-b}
S.~Shaymatov and F.~Atamurotov, \emph{Geodesic circular orbits sharing the same
  orbital frequencies in the black string spacetime},
  \href{https://doi.org/10.3390/galaxies9020040}{\emph{Galaxies} {\bfseries 9}
  (2021) 40} [\href{https://arxiv.org/abs/2007.10793}{{\ttfamily 2007.10793}}].

\bibitem{Shaymatov21c}
S.~{Shaymatov}, B.~{Narzilloev}, A.~{Abdujabbarov} and C.~{Bambi},
  \emph{{Charged particle motion around a magnetized Reissner-Nordstr{\"o}m
  black hole}}, \href{https://doi.org/10.1103/PhysRevD.103.124066}{\emph{Phys.
  Rev. D} {\bfseries 103} (2021) 124066}
  [\href{https://arxiv.org/abs/2105.00342}{{\ttfamily 2105.00342}}].

\bibitem{Fender04mnrs}
R.P.~{Fender}, T.M.~{Belloni} and E.~{Gallo}, \emph{{Towards a unified model
  for black hole X-ray binary jets}},
  \href{https://doi.org/10.1111/j.1365-2966.2004.08384.x}{\emph{Mon. Not. R.
  Astron. Soc.} {\bfseries 355} (2004) 1105}
  [\href{https://arxiv.org/abs/astro-ph/0409360}{{\ttfamily
  astro-ph/0409360}}].

\bibitem{Auchettl17ApJ}
K.~{Auchettl}, J.~{Guillochon} and E.~{Ramirez-Ruiz}, \emph{{New Physical
  Insights about Tidal Disruption Events from a Comprehensive Observational
  Inventory at X-Ray Wavelengths}},
  \href{https://doi.org/10.3847/1538-4357/aa633b}{\emph{Astrophys. J.}
  {\bfseries 838} (2017) 149}
  [\href{https://arxiv.org/abs/1611.02291}{{\ttfamily 1611.02291}}].

\bibitem{IceCube17b}
{The IceCube Collaboration} and et~al., \emph{{Multimessenger observations of a
  flaring blazar coincident with high-energy neutrino IceCube-170922A}},
  \href{https://doi.org/10.1126/science.aat1378}{\emph{Science} {\bfseries 361}
  (2018) eaat1378} [\href{https://arxiv.org/abs/1807.08816}{{\ttfamily
  1807.08816}}].

\bibitem{Banados09}
M.~{Ba{\~n}ados}, J.~{Silk} and S.M.~{West}, \emph{{Kerr Black Holes as
  Particle Accelerators to Arbitrarily High Energy}},
  \href{https://doi.org/10.1103/PhysRevLett.103.111102}{\emph{Phys. Rev. Lett.}
  {\bfseries 103} (2009) 111102}.

\bibitem{Penrose69}
R.~{Penrose}, \emph{{Gravitational Collapse: the Role of General Relativity}},
  {\emph{Riv. Nuovo Cimento} {\bfseries 1} (1969) }.

\bibitem{Grib11}
A.A.~{Grib} and Y.V.~{Pavlov}, \emph{{On particle collisions near rotating
  black holes}},
  \href{https://doi.org/10.1134/S0202289311010099}{\emph{Gravitation and
  Cosmology} {\bfseries 17} (2011) 42}
  [\href{https://arxiv.org/abs/1010.2052}{{\ttfamily 1010.2052}}].

\bibitem{Jacobson10}
T.~{Jacobson} and T.P.~{Sotiriou}, \emph{{Spinning Black Holes as Particle
  Accelerators}},
  \href{https://doi.org/10.1103/PhysRevLett.104.021101}{\emph{Phys. Rev. Lett.}
  {\bfseries 104} (2010) 021101}
  [\href{https://arxiv.org/abs/0911.3363}{{\ttfamily 0911.3363}}].

\bibitem{Harada11b}
T.~{Harada} and M.~{Kimura}, \emph{{Collision of an innermost stable circular
  orbit particle around a Kerr black hole}},
  \href{https://doi.org/10.1103/PhysRevD.83.024002}{\emph{Phys. Rev. D}
  {\bfseries 83} (2011) 024002}
  [\href{https://arxiv.org/abs/1010.0962}{{\ttfamily 1010.0962}}].

\bibitem{Wei10}
S.-W.~{Wei}, Y.-X.~{Liu}, H.~{Guo} and C.-E.~{Fu}, \emph{{Charged spinning
  black holes as particle accelerators}},
  \href{https://doi.org/10.1103/PhysRevD.82.103005}{\emph{Phys. Rev. D}
  {\bfseries 82} (2010) 103005}
  [\href{https://arxiv.org/abs/1006.1056}{{\ttfamily 1006.1056}}].

\bibitem{Zaslavskii11b}
O.B.~{Zaslavskii}, \emph{{Acceleration of particles by nonrotating charged
  black holes?}}, \href{https://doi.org/10.1134/S0021364010210010}{\emph{Soviet
  Journal of Experimental and Theoretical Physics Letters} {\bfseries 92}
  (2011) 571} [\href{https://arxiv.org/abs/1007.4598}{{\ttfamily 1007.4598}}].

\bibitem{Zaslavskii11c}
O.B.~{Zaslavskii}, \emph{{Acceleration of particles by black holes - a general
  explanation}},
  \href{https://doi.org/10.1088/0264-9381/28/10/105010}{\emph{Class. Quantum
  Grav.} {\bfseries 28} (2011) 105010}
  [\href{https://arxiv.org/abs/1011.0167}{{\ttfamily 1011.0167}}].

\bibitem{Kimura11}
M.~{Kimura}, K.-I.~{Nakao} and H.~{Tagoshi}, \emph{{Acceleration of colliding
  shells around a black hole: Validity of the test particle approximation in
  the Banados-Silk-West process}},
  \href{https://doi.org/10.1103/PhysRevD.83.044013}{\emph{Phys. Rev. D}
  {\bfseries 83} (2011) 044013}
  [\href{https://arxiv.org/abs/1010.5438}{{\ttfamily 1010.5438}}].

\bibitem{Banados11}
M.~{Ba{\~n}ados}, B.~{Hassanain}, J.~{Silk} and S.M.~{West}, \emph{{Emergent
  flux from particle collisions near a Kerr black hole}},
  \href{https://doi.org/10.1103/PhysRevD.83.023004}{\emph{Physical Review D}
  {\bfseries 83} (2011) 023004}
  [\href{https://arxiv.org/abs/1010.2724}{{\ttfamily 1010.2724}}].

\bibitem{Frolov12}
V.P.~{Frolov}, \emph{{Weakly magnetized black holes as particle accelerators}},
  \href{https://doi.org/10.1103/PhysRevD.85.024020}{\emph{Phys. Rev. D}
  {\bfseries 85} (2012) 024020}
  [\href{https://arxiv.org/abs/1110.6274}{{\ttfamily 1110.6274}}].

\bibitem{Abdujabbarov13a}
A.A.~{Abdujabbarov}, A.A.~{Tursunov}, B.J.~{Ahmedov} and A.~{Kuvatov},
  \emph{{Acceleration of particles by black hole with gravitomagnetic charge
  immersed in magnetic field}},
  \href{https://doi.org/10.1007/s10509-012-1251-y}{\emph{Astrophys Space Sci}
  {\bfseries 343} (2013) 173}
  [\href{https://arxiv.org/abs/1209.2680}{{\ttfamily 1209.2680}}].

\bibitem{Liu11}
C.~{Liu}, S.~{Chen}, C.~{Ding} and J.~{Jing}, \emph{{Particle acceleration on
  the background of the Kerr-Taub-NUT spacetime}},
  \href{https://doi.org/10.1016/j.physletb.2011.05.070}{\emph{Phys. Lett. B}
  {\bfseries 701} (2011) 285}
  [\href{https://arxiv.org/abs/1012.5126}{{\ttfamily 1012.5126}}].

\bibitem{Atamurotov13a}
F.~{Atamurotov}, B.~{Ahmedov} and S.~{Shaymatov}, \emph{{Formation of black
  holes through BSW effect and black hole-black hole collisions}},
  \href{https://doi.org/10.1007/s10509-013-1527-x}{\emph{Astrophys. Space Sci.}
  {\bfseries 347} (2013) 277}.

\bibitem{Stuchlik11a}
Z.~{Stuchl{\'{\i}}k}, S.~{Hled{\'{\i}}k} and K.~{Truparov{\'a}},
  \emph{{Evolution of Kerr superspinars due to accretion counterrotating thin
  discs}}, \href{https://doi.org/10.1088/0264-9381/28/15/155017}{\emph{Class.
  Quantum Grav.} {\bfseries 28} (2011) 155017}.

\bibitem{Stuchlik12a}
Z.~{Stuchl{\'{\i}}k} and J.~{Schee}, \emph{{Observational phenomena related to
  primordial Kerr superspinars}},
  \href{https://doi.org/10.1088/0264-9381/29/6/065002}{\emph{Class. Quantum
  Grav.} {\bfseries 29} (2012) 065002}.

\bibitem{Igata12}
T.~{Igata}, T.~{Harada} and M.~{Kimura}, \emph{{Effect of a weak
  electromagnetic field on particle acceleration by a rotating black hole}},
  \href{https://doi.org/10.1103/PhysRevD.85.104028}{\emph{Phys. Rev. D}
  {\bfseries 85} (2012) 104028}
  [\href{https://arxiv.org/abs/1202.4859}{{\ttfamily 1202.4859}}].

\bibitem{Shaymatov13}
S.R.~{Shaymatov}, B.J.~{Ahmedov} and A.A.~{Abdujabbarov}, \emph{{Particle
  acceleration near a rotating black hole in a Randall-Sundrum brane with a
  cosmological constant}},
  \href{https://doi.org/10.1103/PhysRevD.88.024016}{\emph{Phys. Rev. D}
  {\bfseries 88} (2013) 024016}.

\bibitem{Tursunov13}
A.~{Tursunov}, M.~{Kolo{\v s}}, A.~{Abdujabbarov}, B.~{Ahmedov} and
  Z.~{Stuchl{\'{\i}}k}, \emph{{Acceleration of particles in spacetimes of black
  string}}, \href{https://doi.org/10.1103/PhysRevD.88.124001}{\emph{Phys. Rev.
  D} {\bfseries 88} (2013) 124001}.

\bibitem{Ghosh:2014mea}
S.G.~Ghosh, P.~Sheoran and M.~Amir, \emph{{Rotating Ay\'on-Beato-Garc\'\i{}a
  black hole as a particle accelerator}},
  \href{https://doi.org/10.1103/PhysRevD.90.103006}{\emph{Phys. Rev. D}
  {\bfseries 90} (2014) 103006}
  [\href{https://arxiv.org/abs/1410.5588}{{\ttfamily 1410.5588}}].

\bibitem{Shaymatov18a}
S.~{Shaymatov}, B.~{Ahmedov}, Z.~{Stuchl{\'\i}k} and A.~{Abdujabbarov},
  \emph{{Effect of an external magnetic field on particle acceleration by a
  rotating black hole surrounded with quintessential energy}},
  \href{https://doi.org/10.1142/S0218271818500888}{\emph{International Journal
  of Modern Physics D} {\bfseries 27} (2018) 1850088}.

\bibitem{Babar2021b}
G.Z.~Babar, F.~Atamurotov, S.~Ul~Islam and S.G.~Ghosh, \emph{Particle
  acceleration around rotating einstein-born-infeld black hole and plasma
  effect on gravitational lensing},
  \href{https://doi.org/10.1103/PhysRevD.103.084057}{\emph{Phys. Rev. D}
  {\bfseries 103} (2021) 084057}
  [\href{https://arxiv.org/abs/2104.00714}{{\ttfamily 2104.00714}}].

\bibitem{Josh:2016a}
M.~Amir, F.~Ahmed and S.~G.~Ghosh, \emph{Collision of two general particles
  around a rotating regular hayward's black holes},
  \href{https://doi.org/10.1140/epjc/s10052-016-4365-5}{\emph{Eur.~Phys.~J.~C.}
  {\bfseries 76} (2016) 532}.

\bibitem{Dadhich18}
N.~{Dadhich}, A.~{Tursunov}, B.~{Ahmedov} and Z.~{Stuchl{\'\i}k}, \emph{{The
  distinguishing signature of magnetic Penrose process}},
  \href{https://doi.org/10.1093/mnrasl/sly073}{\emph{Mon. Not. Roy. Astron.
  Soc.} {\bfseries 478} (2018) L89}
  [\href{https://arxiv.org/abs/1804.09679}{{\ttfamily 1804.09679}}].

\bibitem{Abdujabbarov11}
A.A.~{Abdujabbarov}, B.J.~{Ahmedov}, S.R.~{Shaymatov} and A.S.~{Rakhmatov},
  \emph{{Penrose process in Kerr-Taub-NUT spacetime}},
  \href{https://doi.org/10.1007/s10509-011-0740-8}{\emph{Astrophys Space Sci}
  {\bfseries 334} (2011) 237}
  [\href{https://arxiv.org/abs/1105.1910}{{\ttfamily 1105.1910}}].

\bibitem{Okabayashi20}
K.~{Okabayashi} and K.-i.~{Maeda}, \emph{{Maximal efficiency of the collisional
  Penrose process with a spinning particle. II. Collision with a particle on
  the innermost stable circular orbit}},
  \href{https://doi.org/10.1093/ptep/ptz143}{\emph{Prog. Theor. Exp. Phys.}
  {\bfseries 2020} (2020) 013E01}
  [\href{https://arxiv.org/abs/1907.07126}{{\ttfamily 1907.07126}}].

\bibitem{Ghosh:2013ona}
S.G.~Ghosh and P.~Sheoran, \emph{{Higher dimensional non-Kerr black hole and
  energy extraction}},
  \href{https://doi.org/10.1103/PhysRevD.89.024023}{\emph{Phys. Rev. D}
  {\bfseries 89} (2014) 024023}
  [\href{https://arxiv.org/abs/1309.5519}{{\ttfamily 1309.5519}}].

\bibitem{Jefremov:2015gza}
P.I.~Jefremov, O.Y.~Tsupko and G.S.~Bisnovatyi-Kogan, \emph{{Innermost stable
  circular orbits of spinning test particles in Schwarzschild and Kerr
  space-times}}, \href{https://doi.org/10.1103/PhysRevD.91.124030}{\emph{Phys.
  Rev. D} {\bfseries 91} (2015) 124030}
  [\href{https://arxiv.org/abs/1503.07060}{{\ttfamily 1503.07060}}].

\bibitem{Armaza:2016vfz}
C.~Armaza, S.A.~Hojman, B.~Koch and N.~Zalaquett, \emph{{On the possibility of
  non-geodesic motion of massless spinning tops}},
  \href{https://doi.org/10.1088/0264-9381/33/14/145011}{\emph{Class. Quant.
  Grav.} {\bfseries 33} (2016) 145011}
  [\href{https://arxiv.org/abs/1601.05809}{{\ttfamily 1601.05809}}].

\bibitem{Zhang:2017nhl}
Y.-P.~Zhang, S.-W.~Wei, W.-D.~Guo, T.-T.~Sui and Y.-X.~Liu, \emph{{Innermost
  stable circular orbit of spinning particle in charged spinning black hole
  background}}, \href{https://doi.org/10.1103/PhysRevD.97.084056}{\emph{Phys.
  Rev. D} {\bfseries 97} (2018) 084056}
  [\href{https://arxiv.org/abs/1711.09361}{{\ttfamily 1711.09361}}].

\bibitem{ZHANG2019393}
M.~Zhang and W.-B.~Liu, \emph{Innermost stable circular orbits of charged
  spinning test particles},
  \href{https://doi.org/https://doi.org/10.1016/j.physletb.2018.12.051}{\emph{Physics
  Letters B} {\bfseries 789} (2019) 393}.

\bibitem{PhysRevD.98.084023}
S.~Mukherjee and K.R.~Nayak, \emph{Off-equatorial stable circular orbits for
  spinning particles},
  \href{https://doi.org/10.1103/PhysRevD.98.084023}{\emph{Phys. Rev. D}
  {\bfseries 98} (2018) 084023}.

\bibitem{Hojman:2018evi}
S.A.~Hojman and F.A.~Asenjo, \emph{{Non-geodesic circular motion of massive
  spinning test bodies around a Schwarzschild field in the Lagrangian theory}},
  \href{https://doi.org/10.1140/epjc/s10052-018-6341-8}{\emph{Eur. Phys. J. C}
  {\bfseries 78} (2018) 843}
  [\href{https://arxiv.org/abs/1803.03873}{{\ttfamily 1803.03873}}].

\bibitem{PhysRevD.99.104059}
C.~Conde, C.~Galvis and E.~Larra\~naga, \emph{Properties of the innermost
  stable circular orbit of a spinning particle moving in a rotating
  maxwell-dilaton black hole background},
  \href{https://doi.org/10.1103/PhysRevD.99.104059}{\emph{Phys. Rev. D}
  {\bfseries 99} (2019) 104059}.

\bibitem{2018mgm..conf.3715J}
P.I.~{Jefremov}, O.Y.~{Tsupko} and G.S.~{Bisnovatyi-Kogan}, \emph{{Spin-induced
  changes in the parameters of ISCO in Kerr spacetime}},  in \emph{Fourteenth
  Marcel Grossmann Meeting - MG14}, M.~{Bianchi}, R.T.~{Jansen} and
  R.~{Ruffini}, eds., pp.~3715--3721, Jan., 2018,
  \href{https://doi.org/10.1142/9789813226609\_0486}{DOI}.

\bibitem{101142S0218271820501217}
E.~Larrañaga, \emph{Circular motion and the innermost stable circular orbit
  for spinning particles around a charged hayward black hole background},
  \href{https://doi.org/10.1142/S0218271820501217}{\emph{International Journal
  of Modern Physics D} {\bfseries 29} (2020) 2050121}
  [\href{https://arxiv.org/abs/https://doi.org/10.1142/S0218271820501217}{{\ttfamily
  https://doi.org/10.1142/S0218271820501217}}].

\bibitem{Toshmatov:2020wky}
B.~Toshmatov, O.~Rahimov, B.~Ahmedov and D.~Malafarina, \emph{{Motion of
  spinning particles in non asymptotically flat spacetimes}},
  \href{https://doi.org/10.1140/epjc/s10052-020-8254-6}{\emph{Eur. Phys. J. C}
  {\bfseries 80} (2020) 675}
  [\href{https://arxiv.org/abs/2003.09227}{{\ttfamily 2003.09227}}].

\bibitem{Toshmatov:2019bda}
B.~Toshmatov and D.~Malafarina, \emph{{Spinning test particles in the $\gamma$
  spacetime}}, \href{https://doi.org/10.1103/PhysRevD.100.104052}{\emph{Phys.
  Rev. D} {\bfseries 100} (2019) 104052}
  [\href{https://arxiv.org/abs/1910.11565}{{\ttfamily 1910.11565}}].

\bibitem{Nucamendi:2019qsn}
U.~Nucamendi, R.~Becerril and P.~Sheoran, \emph{{Bounds on spinning particles
  in their innermost stable circular orbits around rotating braneworld black
  hole}}, \href{https://doi.org/10.1140/epjc/s10052-019-7584-8}{\emph{Eur.
  Phys. J. C} {\bfseries 80} (2020) 35}
  [\href{https://arxiv.org/abs/1910.00156}{{\ttfamily 1910.00156}}].

\bibitem{Zhang:2020qew}
Y.-P.~Zhang, S.-W.~Wei and Y.-X.~Liu, \emph{{Spinning Test Particle in
  Four-Dimensional Einstein\textendash{}Gauss\textendash{}Bonnet Black Holes}},
  \href{https://doi.org/10.3390/universe6080103}{\emph{Universe} {\bfseries 6}
  (2020) 103} [\href{https://arxiv.org/abs/2003.10960}{{\ttfamily
  2003.10960}}].

\bibitem{Armaza:2015eha}
C.~Armaza, M.~Ba\~nados and B.~Koch, \emph{{Collisions of spinning massive
  particles in a Schwarzschild background}},
  \href{https://doi.org/10.1088/0264-9381/33/10/105014}{\emph{Class. Quant.
  Grav.} {\bfseries 33} (2016) 105014}
  [\href{https://arxiv.org/abs/1510.01223}{{\ttfamily 1510.01223}}].

\bibitem{Zhang:2016btg}
Y.-P.~Zhang, B.-M.~Gu, S.-W.~Wei, J.~Yang and Y.-X.~Liu, \emph{{Charged
  spinning black holes as accelerators of spinning particles}},
  \href{https://doi.org/10.1103/PhysRevD.94.124017}{\emph{Phys. Rev. D}
  {\bfseries 94} (2016) 124017}
  [\href{https://arxiv.org/abs/1608.08705}{{\ttfamily 1608.08705}}].

\bibitem{Zaslavskii:2016dfh}
O.B.~Zaslavskii, \emph{{Schwarzschild black hole as particle accelerator of
  spinning particles}},
  \href{https://doi.org/10.1209/0295-5075/114/30003}{\emph{EPL} {\bfseries 114}
  (2016) 30003} [\href{https://arxiv.org/abs/1603.09353}{{\ttfamily
  1603.09353}}].

\bibitem{Guo:2016vbt}
M.~Guo and S.~Gao, \emph{{Kerr black holes as accelerators of spinning test
  particles}}, \href{https://doi.org/10.1103/PhysRevD.93.084025}{\emph{Phys.
  Rev. D} {\bfseries 93} (2016) 084025}
  [\href{https://arxiv.org/abs/1602.08679}{{\ttfamily 1602.08679}}].

\bibitem{Zhang:2020cpu}
M.~Zhang and J.~Jiang, \emph{{Revisiting collisional Penrose processes in terms
  of escape probabilities for spinning particles}},
  \href{https://doi.org/10.1103/PhysRevD.102.044050}{\emph{Phys. Rev. D}
  {\bfseries 102} (2020) 044050}
  [\href{https://arxiv.org/abs/2008.05696}{{\ttfamily 2008.05696}}].

\bibitem{Sheoran:2020kmn}
P.~Sheoran, H.~Nandan, E.~Hackmann, U.~Nucamendi and A.~Abebe,
  \emph{{Schwarzschild black hole surrounded by quintessential matter field as
  an accelerator for spinning particles}},
  \href{https://doi.org/10.1103/PhysRevD.102.064046}{\emph{Phys. Rev. D}
  {\bfseries 102} (2020) 064046}
  [\href{https://arxiv.org/abs/2006.15286}{{\ttfamily 2006.15286}}].

\bibitem{Yuan:2019dih}
X.~Yuan, Y.~Liu and X.~Zhang, \emph{{Collision of spinning particles near BTZ
  black holes}},
  \href{https://doi.org/10.1088/1674-1137/44/6/065104}{\emph{Chin. Phys. C}
  {\bfseries 44} (2020) 065104}
  [\href{https://arxiv.org/abs/1912.13177}{{\ttfamily 1912.13177}}].

\bibitem{Liu:2019wvp}
Y.~Liu and X.~Zhang, \emph{{Maximal efficiency of the collisional Penrose
  process with spinning particles in Kerr-Sen black hole}},
  \href{https://doi.org/10.1140/epjc/s10052-019-7605-7}{\emph{Eur. Phys. J. C}
  {\bfseries 80} (2020) 31} [\href{https://arxiv.org/abs/1910.01872}{{\ttfamily
  1910.01872}}].

\bibitem{Okabayashi:2019wjs}
K.~Okabayashi and K.-i.~Maeda, \emph{{Maximal efficiency of the collisional
  Penrose process with a spinning particle. II. Collision with a particle on
  the innermost stable circular orbit}},
  \href{https://doi.org/10.1093/ptep/ptz143}{\emph{PTEP} {\bfseries 2020}
  (2020) 013E01} [\href{https://arxiv.org/abs/1907.07126}{{\ttfamily
  1907.07126}}].

\bibitem{Zhang:2018ocv}
S.~Zhang, Y.~Liu and X.~Zhang, \emph{{Kerr-de Sitter and Kerr-anti-de Sitter
  black holes as accelerators for spinning particles}},
  \href{https://doi.org/10.1103/PhysRevD.99.064022}{\emph{Phys. Rev. D}
  {\bfseries 99} (2019) 064022}
  [\href{https://arxiv.org/abs/1812.10702}{{\ttfamily 1812.10702}}].

\bibitem{Zhang:2018gpn}
M.~Zhang, J.~Jiang, Y.~Liu and W.-B.~Liu, \emph{{Collisional Penrose process of
  charged spinning particles}},
  \href{https://doi.org/10.1103/PhysRevD.98.044006}{\emph{Phys. Rev. D}
  {\bfseries 98} (2018) 044006}.

\bibitem{Maeda:2018hfi}
K.-I.~Maeda, K.~Okabayashi and H.~Okawa, \emph{{Maximal efficiency of the
  collisional Penrose process with spinning particles}},
  \href{https://doi.org/10.1103/PhysRevD.98.064027}{\emph{Phys. Rev. D}
  {\bfseries 98} (2018) 064027}
  [\href{https://arxiv.org/abs/1804.07264}{{\ttfamily 1804.07264}}].

\bibitem{Liu:2018myg}
Y.~Liu and W.-B.~Liu, \emph{{Energy extraction of a spinning particle via the
  super Penrose process from an extremal Kerr black hole}},
  \href{https://doi.org/10.1103/PhysRevD.97.064024}{\emph{Phys. Rev. D}
  {\bfseries 97} (2018) 064024}.

\bibitem{Mukherjee:2018kju}
S.~Mukherjee, \emph{{Collisional Penrose process with spinning particles}},
  \href{https://doi.org/10.1016/j.physletb.2018.01.003}{\emph{Phys. Lett. B}
  {\bfseries 778} (2018) 54}.

\bibitem{Zalaquett:2014eia}
N.~Zalaquett, S.A.~Hojman and F.A.~Asenjo, \emph{{Spinning massive test
  particles in cosmological and general static spherically symmetric
  spacetimes}},
  \href{https://doi.org/10.1088/0264-9381/31/8/085011}{\emph{Class. Quant.
  Grav.} {\bfseries 31} (2014) 085011}
  [\href{https://arxiv.org/abs/1308.4435}{{\ttfamily 1308.4435}}].

\bibitem{Zhang:2018omr}
Y.-P.~Zhang, S.-W.~Wei, P.~Amaro-Seoane, J.~Yang and Y.-X.~Liu, \emph{{Motion
  deviation of test body induced by spin and cosmological constant in extreme
  mass ratio inspiral binary system}},
  \href{https://arxiv.org/abs/1812.06345}{{\ttfamily 1812.06345}}.

\bibitem{Chakraborty:2018tvy}
C.~Chakraborty and P.~Majumdar, \emph{{Spinning Gyroscope in an Acoustic Black
  Hole : Precession Effects and Observational Aspects}},
  \href{https://doi.org/10.1140/epjc/s10052-020-8060-1}{\emph{Eur. Phys. J. C}
  {\bfseries 80} (2020) 493}
  [\href{https://arxiv.org/abs/1811.08349}{{\ttfamily 1811.08349}}].

\bibitem{Deriglazov:2017jub}
A.A.~Deriglazov and W.~Guzm\'an~Ram\'\i{}rez, \emph{{Recent progress on the
  description of relativistic spin: vector model of spinning particle and
  rotating body with gravimagnetic moment in General Relativity}},
  \href{https://doi.org/10.1155/2017/7397159}{\emph{Adv. Math. Phys.}
  {\bfseries 2017} (2017) 7397159}
  [\href{https://arxiv.org/abs/1710.07135}{{\ttfamily 1710.07135}}].

\bibitem{Chakraborty:2016mhx}
C.~Chakraborty, M.~Patil, P.~Kocherlakota, S.~Bhattacharyya, P.S.~Joshi and
  A.~Kr\'olak, \emph{{Distinguishing Kerr naked singularities and black holes
  using the spin precession of a test gyro in strong gravitational fields}},
  \href{https://doi.org/10.1103/PhysRevD.95.084024}{\emph{Phys. Rev. D}
  {\bfseries 95} (2017) 084024}
  [\href{https://arxiv.org/abs/1611.08808}{{\ttfamily 1611.08808}}].

\bibitem{Hojman:2016mox}
S.A.~Hojman and F.A.~Asenjo, \emph{{Spinning particles coupled to gravity and
  the validity of the universality of free fall}},
  \href{https://doi.org/10.1088/1361-6382/aa6ca2}{\emph{Class. Quant. Grav.}
  {\bfseries 34} (2017) 115011}
  [\href{https://arxiv.org/abs/1610.08719}{{\ttfamily 1610.08719}}].

\bibitem{Deriglazov:2016mhk}
A.A.~Deriglazov and A.M.~Pupasov-Maksimov, \emph{{Relativistic corrections to
  the algebra of position variables and spin-orbital interaction}},
  \href{https://doi.org/10.1016/j.physletb.2016.08.034}{\emph{Phys. Lett. B}
  {\bfseries 761} (2016) 207}
  [\href{https://arxiv.org/abs/1609.00043}{{\ttfamily 1609.00043}}].

\bibitem{Deriglazov:2015wde}
A.A.~Deriglazov and W.G.~Ram\'\i{}rez, \emph{{Ultrarelativistic Spinning
  Particle and a Rotating Body in External Fields}},
  \href{https://doi.org/10.1155/2016/1376016}{\emph{Adv. High Energy Phys.}
  {\bfseries 2016} (2016) 1376016}
  [\href{https://arxiv.org/abs/1511.00645}{{\ttfamily 1511.00645}}].

\bibitem{Jafarzade_2021}
K.~Jafarzade, M.~Kord~Zangeneh and F.S.~Lobo, \emph{Shadow, deflection angle
  and quasinormal modes of born-infeld charged black holes},
  \href{https://doi.org/10.1088/1475-7516/2021/04/008}{\emph{Journal of
  Cosmology and Astroparticle Physics} {\bfseries 2021} (2021) 008}.

\bibitem{Dadhich07gb}
N.~{Dadhich}, \emph{{On the Gauss-Bonnet Gravity}},  in \emph{Mathematical
  Physics}, M.J.~{Aslam}, F.~{Hussain}, A.~{Qadir}, {Riazuddin} and
  H.~{Saleem}, eds., pp.~331--340, Apr., 2007,
  \href{https://doi.org/10.1142/9789812770523_0032}{DOI}.

\bibitem{Torii05}
T.~{Torii} and H.~{Maeda}, \emph{{Spacetime structure of static solutions in
  Gauss-Bonnet gravity: Neutral case}},
  \href{https://doi.org/10.1103/PhysRevD.71.124002}{\emph{Phys. Rev. D}
  {\bfseries 71} (2005) 124002}
  [\href{https://arxiv.org/abs/hep-th/0504127}{{\ttfamily hep-th/0504127}}].

\bibitem{Yang20b}
S.-J.~{Yang}, J.-J.~{Wan}, J.~{Chen}, J.~{Yang} and Y.-Q.~{Wang}, \emph{{Weak
  cosmic censorship conjecture for the novel 4D charged Einstein-Gauss-Bonnet
  black hole with test scalar field and particle}},
  \href{https://doi.org/10.1140/epjc/s10052-020-08511-9}{\emph{Eur. Phys. J. C}
  {\bfseries 80} (2020) 937}
  [\href{https://arxiv.org/abs/2004.07934}{{\ttfamily 2004.07934}}].

\bibitem{Misner73}
C.W.~Misner, K.S.~Thorne and J.A.~Wheeler, \emph{Gravitation}, W. H. Freeman,
  San Francisco (1973).

\bibitem{Baushev:2008yz}
A.N.~Baushev, \emph{{Dark matter annihilation in the gravitational field of a
  black hole}}, \href{https://doi.org/10.1142/S0218271809014509}{\emph{Int. J.
  Mod. Phys. D} {\bfseries 18} (2009) 1195}
  [\href{https://arxiv.org/abs/0805.0124}{{\ttfamily 0805.0124}}].

\bibitem{Grib:2010bs}
A.A.~Grib and Y.V.~Pavlov, \emph{{On Particle Collisions and Extraction of
  Energy from a Rotating Black Hole}},
  \href{https://doi.org/10.1134/S0021364010150014}{\emph{JETP Lett.} {\bfseries
  92} (2010) 125} [\href{https://arxiv.org/abs/1004.0913}{{\ttfamily
  1004.0913}}].

\bibitem{Mathisson:1937zz}
M.~Mathisson, \emph{{Neue mechanik materieller systemes}}, {\emph{Acta Phys.
  Polon.} {\bfseries 6} (1937) 163}.

\bibitem{Papapetrou:1951pa}
A.~Papapetrou, \emph{{Spinning test particles in general relativity. 1.}},
  \href{https://doi.org/10.1098/rspa.1951.0200}{\emph{Proc. Roy. Soc. Lond. A}
  {\bfseries 209} (1951) 248}.

\bibitem{Corinaldesi:1951pb}
E.~Corinaldesi and A.~Papapetrou, \emph{{Spinning test particles in general
  relativity. 2.}}, \href{https://doi.org/10.1098/rspa.1951.0201}{\emph{Proc.
  Roy. Soc. Lond. A} {\bfseries 209} (1951) 259}.

\bibitem{Tulczyjew:1959a}
W.~Tulczyjew, \emph{{Motion of multipole particles in general relativity
  theory}}, {\emph{Acta Phys.Polon.} {\bfseries 18} (1959) 393}.

\bibitem{doi:10.1063/1.1704055}
A.H.~Taub, \emph{Motion of test bodies in general relativity},
  \href{https://doi.org/10.1063/1.1704055}{\emph{Journal of Mathematical
  Physics} {\bfseries 5} (1964) 112}
  [\href{https://arxiv.org/abs/https://doi.org/10.1063/1.1704055}{{\ttfamily
  https://doi.org/10.1063/1.1704055}}].

\bibitem{Dixon:1964aa}
W.G.~Dixon, \emph{{}}, {\emph{Riv. Nuovo Cimento Soc. Ital. Fis.} {\bfseries
  34} (1964) 317}.

\bibitem{Hanson:1974qy}
A.J.~Hanson and T.~Regge, \emph{{The Relativistic Spherical Top}},
  \href{https://doi.org/10.1016/0003-4916(74)90046-3}{\emph{Annals Phys.}
  {\bfseries 87} (1974) 498}.

\bibitem{Hojmanphdthesis:1975}
S.A.~Hojman, \emph{Electromagnetic and Gravitational Interactions of a
  Spherical Relativistic Top}, Ph.D. thesis, Princeton University, 1975,
  unpublished.

\bibitem{Hojman:1976kn}
R.~Hojman and S.~Hojman, \emph{{Spinning Charged Test Particles in a
  Kerr-Newman Background}},
  \href{https://doi.org/10.1103/PhysRevD.15.2724}{\emph{Phys. Rev. D}
  {\bfseries 15} (1977) 2724}.

\bibitem{Costa:2014nta}
L.F.O.~Costa and J.~Nat\'ario, \emph{{Center of mass, spin supplementary
  conditions, and the momentum of spinning particles}},
  \href{https://doi.org/10.1007/978-3-319-18335-0_6}{\emph{Fund. Theor. Phys.}
  {\bfseries 179} (2015) 215}
  [\href{https://arxiv.org/abs/1410.6443}{{\ttfamily 1410.6443}}].

\end{thebibliography}\endgroup

\end{document}